\documentclass{JFM-FLM_Au}

\usepackage{xcolor}
\usepackage{amsmath}
\usepackage{siunitx}
\usepackage{microtype}
\usepackage{placeins}

\makeatletter
\providecommand{\headerps@out}[1]{}

\def\preprint@pagenumber{%
  \hbox to \textwidth{\hfil\normalfont\thepage\hfil}%
}
\def\ps@headings{%
  \let\@mkboth\markboth
  \def\@oddhead{\hfill{\itshape\@righttitle}\hfill}%
  \def\@evenhead{\hfill{\itshape\@lefttitle}\hfill}%
  \def\@oddfoot{\preprint@pagenumber}%
  \def\@evenfoot{\preprint@pagenumber}%
  \def\sectionmark##1{\markboth{##1}{}}%
  \def\subsectionmark##1{\markright{##1}}%
}
\def\ps@titlepage{%
  \leftskip\z@
  \let\@mkboth\@gobbletwo
  \def\@oddhead{}%
  \def\@evenhead{}%
  \def\@oddfoot{\preprint@pagenumber}%
  \def\@evenfoot{\preprint@pagenumber}%
  \def\sectionmark##1{}%
  \def\subsectionmark##1{}%
}

\def\@maketitle{%
  \newpage
  \vspace*{10\p@}%
  {\flushleft
    {\titlefont\@title\par}%
    \vspace*{23.5\p@}%
    {\normalfont\authorfont\fontswitch\bfseries\baselineskip=12\p@
      \lowercase{\@author}\par}%
    \vspace*{2\p@}%
    {\normalfont\small\@affiliation\par\vspace*{2\p@}}%
    \ifx\@corresau\empty\else
      {\normalfont\small\@corresau\vskip7\p@}%
    \fi
  \par}%
}
\makeatother

\hypersetup{
  linkcolor=blue,
  pdftitle={Revisit the simplified lattice Boltzmann method: dissipation, dispersion and stability},
  pdfauthor={Zhengwei He and Zhen Chen},
  pdfsubject={arXiv preprint},
  pdfkeywords={Computational methods; Kinetic theory}
}

\graphicspath{{figures/}}
\DeclareGraphicsExtensions{.pdf}
\newcommand{\figref}[1]{figure~\ref{#1}}
\newcommand{\figpanelref}[2]{figure~\ref{#1}(#2)}
\newcommand{\tabref}[1]{table~\ref{#1}}
\newcommand{\eqnref}[1]{Eq.~\eqref{#1}}
\newcommand{\eqnsref}[2]{Eqs.~\eqref{#1}--\eqref{#2}}
\newcommand{\secref}[1]{\hyperref[#1]{Section~\ref*{#1}}}
\newcommand{\secrefs}[2]{Sections~\hyperref[#1]{\ref*{#1}} and~\hyperref[#2]{\ref*{#2}}}
\newcommand{\appref}[1]{\hyperref[#1]{Appendix~\ref*{#1}}}
\newcommand{\apprefs}[2]{Appendices~\hyperref[#1]{\ref*{#1}} and~\hyperref[#2]{\ref*{#2}}}

\lefttitle{Z. He and Z. Chen}
\righttitle{Revisit the simplified lattice Boltzmann method}
\title{Revisit the simplified lattice Boltzmann method: dissipation, dispersion and stability}
\author{Zhengwei He\aff{1,2} \and Zhen Chen\aff{1,2}}
\affiliation{\aff{1}Marine Numerical Experiment Center, State Key Laboratory of Ocean Engineering, Shanghai 200240, China
\aff{2}School of Naval Architecture, Ocean and Civil Engineering, Shanghai Jiao Tong University, Shanghai 200240, China}
\corresau{Zhen Chen, \email{zhen.chen@sjtu.edu.cn}}

\begin{document}

\maketitle

\begin{abstract}
The simplified lattice Boltzmann method (SLBM) is a recent development in the lattice Boltzmann method (LBM) community, addressing the intrinsic limitations of the traditional LBM by directly evolving macroscopic quantities and maintaining numerical stability in high-Reynolds-number simulations. However, fundamental understanding of the numerical dissipation and dispersion of SLBM is still lacking, and the origin of its good numerical stability is not fully resolved. In this work, a generalized formulation is developed, revealing that the SLBM recovers modified macroscopic equations containing both intrinsic physical deviations and numerical truncation errors. To remove these deviations, the macroscopic equation derived from the standard Bhatnagar--Gross--Krook lattice Boltzmann method (BGK-LBM) is adopted as a reference model and solved by the predictor-corrector strategy, which constitutes the reformulated SLBM. The proposed method uses the generalized SLBM formulation in the predictor step with tunable high-order parameters, while the corrector step is realized by the finite-difference discretization. Linear stability analysis, together with linear-wave validation, clarifies the roles of these parameters in controlling numerical dissipation and dispersion, which is then examined in more complicated numerical examples. It is demonstrated that the reformulated SLBM preserves the second-order accuracy, improves the dispersion-dissipation performance, remains stable in under-resolved cases, and resolves fine vortex structures on relatively coarse grids. Thus, the proposed method combines improved numerical properties with the simplicity of SLBM, offering a high-fidelity and stable scheme for incompressible flow simulations.
\end{abstract}



\section{Introduction}\label{sec:introduction}
The lattice Boltzmann method (LBM) \citep{mcnamara1988,qian1992,shan2006} is a mesoscopic numerical approach for simulating flow problems \citep{guo2002,li2016,zhang2024,hu2025,kusano2025}. Over decades of development, the numerical characteristics of LBM have been well recognized, and its popularity in the computational mechanics community has been growing due to its kinetic nature, simplicity, and explicitness. However, intrinsic limitations such as high cost of virtual memory and inconvenient implementation of physical boundary conditions remain. In addition, the standard single-relaxation-time (SRT) LBM can be unstable at low lattice viscosity, which can be alleviated by the multiple-relaxation-time (MRT) schemes \citep{lallemand2000}, the entropic LBM (ELBM) \citep{karlin1998} and others.

The simplified lattice Boltzmann method (SLBM) was originally proposed by \citet{chen2017slbm} as a robust alternative for addressing the aforementioned limitations of the standard LBM \citep[see also][]{chen2020}. Unlike the standard LBM, SLBM directly evolves macroscopic quantities using the predictor-corrector strategy, which circumvents the storage of distribution functions, reduces the memory consumption and improves numerical stability \citep{chen2017,chen2018}. Customized solvers were then devised in the SLBM framework to tackle multiple types of flow problems including thermal flows \citep{chen201714,chen201715}, multiphase flows \citep{chen201816}, phase-change modelling \citep{chen2021}, and non-Newtonian fluids \citep{chen202018}. In the meantime, to accommodate complex geometries, SLBM has been integrated with non-uniform grids \citep{chen201819} and immersed boundary methods \citep{chen201820}. Furthermore, owing to its numerical stability and memory-efficient structure, SLBM has been extended to a broad range of physical problems. Representative applications include magnetohydrodynamics (MHD) \citep{derosis2021}, ferrofluid droplet deformation \citep{khan2020}, nanofluid convection \citep{ma2020}, and GPU-accelerated implementations \citep{delgadogutierrez2020}.

Over the past decade, algorithmic developments of the SLBM have followed two complementary directions: incorporating additional physics and refining the time-marching procedure to improve efficiency. On the physical side, \citet{dai2020} introduced an enthalpy-based version (SELBM) with an entropy-based source term for phase-change simulations \citep{dai2024}, while \citet{gao2021} reformulated the equilibrium distribution to achieve a strictly incompressible formulation (IC-SHSLBM). Concurrently, efforts to streamline time integration have produced a family of single-step methods, including the SS-SLBM for turbulence proposed by \citet{delgadogutierrez2021} and its extension to 3D MHD \citep{derosis202129}, as well as the one-step methods (OSLBM) proposed by \citet{qin2022,qin2023} and its improved version NOSLBM for large-density-ratio multiphase flows \citep{qin202332}.

Despite its well-documented numerical stability presented by the above works and previous studies \citep{chen2017,chen2018,he2025}, numerical intuition suggests that the improved stability could come at the cost of extra dissipation. Some practices have confirmed this, and techniques like local mesh refinement \citep{jiang2022,he2024} or high-order interpolation (HSLBM \citep{chen201836}) are required to recover delicate flow structures on comparable mesh sizes. Several attempts have been made to restore this trade-off. \citet{chen202037} reduced numerical dissipation by reformulating the predictor step, though at the expense of stability. \citet{gao202138} proposed a consistent forcing scheme, in which the predictor-corrector equations are reconstructed based on the GZS forcing model in LBM \citep{guo200239}. The resulting scheme enables SLBM to interpret external forcing effects with better accuracy, especially in the circumstances with non-uniform forcing fields. They further introduced an additional correction step \citep{gao202140} (CSLBM) to enhance accuracy. In the context of multiphase flow simulations, \citet{li2023} modified the phase-field solver by solving the Allen-Cahn equation instead of the Cahn-Hilliard equation in the phase-field model of \citet{chen201816}, which improves the interfacial sharpness.

While these approaches differ in implementation, one important insight emerges: numerical performance of the SLBM is fundamentally associated with the macroscopic equations it reconstructs. This viewpoint was further supported by \citet{lu2021}, who showed that the SLBM essentially reconstructs solutions to the macroscopic equations with additional terms responsible for its strong numerical stability. These investigations suggest that improving SLBM accuracy without extra mesh refinement or high-order techniques requires a deeper understanding of the SLBM's equivalent macroscopic equations and their numerical implications.

In this paper, we present a generalized formulation that enables a more comprehensive theoretical analysis of the SLBM by recasting it as a specific reconstruction of the macroscopic equations. Using macroscopic equations derived from the Bhatnagar--Gross--Krook lattice Boltzmann method (BGK-LBM) \citep{lu2020} as a reference baseline, we show that the SLBM solves the equations containing both intrinsic physical deviations and numerical truncation errors. Motivated by this finding, we replace the macroscopic equations reconstructed by the original SLBM with the physically consistent reference model and propose a new predictor-corrector scheme. The predictor step is reconstructed from equilibrium moments via the generalized SLBM, while the corrector step employs the finite-difference discretization. The equilibrium-based reconstruction introduces tunable parameters associated with high-order derivatives in the macroscopic equations, allowing flexible adjustments of numerical dissipation, dispersion, and stability. Systematic linear stability analysis, linear-wave validation and numerical benchmarks demonstrate the advantages of the present method over the original SLBM and validate the dissipation inherent in the original SLBM. In general, the proposed method provides a systematic framework for constructing equilibrium-based predictor-corrector schemes that retains the low memory cost, macroscopic boundary treatment and good numerical stability of the SLBM while largely reducing numerical dissipation.

The remainder of this paper is organized as follows. \secref{sec:methodology} reviews the principles of the SLBM with theoretical analysis and derives the proposed method with mathematical details. \secref{sec:linear_stability_analysis} presents the linear stability analysis and validates the shear-mode predictions using linear-wave simulations. \secref{sec:numerical_tests} examines how these linear predictions manifest in nonlinear benchmarks and demonstrates the performance of the new method. Finally, conclusions are drawn in \secref{sec:conclusion}.


\section{Methodology}\label{sec:methodology}

\subsection{Revisit of the simplified lattice Boltzmann method}\label{sec:slbm_revisit}

\subsubsection{The simplified lattice Boltzmann method (SLBM)}\label{sec:slbm_original}

 The SLBM \citep{chen2017slbm} reformulates the conventional lattice Boltzmann method by directly evolving macroscopic variables through a predictor-corrector scheme, where only equilibrium distribution functions are involved. For isothermal D2Q9 flows, the formulations of SLBM can be written as follows:

Predictor:
\begin{subequations}
\begin{align}
\rho^\ast
  &=\sum_i f_i^{\mathrm{eq}}(\boldsymbol{r}-\boldsymbol{e}_i\delta t,t-\delta t),
\\
(\rho\boldsymbol{u})^\ast
  &=\sum_i\boldsymbol{e}_i f_i^{\mathrm{eq}}(\boldsymbol{r}-\boldsymbol{e}_i\delta t,t-\delta t).
\end{align}
\end{subequations}
Corrector:
\begin{subequations}
\begin{align}
\rho
  &=\rho^\ast,
\\
\rho\boldsymbol{u}
  &= (\rho\boldsymbol{u})^\ast
   +(\tau-1)\left[
\sum_i\boldsymbol{e}_i f_i^{\mathrm{eq}}(\boldsymbol{r}+\boldsymbol{e}_i\delta t,t)
      -\rho(\boldsymbol{r},t-\delta t)\boldsymbol{u}(\boldsymbol{r},t-\delta t)
\right].
\end{align}
\end{subequations}
Here $\boldsymbol{r}$ and $t$ denote the spatial location and time level, $\rho$ and $\boldsymbol{u}$ represent the density and the velocity vector, respectively; $\delta t$ is the streaming time step. The relaxation parameter $\tau$ is related to the kinematic viscosity $\nu$ by
\begin{equation}
\nu = c_s^2\left(\tau-\frac{1}{2}\right)\delta t.
\end{equation}
The equilibrium distribution is
\begin{equation}
f_i^{\mathrm{eq}}(\boldsymbol{r},t)
  = w_i\rho\left[
      1+\frac{\boldsymbol{e}_i\cdot\boldsymbol{u}}{c_s^2}
      +\frac{(\boldsymbol{e}_i\cdot\boldsymbol{u})^2}{2c_s^4}
      -\frac{\boldsymbol{u}\cdot\boldsymbol{u}}{2c_s^2}
\right],
\end{equation}
with weights $w_0=4/9$, $w_i=1/9$ for $i=1,\ldots,4$, and $w_i=1/36$ for $i=5,\ldots,8$. The sound speed is $c_s=(\delta x/\delta t)/\sqrt{3}$, and the D2Q9 velocities consist of one rest velocity, four axial velocities $c(\pm1,0)$ and $c(0,\pm1)$, and four diagonal velocities $c(\pm1,\pm1)$, where $c=\delta x/\delta t$.

In summary, SLBM evolves the macroscopic variables through a predictor-corrector scheme in which both steps are constructed using equilibrium distribution functions within the lattice Boltzmann framework. This formulation preserves the simplicity of lattice-based schemes and provides a concise and memory-efficient numerical procedure, which serves as the baseline for the method developed in the following subsections.

\subsubsection{The generalized SLBM formulation and theoretical analysis}\label{sec:gslbm_theory}

Although the SLBM is simple and memory efficient, its robustness cannot be understood from the algorithmic form alone. A deeper theoretical understanding is required to clarify the origin of its numerical performance. We therefore use a generalized SLBM formulation to identify the macroscopic equation reconstructed by the original SLBM and to separate physical deviations from numerical truncation errors.

The generalized formulation abstracts the equilibrium-moment reconstruction used by SLBM from the particular coefficients and stencil of the original scheme. For clarity, we focus on the momentum equation. The generalized macroscopic momentum update is written as
\begin{equation}
(\rho u_\alpha)_{\mathrm{GSLBM}}(\boldsymbol{r},t)
  = g_{m,\alpha}+M g^\ast_{m,\alpha},
\end{equation}
where $g_{m,\alpha}$ and $g^\ast_{m,\alpha}$ are generalized predictor and corrector operators, respectively. They are linear combinations of equilibrium distribution functions with free coefficients; their explicit forms and coefficient definitions are given in \appref{app:gslbm_derivation}. Substituting the coefficients of the original SLBM into this framework and using Taylor expansion gives the following compact classification of the recovered momentum equation:
\begin{equation}
\left\{
\begin{aligned}
\partial_t(\rho u_\alpha)
={}&\mathcal{M}_\alpha^{\mathrm{MAME}}
  +\mathcal{D}_\alpha^{\mathrm{phys}}
  +\mathcal{E}_\alpha^{\mathrm{num}},\\
\mathcal{M}_\alpha^{\mathrm{MAME}}
={}&-\partial_\beta\Pi_{\alpha\beta}^{\mathrm{eq}}
 +\frac{\nu_{\mathrm{phy}}}{c_s^2}
\partial_\beta\partial_\gamma\Pi_{\alpha\beta\gamma}^{\mathrm{eq}}
 +\left(\frac{\nu_{\mathrm{phy}}}{c_s^2}-\frac{1}{2}\delta t\right)
\partial_t\partial_\beta\Pi_{\alpha\beta}^{\mathrm{eq}},\\
\mathcal{D}_\alpha^{\mathrm{phys}}
={}&\frac{\nu_{\mathrm{virtual}}}{c_s^2}
\partial_\beta\partial_\gamma\Pi_{\alpha\beta\gamma}^{\mathrm{eq}}
+\left(\frac{\nu_{\mathrm{phy}}}{c_s^2}-\frac{1}{2}\delta t\right)\mathcal{T}_{\alpha}^{\ast} .
\end{aligned}
\right.
\label{eq:slbm_error_decomposition}
\end{equation}
Here $\mathcal{M}_\alpha^{\mathrm{MAME}}$ is the physically consistent more-actual macroscopic equation (MAME) derived from the BGK-LBE \citep{lu2020}, $\mathcal{D}_\alpha^{\mathrm{phys}}$ contains the physical deviations of the original SLBM, and $\mathcal{E}_\alpha^{\mathrm{num}}$ collects the remaining numerical-error terms. The full algebraic expansion, including the second- and third-derivative corrector residuals contained in $\mathcal{E}_\alpha^{\mathrm{num}}$, is given in \apprefs{app:gslbm_derivation}{app:slbm_analysis}. The superscript $*$ denotes quantities associated with the predictor step.

The first physical deviation in $\mathcal{D}_\alpha^{\mathrm{phys}}$ can be interpreted as a predictor-level virtual viscosity. The effective viscosity reconstructed by the SLBM predictor satisfies
\begin{equation}
\nu_{\mathrm{eff}}=\nu_{\mathrm{phy}}+\nu_{\mathrm{virtual}},
\qquad
\nu_{\mathrm{virtual}}
  = c_s^2(1-\tau)\delta t.
\end{equation}
Thus $\nu_{\mathrm{virtual}}>0$ for $\tau<1$ and $\nu_{\mathrm{virtual}}<0$ for $\tau>1$, indicating that the original SLBM predictor is respectively more dissipative or anti-dissipative than the MAME-consistent predictor. This interpretation is only stage-level: $\nu_{\mathrm{virtual}}$ is not the leading viscosity of the complete predictor--corrector SLBM update. As shown later by the linear stability analysis, the corrector compensates the leading long-wave effect of this predictor-level deviation, while finite-wavenumber residuals remain responsible for the observed damping and robustness trends, consistent with previous observations and analyses \citep{he2025}.

The second physical deviation is the term proportional to $\mathcal{T}_{\alpha}^{\ast}$ in $\mathcal{D}_\alpha^{\mathrm{phys}}$, which measures the difference between the MAME-consistent predictor and the predictor used in the original SLBM. These two terms are therefore inconsistencies with the target macroscopic model rather than conventional discretization errors. The distinction is made at the equation-reconstruction level rather than the discretization level.

Beyond these physical deviations, $\mathcal{E}_\alpha^{\mathrm{num}}$ contains numerical errors inherent to the generalized SLBM reconstruction. In the predictor step, the original SLBM yields nonzero higher-order coefficients $t_3$ and $t_4$ in \appref{app:gslbm_derivation}, corresponding to third- and fourth-order spatial derivatives. Additional corrector errors contain second- and fourth-order derivative residuals that cannot be removed, reflecting intrinsic structural limitations of the equilibrium-based predictor--corrector formulation. Their orders with respect to $\delta x$ must be assessed after the physical-deviation and numerical-error terms in the complete recovered equation are combined and the scaling strategy is specified, as shown in \appref{app:slbm_error_scaling}.

It is therefore inappropriate to attribute the numerical behaviour of the SLBM solely to the predictor-level virtual viscosity. The corrector residual partially offsets the leading predictor deviation, leaving a $O(\delta t)$ residual (see \appref{app:slbm_analysis_slbm})
\begin{equation}
\frac{1}{2}\partial_\beta\partial_\gamma
\left(
\Pi_{\alpha\beta\gamma}^{\mathrm{eq},\ast,\mathrm{SLBM}}
  -\Pi_{\alpha\beta\gamma}^{\mathrm{eq}}
\right)
\propto O(\delta t).
\end{equation}
However, this relation does not by itself determine how the complete contribution varies with $\delta x$. That assessment must also include the prefactors multiplying this residual and the specific scaling strategy \citep{kruger2017}, as discussed in \appref{app:gslbm_derivation} and \appref{app:slbm_error_scaling}. This compensation explains why the original SLBM often recovers the correct leading-order flow behaviour despite the predictor-level inconsistency. At finite wavenumbers, however, the compensation is not exact, leaving residual damping or over-correction effects that will be quantified by the linear stability analysis in \secref{sec:linear_stability_analysis}.

This classification differs from the analysis of \citet{lu2021}, which attributed the SLBM dissipation primarily to numerical discretization effects. Here the approximation $\Phi^\ast\approx\Phi+(\partial\Phi/\partial t)\delta t+O(\delta t^2)$ is not invoked; instead, the spatial Taylor expansion is applied to the full predictor-corrector structure, allowing the virtual viscosity to be explicitly identified and separated from numerical errors.

Finally, a particularly illuminating situation arises in the special case of $\tau=1$. In this condition, the virtual viscosity and the leading low-order truncation errors vanish, and the predictor-corrector formulation degenerates to a second-order single-step update. Consequently, the original SLBM converges to the MAME at this order and becomes formally identical to the standard BGK-LBM, which elucidates the theoretical connections among the SLBM (and the SS-SLBM), the MAME, and the BGK-LBM (see \appref{app:slbm_analysis}).

In summary, the numerical stability, dissipation, dispersion, and accuracy of the SLBM are jointly determined by both the physical deviation terms and the numerical error terms. For the same physical problem at fixed $\mathrm{Re}$, the accuracy under mesh refinement depends on the scaling. Under diffusive scaling, where $\tau$ is fixed, the original SLBM is formally second-order accurate in smooth bulk regions. Under acoustic scaling at fixed $\mathrm{Re}$, the original SLBM does not generally retain the strict second-order consistency with the desired macroscopic equations. At this situation, the apparent second-order accuracy often observed in practice could be the result of a consequence that these error terms tend to be small. The detailed order analysis is given in \appref{app:slbm_error_scaling}.

\subsection{Reformulated simplified lattice Boltzmann method (RSLBM)}\label{sec:rslbm}

As discussed in \secref{sec:slbm_revisit}, the simplified lattice Boltzmann method is shown to recover the macroscopic equations that deviate from the physically consistent ones. This observation implies that the employment of physically consistent macroscopic equations would benefit the construction of improved numerical schemes, which constitutes the motivation of the reformulated simplified lattice Boltzmann method proposed in this work. The more actual macroscopic equations (MAME), as shown in \eqnref{eq:mame_continuity} and \eqnref{eq:mame_momentum}, provides such a physically consistent description by remaining directly rooted in the BGK lattice Boltzmann equation. In the present work, the MAME are adopted as the physically consistent target equations.
\begin{subequations}
\begin{align}
\partial_t\rho
  &\approx\frac{\rho(\boldsymbol{r},t)-\rho(\boldsymbol{r},t-\delta t)}{\delta t}
   =-\partial_\alpha(\rho u_\alpha)
    +\frac{1}{2}\delta t\,\partial_\alpha\partial_\beta
      (\rho u_\alpha u_\beta+\rho c_s^2\delta_{\alpha\beta}),
\label{eq:mame_continuity}\\
\partial_t(\rho u_\alpha)
  &\approx
\frac{(\rho u_\alpha)(\boldsymbol{r},t)-(\rho u_\alpha)(\boldsymbol{r},t-\delta t)}{\delta t}
\nonumber\\
  &=-\partial_\beta(\rho u_\alpha u_\beta+\rho c_s^2\delta_{\alpha\beta})
    +\nu\partial_\beta\left[
\partial_\beta(\rho u_\alpha)
      +2\partial_\gamma(\rho u_\gamma)\delta_{\alpha\beta}
\right]\nonumber\\
  &\quad
    +\left(\frac{\nu}{c_s^2}-\frac{1}{2}\delta t\right)
\partial_t\partial_\beta
      (\rho u_\alpha u_\beta+\rho c_s^2\delta_{\alpha\beta}).
\label{eq:mame_momentum}
\end{align}
\end{subequations}
The viscous contribution is written in this form because the D2Q9 second-order equilibrium gives $\Pi_{\alpha\beta\gamma}^{\mathrm{eq}}=\rho c_s^2 (u_\alpha\delta_{\beta\gamma}+u_\beta\delta_{\alpha\gamma}+u_\gamma\delta_{\alpha\beta})$ under the low-Mach-number approximation adopted throughout this work, from which the above second-derivative structure follows. To facilitate time marching, the predictor-corrector approximation of the temporal derivative is adopted. For convenience, this formulation is referred to as the semi-discrete MAME, consisting of the predictor equations \eqnsref{eq:mame_predictor_density}{eq:mame_predictor_momentum} and the corrector equations \eqnsref{eq:mame_corrector_density}{eq:mame_corrector_momentum}:

\noindent \textbf{Predictor:}
\begin{subequations}
\begin{align}
\rho^\ast
  &=\rho^{t-\delta t}
   -\delta t\,\partial_\alpha(\rho u_\alpha)^{t-\delta t}
   +\frac{1}{2}\delta t^2\,\partial_\alpha\partial_\beta
      (\rho u_\alpha u_\beta+\rho c_s^2\delta_{\alpha\beta})^{t-\delta t},
\label{eq:mame_predictor_density}\\
(\rho u_\alpha)^\ast
  &= (\rho u_\alpha)^{t-\delta t}
   -\delta t\,\partial_\beta
      (\rho u_\alpha u_\beta+\rho c_s^2\delta_{\alpha\beta})^{t-\delta t}
\nonumber\\
  &\quad
   +\nu\delta t \partial_\beta
\left[
\partial_\beta(\rho u_\alpha)
      +2\partial_\gamma(\rho u_\gamma)\delta_{\alpha\beta}
\right]^{t-\delta t}.
\label{eq:mame_predictor_momentum}
\end{align}
\end{subequations}
\textbf{Corrector:}
\begin{subequations}
\begin{align}
\rho(\boldsymbol{r},t)
  &=\rho^\ast,
\label{eq:mame_corrector_density}\\
(\rho u_\alpha)(\boldsymbol{r},t)
  &= (\rho u_\alpha)^\ast
   +\left(\frac{\nu}{c_s^2}-\frac{1}{2}\delta t\right)
\left[
\partial_\beta(\rho u_\alpha u_\beta+\rho c_s^2\delta_{\alpha\beta})^\ast
      -
\partial_\beta(\rho u_\alpha u_\beta+\rho c_s^2\delta_{\alpha\beta})^{t-\delta t}
\right].
\label{eq:mame_corrector_momentum}
\end{align}
\end{subequations}
which can be rewritten into a more concise form:
\begin{subequations}
\begin{align}
\rho_{\mathrm{MAME}}(\boldsymbol{r},t)
  &=\rho^{t-\delta t}
   -\delta t\,\partial_\alpha(\rho u_\alpha)^{t-\delta t}
   +\frac{1}{2}\delta t^2\,\partial_\alpha\partial_\beta
      (\rho u_\alpha u_\beta+\rho c_s^2\delta_{\alpha\beta})^{t-\delta t},
\\
(\rho u_\alpha)_{\mathrm{MAME}}(\boldsymbol{r},t)
  &= (\rho u_\alpha)^{t-\delta t}
   -\delta t\,\partial_\beta
      (\rho u_\alpha u_\beta+\rho c_s^2\delta_{\alpha\beta})^{t-\delta t}
\nonumber\\
  &\quad
   +\nu\delta t \partial_\beta
\left[
\partial_\beta(\rho u_\alpha)
      +2\partial_\gamma(\rho u_\gamma)\delta_{\alpha\beta}
\right]^{t-\delta t}\nonumber\\
  &\quad
   +\left(\frac{\nu}{c_s^2}-\frac{1}{2}\delta t\right)
\left[
\partial_\beta(\rho u_\alpha u_\beta+\rho c_s^2\delta_{\alpha\beta})^{\ast,\mathrm{MAME}}
      -
\partial_\beta(\rho u_\alpha u_\beta+\rho c_s^2\delta_{\alpha\beta})^{t-\delta t}
\right].
\end{align}
\end{subequations}

Here, $\rho_{\mathrm{MAME}}$ and $(\rho u_\alpha)_{\mathrm{MAME}}$ denote the exact solution of semi-discrete MAME without numerical error, and the subscript MAME means that this term is computed by the intermediate values given by the predictor step of the semi-discrete MAME. Owing to its explicitness, this semi-discrete form provides a natural basis for constructing efficient numerical schemes, and it is therefore taken as the reference framework for the present reformulated SLBM.

\subsubsection{Equilibrium-based formulation}\label{sec:rslbm_equilibrium_formulation}

As discussed in \secref{sec:gslbm_theory}, the generalized SLBM formulation is an extension of the SLBM. Since the MAME is chosen as the physically consistent target equations, the objective of this subsection is to recover the semi-discrete MAME with second-order accuracy. In the present method, the generalized SLBM is adopted to reconstruct the predictor step, while the corrector step is essentially a finite-difference discretization.

To this end, we begin by recalling that the equilibrium distribution functions $f_i^{\mathrm{eq}}$ uniquely determine the macroscopic variables and fluxes (i.e. $\rho u_\alpha$, $\rho u_\alpha u_\beta$,\ldots) through moments. In particular, the zeroth-, the first-, and the second-order moments of the equilibrium distribution yield the density, the momentum, and the momentum flux tensor, respectively.
\begin{equation}
\begin{aligned}
\sum_i f_i^{\mathrm{eq}} &=\rho,\\
\sum_i e_{i\alpha} f_i^{\mathrm{eq}} &=\rho u_\alpha,\\
\sum_i e_{i\alpha}e_{i\beta} f_i^{\mathrm{eq}}
  &=\Pi_{\alpha\beta}^{\mathrm{eq}}
   =\rho u_\alpha u_\beta+\rho c_s^2\delta_{\alpha\beta},\\
\sum_i e_{i\alpha}e_{i\beta}e_{i\gamma} f_i^{\mathrm{eq}}
  &=\Pi_{\alpha\beta\gamma}^{\mathrm{eq}}
   =\rho c_s^2
      (u_\alpha\delta_{\beta\gamma}
       +u_\beta\delta_{\alpha\gamma}
       +u_\gamma\delta_{\alpha\beta}).
\end{aligned}
\end{equation}
Using these equilibrium moments, the semi-discrete form of the MAME can be rewritten in terms of the moment relations of $f_i^{\mathrm{eq}}$:
\begin{subequations}
\begin{align}
\rho_{\mathrm{MAME}}(\boldsymbol{r},t)
  &=\rho^{t-\delta t}
   -\delta t\,\partial_\alpha(\rho u_\alpha)^{t-\delta t}
   +\frac{1}{2}\delta t^2
\partial_\alpha\partial_\beta
      (\Pi_{\alpha\beta}^{\mathrm{eq}})^{t-\delta t},
\\
(\rho u_\alpha)_{\mathrm{MAME}}(\boldsymbol{r},t)
  &= (\rho u_\alpha)^{t-\delta t}
   -\delta t\,\partial_\beta(\Pi_{\alpha\beta}^{\mathrm{eq}})^{t-\delta t}
   +\frac{\nu}{c_s^2{\delta t}}{\delta t}^2\,
\partial_\beta\partial_\gamma
      (\Pi_{\alpha\beta\gamma}^{\mathrm{eq}})^{t-\delta t}
\nonumber\\
  &\quad
   +\left(\frac{\nu}{c_s^2\delta t}-\frac{1}{2}\right)
\delta t\left[
\partial_\beta\Pi_{\alpha\beta}^{\mathrm{eq},\ast,\mathrm{MAME}}
      -\partial_\beta(\Pi_{\alpha\beta}^{\mathrm{eq}})^{t-\delta t}
\right].
\end{align}
\end{subequations}

By adopting the predictor operator in \secref{sec:gslbm_theory}, we explicitly construct the following scheme with $\rho=\rho(\boldsymbol{r},t-\delta t)$, $\rho u_\alpha=\rho u_\alpha(\boldsymbol{r},t-\delta t)$, and $\Pi^{\mathrm{eq}}_{\alpha\beta}=\Pi^{\mathrm{eq}}_{\alpha\beta}(\boldsymbol{r},t-\delta t)$ for simplicity:
\begin{subequations}
\begin{align}
\rho^\ast
  &= A_\rho\sum_i f_i^{\mathrm{eq}}(\boldsymbol{r}-\boldsymbol{e}_i\delta t,t-\delta t)
   + B_\rho\sum_i f_i^{\mathrm{eq}}(\boldsymbol{r}-2\boldsymbol{e}_i\delta t,t-\delta t)
\nonumber\\
  &\quad
   + C_\rho\sum_i f_i^{\mathrm{eq}}(\boldsymbol{r}+\boldsymbol{e}_i\delta t,t-\delta t)
   + D_\rho\sum_i f_i^{\mathrm{eq}}(\boldsymbol{r}+2\boldsymbol{e}_i\delta t,t-\delta t)
   -N_\rho\rho,
\\
(\rho u_\alpha)^\ast
  &= A_m\sum_i e_{i\alpha}f_i^{\mathrm{eq}}(\boldsymbol{r}-\boldsymbol{e}_i\delta t,t-\delta t)
   + B_m\sum_i e_{i\alpha}f_i^{\mathrm{eq}}(\boldsymbol{r}-2\boldsymbol{e}_i\delta t,t-\delta t)
\nonumber\\
  &\quad
   + C_m\sum_i e_{i\alpha}f_i^{\mathrm{eq}}(\boldsymbol{r}+\boldsymbol{e}_i\delta t,t-\delta t)
   + D_m\sum_i e_{i\alpha}f_i^{\mathrm{eq}}(\boldsymbol{r}+2\boldsymbol{e}_i\delta t,t-\delta t)
   -N_m(\rho u_\alpha).
\end{align}
\end{subequations}
By performing the Taylor expansion, the above scheme corresponds to the following equations:
\begin{subequations}
\begin{align}
\rho^\ast
  &=\tilde{t}_0\rho
   +\tilde{t}_1\delta t\,\partial_\alpha(\rho u_\alpha)
   +\tilde{t}_2\delta t^2\,\partial_\alpha\partial_\beta
\Pi_{\alpha\beta}^{\mathrm{eq}}
   +\tilde{t}_3\delta t^3\partial^3
   +\tilde{t}_4\delta t^4\partial^4
   +O(\delta t^5),
\\
(\rho u_\alpha)^\ast
  &= t_0\rho u_\alpha
   +t_1\delta t\,\partial_\beta\Pi_{\alpha\beta}^{\mathrm{eq}}
   +t_2\delta t^2\,\partial_\beta\partial_\gamma
\Pi_{\alpha\beta\gamma}^{\mathrm{eq}}
   +t_3\delta t^3\partial^3
   +t_4\delta t^4\partial^4
   +O(\delta t^5).
\end{align}
\end{subequations}
Here, $\partial^{n}$ denotes the corresponding nth-order derivative tensor. The coefficients A,B,C,D,N appearing in the linear combination of equilibrium distribution functions are related to the parameters $\{\tilde{t}_n,t_n\}$ through the common linear system obtained by Taylor matching; the system is given in \eqnref{eq:gslbm_predictor_coefficient_system} of \appref{app:gslbm_derivation}. In the present formulation, these coefficients define the numerical scheme, while $n$ denotes the order of the associated spatial derivative. Therefore, the parameters $\{\tilde{t}_n,t_n\}$ possess clear physical meanings associated with the corresponding differential operators.

For $n\le 2$, these parameters are uniquely determined by the mathematical form of the MAME and are explicitly substituted by their physically consistent expressions in the continuity and the momentum equations:
\begin{subequations}
\begin{align}
\left \{
\tilde{t}_0=1,\quad
\tilde{t}_1=-1,\quad
\tilde{t}_2=\frac{1}{2}
\right\}_{\mathrm{MAME}},
\\
\left \{
t_0=1,\quad
t_1=-1,\quad
t_2=\frac{\nu}{c_s^2\delta t}=\tau-\frac{1}{2}
\right\}_{\mathrm{MAME}}.
\end{align}
\end{subequations}
As a result, the predictor step recovers the semi-discrete MAME with second-order accuracy as
\begin{subequations}
\begin{align}
\rho^\ast
  &=\rho-\delta t\,\partial_\alpha(\rho u_\alpha)
    +\frac{1}{2}\delta t^2
\partial_\alpha\partial_\beta\Pi_{\alpha\beta}^{\mathrm{eq}}
    +O(\delta t^3,\partial^l\mid l\ge 3)_{\mathrm{GSLBM}}
\nonumber\\
  &=\rho^\ast_{\mathrm{MAME}}
    +O(\delta t^3,\partial^l\mid l\ge 3)_{\mathrm{GSLBM}},
\\
(\rho u_\alpha)^\ast
  &=\rho u_\alpha
    -\delta t\,\partial_\beta\Pi_{\alpha\beta}^{\mathrm{eq}}
    +\frac{\nu}{c_s^2{\delta t}}{\delta t}^2\,
\partial_\beta\partial_\gamma
\Pi_{\alpha\beta\gamma}^{\mathrm{eq}}
    +O(\delta t^3,\partial^l\mid l\ge 3)_{\mathrm{GSLBM}}
\nonumber\\
  &= (\rho u_\alpha)^\ast_{\mathrm{MAME}}
    +O(\delta t^3,\partial^l\mid l\ge 3)_{\mathrm{GSLBM}}.
\end{align}
\end{subequations}
where $\rho^\ast_{\mathrm{MAME}}$ and $(\rho u_\alpha)^\ast_{\mathrm{MAME}}$ denote the exact solution of the predictor step of the semi-discrete MAME without numerical error. The subscript GSLBM denotes the error term introduced by the generalized SLBM scheme.

In contrast, the parameters associated with higher-order derivatives ($n\ge 3$) arise purely from the discretization and the reconstruction procedures and thus represent higher-order truncation errors. Errors of the same formal order are also introduced when the MAME is discretized using finite-difference schemes, although their algebraic forms differ.
\begin{subequations}
\begin{align}
\rho^\ast_{\mathrm{FD}}
  &=\rho^\ast_{\mathrm{MAME}}
   +O(\delta t^3,\partial^l\mid l\ge 3)_{\mathrm{FD}},
\\
(\rho u_\alpha)^\ast_{\mathrm{FD}}
  &= (\rho u_\alpha)^\ast_{\mathrm{MAME}}
   +O(\delta t^3,\partial^l\mid l\ge 3)_{\mathrm{FD}}.
\end{align}
\end{subequations}
While these higher-order terms do not affect the recovery of the target macroscopic equations in second-order accuracy, they play a crucial role in determining the numerical dissipation, dispersion, and stability of the scheme. For the sake of clarity and focus, the present study does not examine the higher-order coefficients in the continuity equation; accordingly, the corresponding parameters $\tilde{t}_3$ and $\tilde{t}_4$ are set to zero, and the subsequent analysis is concentrated on the momentum equation. In particular, the third- and the fourth-order parameters in the momentum equation, denoted by $t_3$ and $t_4$, have a dominant influence on the numerical behaviours, which will be systematically investigated in \secrefs{sec:linear_stability_analysis}{sec:numerical_tests}. Beyond these leading contributions, numerical experiments further indicate that higher-order terms (of order greater than four) can provide additional stabilizing effects, which will be discussed in \secref{sec:dsl_underresolved}.

In principle, a fully self-consistent numerical formulation would discretize both the predictor and the corrector steps within a unified equilibrium-based framework. As outlined in \appref{app:mame_unified_scheme}, the generalized SLBM can indeed be extended to such a unified formulation by introducing additional correction terms. However, this extension requires the solution of extra linear systems and leads to a substantial increase in both computational cost and memory usage. Given that the primary objective of the present study is to preserve the memory efficiency of SLBM while achieving physical consistency and explicit control over numerical dissipation, dispersion, and stability, a hybrid strategy is adopted. Specifically, the predictor step is treated by the generalized SLBM formulation, whereas the corrector step is constructed using the finite-difference discretization. This choice is a compromise among accuracy, efficiency, and practical implementation, and allows the influence of the predictor formulation on numerical behaviour to be examined in a clearer manner.

\begin{figure}[!htbp]
\centering
\includegraphics[width=0.52\linewidth]{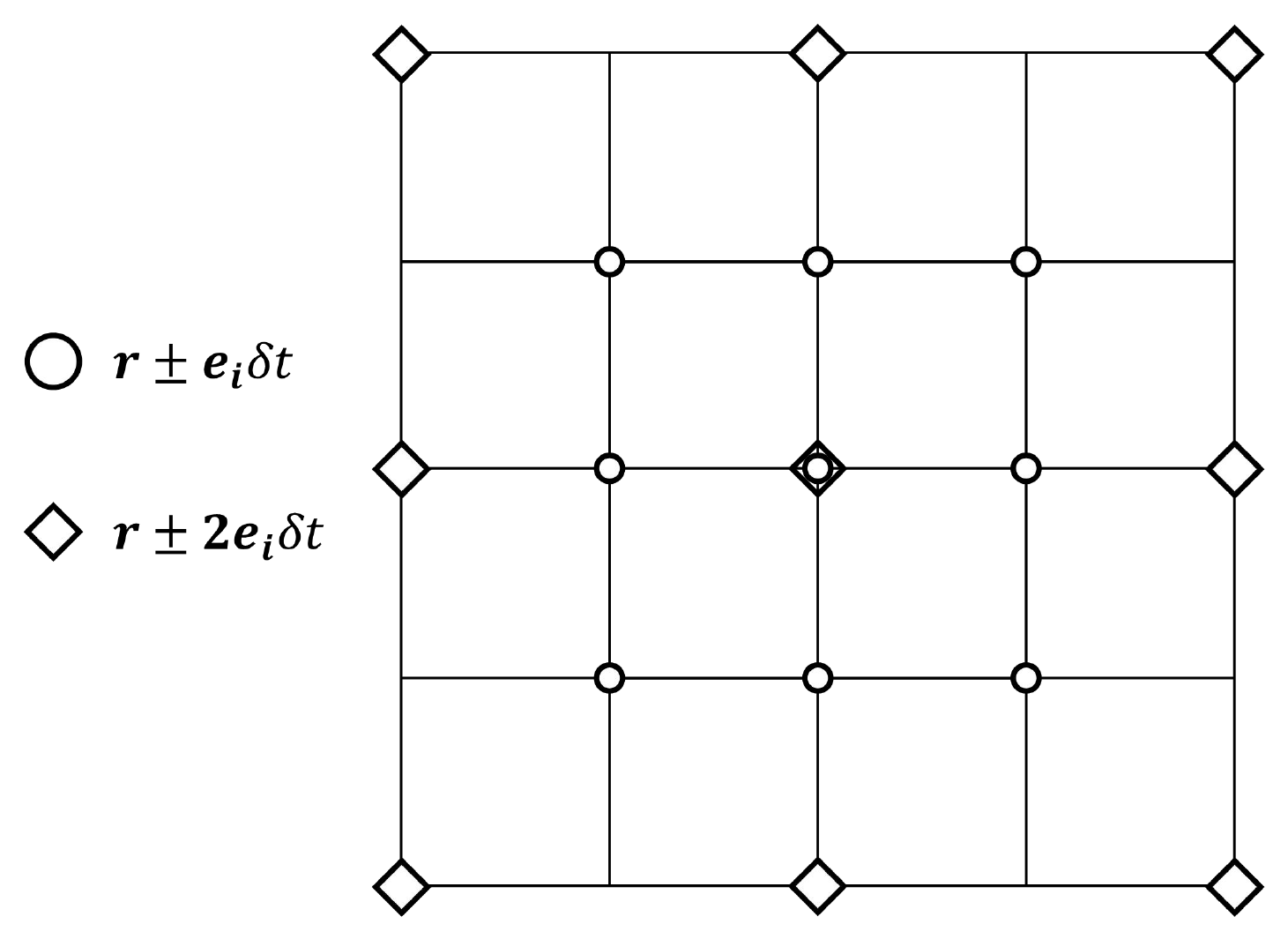}
\caption{Five-point stencil used in the reformulated SLBM. Circles and diamonds denote the locations $\boldsymbol{r}\pm\boldsymbol{e}_i\delta t$ and $\boldsymbol{r}\pm2\boldsymbol{e}_i\delta t$, respectively.}
\label{fig:five_point_stencil}
\end{figure}

In summary, the reformulated SLBM is presented as follows:

\noindent \textbf{Predictor:}
\begin{equation}
\begin{aligned}
\rho^\ast
  &= A_\rho\sum_i f_i^{\mathrm{eq}}(\boldsymbol{r}-\boldsymbol{e}_i\delta t,t-\delta t)
   + B_\rho\sum_i f_i^{\mathrm{eq}}(\boldsymbol{r}-2\boldsymbol{e}_i\delta t,t-\delta t)\\
  &\quad
   + C_\rho\sum_i f_i^{\mathrm{eq}}(\boldsymbol{r}+\boldsymbol{e}_i\delta t,t-\delta t)
   + D_\rho\sum_i f_i^{\mathrm{eq}}(\boldsymbol{r}+2\boldsymbol{e}_i\delta t,t-\delta t)
   -N_\rho\rho^{t-\delta t},
\end{aligned}
\label{eq:rslbm_predictor_density}
\end{equation}
\begin{equation}
\begin{aligned}
(\rho u_\alpha)^\ast
  &= A_m\sum_i e_{i\alpha}f_i^{\mathrm{eq}}(\boldsymbol{r}-\boldsymbol{e}_i\delta t,t-\delta t)
   + B_m\sum_i e_{i\alpha}f_i^{\mathrm{eq}}(\boldsymbol{r}-2\boldsymbol{e}_i\delta t,t-\delta t)\\
  &\quad
   + C_m\sum_i e_{i\alpha}f_i^{\mathrm{eq}}(\boldsymbol{r}+\boldsymbol{e}_i\delta t,t-\delta t)
   + D_m\sum_i e_{i\alpha}f_i^{\mathrm{eq}}(\boldsymbol{r}+2\boldsymbol{e}_i\delta t,t-\delta t)\\
  &\quad
   -N_m(\rho u_\alpha)^{t-\delta t}.
\end{aligned}
\label{eq:rslbm_predictor_momentum}
\end{equation}
\textbf{Corrector:}
\begin{equation}
\rho(\boldsymbol{r},t)=\rho^\ast,
\label{eq:rslbm_corrector_density}
\end{equation}
\begin{equation}
\begin{aligned}
(\rho u_\alpha)(\boldsymbol{r},t)
  &= (\rho u_\alpha)^\ast
   +\left(\frac{\nu}{c_s^2}-\frac{1}{2}\delta t\right)
\left[
\partial_\beta(\rho u_\alpha u_\beta+\rho c_s^2\delta_{\alpha\beta})^\ast
      -
\partial_\beta(\rho u_\alpha u_\beta+\rho c_s^2\delta_{\alpha\beta})^{t-\delta t}
\right]_{\mathrm{FD}}.
\end{aligned}
\label{eq:rslbm_corrector_momentum}
\end{equation}
It is clear that our method is a 5-point-stencil as illustrated in \figref{fig:five_point_stencil}. So, to maintain consistency, the spatial derivatives in \eqnref{eq:rslbm_corrector_momentum} are discretized by the 5-point-stencil fourth-order finite difference method.

\subsubsection{Special parameter selections and their numerical implications}\label{sec:rslbm_parameter_selection}

As discussed in \secref{sec:rslbm_equilibrium_formulation}, the parameters associated with derivatives up to second order, i.e. $t_n$ ($n\le 2$), are uniquely determined by the MAME and must be preserved in the predictor step to ensure physical consistency. However, the parameters $t_3$ and $t_4$ are associated with numerical effects beyond second-order accuracy, which are higher-order truncation effects introduced by the equilibrium-based reconstruction. This provides controllable degrees of freedom for tuning numerical dissipation and dispersion as shown in \secref{sec:linear_stability_analysis}.

To endow interpretable properties to these higher-order parameters, we consider several typical constructions based on different stencil choices and moment constraints. These selections are not exhaustive, but they cover the main numerical behaviours observed in practice and thus can be the reference cases for the subsequent theoretical and numerical analysis.

(1) Elimination of both the third- and the fourth-order errors

A natural starting point is to completely suppress the leading high-order truncation terms by enforcing
\[
  t_3=0,\quad t_4=0
\]
With this choice, the predictor step recovers the semi-discrete MAME up to fourth order without additional dispersive or dissipative corrections. And it can be used as a useful baseline for showcasing effects of higher-order terms.

(2) One-layer stencil construction.

In the second configuration, only a single layer of neighbouring lattice points are used in the equilibrium reconstruction, corresponding to setting the coefficients associated with the outer stencil to zero (i.e., B=D=0). With this restriction, the remaining parameters can be determined uniquely from the common coefficient system in \eqnref{eq:gslbm_predictor_coefficient_system}, yielding explicit expressions for $t_3$ and $t_4$ as
\[
  t_3=-\frac{1}{6},\quad t_4=\frac{\tau}{12}-\frac{1}{24}
\]
This choice leads to nonzero values of both parameters and represents the simplest nontrivial equilibrium-based construction beyond the minimal stencil. The numerical behaviour is then dictated by the competition between the third- and the fourth-order terms.

(3) Asymmetric three-point stencil with $t_3=0$: backward extension.

To isolate the effect of the fourth-order dissipation term while eliminating the leading dispersive error, we further consider a three-point stencil composed of
\[
(\boldsymbol{r}-\boldsymbol{e}_i\delta t,t-\delta t),\quad
(\boldsymbol{r}+\boldsymbol{e}_i\delta t,t-\delta t),\quad
(\boldsymbol{r}-2\boldsymbol{e}_i\delta t,t-\delta t).
\]
By enforcing the condition $t_3=0$, the remaining coefficients can be uniquely determined, leading to the specific reference value of $t_4=\tau/12-1/8$. This construction introduces an asymmetric stencil and allows showcasing only the high-order dissipative effects.

(4) Asymmetric three-point stencil with $t_3=0$: forward extension.

The counterpart configuration of the previous case is the forward extension, whose stencil reads
\[
(\boldsymbol{r}-\boldsymbol{e}_i\delta t,t-\delta t),\quad
(\boldsymbol{r}+\boldsymbol{e}_i\delta t,t-\delta t),\quad
(\boldsymbol{r}+2\boldsymbol{e}_i\delta t,t-\delta t).
\]
In this case, the fourth-order coefficient is solved as:
\[
  t_3=0,\quad t_4=\frac{\tau}{12}+\frac{1}{24}
\]
However, in our tests, this setting always tends to diverge, implying the association between the fourth-order derivative and the numerical stability. Therefore, we list this stencil choice here just for completeness, but will not use it in the rest of this article.

In summary, the above selections generate two typical values of $t_3$ and three typical values of $t_4$, which will be combined to constitute cases A-E as listed in \tabref{tab:high_order_parameters}. These reference cases provide the parameter basis for the linear-wave validation in \secref{sec:linear_wave_validation} and the numerical tests in \secref{sec:numerical_tests}. By anchoring all parameter choices to explicit stencil constructions and moment constraints, the present approach avoids heuristic tuning and provides a transparent framework for interpreting the dissipation-dispersion characteristics observed in subsequent analyses.

\begin{table}
\centering
\renewcommand{\arraystretch}{1.22}
\begin{tabular*}{0.88\textwidth}{@{\extracolsep{\fill}}ccc@{}}
\toprule
    Case & $t_3$ & $t_4$\\
\midrule
    A & $-\frac{1}{6}$ & $\frac{\tau}{12}-\frac{1}{24}$\\
    B & $-\frac{1}{6}$ & $0$\\
    C & $0$ & $0$\\
    D & $0$ & $\frac{\tau}{12}-\frac{1}{8}$\\
    E & $0$ & $\frac{\tau}{12}-\frac{1}{24}$\\
\bottomrule
\end{tabular*}
\smallskip
\caption{Parameter settings for the third- and fourth-order predictor contributions of RSLBM.}
\label{tab:high_order_parameters}
\end{table}

\subsubsection{Implementation procedure}\label{sec:rslbm_implementation}

The practical implementation of the reformulated SLBM proceeds as follows:

(1) Specify the mesh spacing $\delta x$ and time step $\delta t$. Determine the single relaxation parameter according to the physical setting for the flow problem.

(2) Specify the parameters $\tilde{t}_n$ and $t_n$ according to the target macroscopic equations and the desired treatment of higher-order derivative terms. In this work,
\[
\{t_0=1,\quad t_1=-1,\quad t_2=\frac{\nu}{c_s^2\delta t}=\tau-\frac{1}{2}\},
\]
while $t_3$, $t_4$ are selected following the strategies introduced in \secref{sec:rslbm_parameter_selection}.

(3) Obtain coefficients \{$A_\rho$, $B_\rho$, $C_\rho$, $D_\rho$, $N_\rho$\} and \{$A_m$, $B_m$, $C_m$, $D_m$, $N_m$\} by solving the common coefficient system in \eqnref{eq:gslbm_predictor_coefficient_system}. This procedure is performed once prior to the simulation.

(4) Initialize the macroscopic flow variables according to the prescribed initial conditions.

(5) Predictor step. Compute the intermediate macroscopic quantities using \eqnref{eq:rslbm_predictor_density} and \eqnref{eq:rslbm_predictor_momentum}.

(6) Apply boundary conditions to the intermediate macroscopic variables.

(7) Corrector step. Discretize the spatial derivatives in \eqnref{eq:rslbm_corrector_momentum} using the finite-difference formula, and update the macroscopic variables to the next time level.

(8) Apply boundary conditions to the updated macroscopic variables.

(9) Repeat steps (5)--(8) until the simulation reaches the prescribed final time or convergence criterion.

Since the predictor step involves only equilibrium distribution functions, the reformulated SLBM still avoids storing distribution functions and therefore retains a memory footprint comparable to that of the original SLBM. The coefficient determination is performed once before the simulation and therefore introduces no per-step coefficient-solution overhead. This equilibrium-based reconstruction further enables direct control of contributions from higher-order derivatives through the parameters $t_3$ and $t_4$, which, as will be demonstrated in \secrefs{sec:linear_stability_analysis}{sec:numerical_tests}, play a central role in controlling numerical dissipation, dispersion, and stability. With proper parametric choices, the reformulated SLBM achieves accuracy comparable to the finite-difference solvers of the MAME (\secref{sec:taylor_green_vortex}), while higher-order truncation terms beyond the fourth order are found to provide additional numerical stabilization in under-resolved conditions (\secref{sec:double_shear_layer}). Boundary conditions are imposed directly on macroscopic variables as in the original SLBM \citep{chen2018}. However, near-boundary mesh nodes require special treatments due to stencil reduction, and the adopted strategy together with its numerical implications will be discussed in \secref{sec:lid_driven_cavity}.


\section{Linear Stability Analysis}\label{sec:linear_stability_analysis}

In the previous section, the SLBM and the RSLBM were interpreted from the viewpoint of equivalent macroscopic equations recovered by Taylor expansion. That analysis separates the physical deviations and truncation-error terms embedded in the predictor-corrector reconstruction, and clarifies how the high-order parameters of RSLBM enter the recovered equations. We now examine the same schemes in Fourier space. By linearizing both methods around a uniform base flow and writing one time step as an amplification problem for the conserved variables $(\rho,j_x,j_y)^{\mathrm{T}}$, the dispersion, dissipation and stability of individual numerical modes can be assessed directly.

Accordingly, \secref{sec:lsa_formulation} formulates the amplification matrices of SLBM and RSLBM and derives the small-wavenumber shear-mode expansions, thereby connecting the spectral quantities with the equivalent-equation results of \secref{sec:methodology}. \secref{sec:lsa_wavenumber_maps} evaluates these amplification matrices in wavenumber space and visualizes the modal dissipation, dispersion and stability properties, with the shear and acoustic branches identified through their eigenvectors. \secref{sec:linear_wave_validation} then compares the linear stability analysis (LSA) predictions with transverse linear-wave numerical simulations at the discrete wavenumbers allowed by finite periodic domains.

\subsection{Linear stability formulation}\label{sec:lsa_formulation}

We first consider small perturbations around a spatially uniform base state in lattice units. The macroscopic conserved variables are
\begin{equation}
\boldsymbol{Q}=(\rho,j_x,j_y)^{\mathrm{T}},
\qquad
j_x=\rho u_x,\qquad j_y=\rho u_y.
\label{eq:lsa_macroscopic_variables}
\end{equation}
For a base state $\boldsymbol{Q}_0=(\rho_0,\rho_0U_0,\rho_0V_0)^{\mathrm{T}}$, the Fourier perturbation is written as
\begin{equation}
\delta\boldsymbol{Q}^n(\boldsymbol{r})
=
\widehat{\boldsymbol{Q}}^{\,n}
\exp(\mathrm{i}\boldsymbol{k}\cdot\boldsymbol{r}).
\label{eq:lsa_fourier_perturbation}
\end{equation}
Here $\boldsymbol{r}$, $\boldsymbol{k}$, the macroscopic variables and the viscosity are expressed in lattice units. Thus, for a one-dimensional wave whose wavelength $\lambda_w$ is also measured in lattice units, the corresponding lattice-unit wavenumber is $k=2\pi/\lambda_w$.
After linearization, one step of the numerical method can be written as
\begin{equation}
\widehat{\boldsymbol{Q}}^{\,n+1}
=
\mathsfbi{G}(\boldsymbol{k})\widehat{\boldsymbol{Q}}^{\,n},
\label{eq:lsa_amplification_problem}
\end{equation}
where $\mathsfbi{G}(\boldsymbol{k})$ is the amplification matrix. Its eigenvalues determine the amplification of individual numerical modes, while the corresponding eigenvectors are used to identify their physical content.

\subsubsection{Amplification matrix and modal identification}\label{sec:lsa_amplification_modal}

For RSLBM, the predictor and corrector contributions can be written compactly as
\begin{equation}
\mathsfbi{G}_{\mathrm{RSLBM}}(\boldsymbol{k})
=
\mathsfbi{P}_{\mathrm{RSLBM}}(\boldsymbol{k})
+
\mathsfbi{R}_{\mathrm{RSLBM}}(\boldsymbol{k})
\left[
\mathsfbi{P}_{\mathrm{RSLBM}}(\boldsymbol{k})-\mathsfbi{I}
\right].
\label{eq:lsa_rslbm_amplification}
\end{equation}
Here $\mathsfbi{P}_{\mathrm{RSLBM}}$ is obtained by applying the Fourier symbols of the equilibrium-based predictor stencil to the linearized equilibrium distribution, while $\mathsfbi{R}_{\mathrm{RSLBM}}$ represents the finite-difference corrector. The residual $\mathsfbi{P}_{\mathrm{RSLBM}}(\boldsymbol{k})-\mathsfbi{I}$ indicates that the corrector acts on the difference between the intermediate state and the current state.

The original SLBM can be written in a related predictor-corrector form, although the corrector residual is assembled differently. We use
\begin{equation}
\mathsfbi{G}_{\mathrm{SLBM}}(\boldsymbol{k})
=
\mathsfbi{P}_{\mathrm{SLBM}}(\boldsymbol{k})
+
\sigma_\nu \mathsfbi{S}
\left[
\mathsfbi{C}_{\mathrm{SLBM}}(\boldsymbol{k})
\mathsfbi{P}_{\mathrm{SLBM}}(\boldsymbol{k})-\mathsfbi{I}
\right],
\label{eq:lsa_slbm_amplification}
\end{equation}
with
\begin{equation}
\sigma_\nu=3\nu-\frac{1}{2}=\tau-1,
\qquad
\mathsfbi{S}=\operatorname{diag}(0,1,1).
\label{eq:lsa_slbm_corrector_coefficient}
\end{equation}
Here $\nu$ is the lattice kinematic viscosity, so that $\nu=(\tau-1/2)/3$ in lattice units. The selector $\mathsfbi{S}$ indicates that the SLBM corrector updates the momentum components but not the density. The matrix $\mathsfbi{P}_{\mathrm{SLBM}}$ corresponds to the backward equilibrium reconstruction used in the predictor, whereas $\mathsfbi{C}_{\mathrm{SLBM}}$ corresponds to the forward reconstruction used in the corrector.

For each eigenvalue $\lambda_m(\boldsymbol{k})$ of $\mathsfbi{G}(\boldsymbol{k})$, the numerical angular frequency is defined by
\begin{equation}
\lambda_m=\exp(-\mathrm{i}\omega_m),
\qquad
\omega_m=\mathrm{i}\log\lambda_m.
\label{eq:lsa_lambda_omega_convention}
\end{equation}
With this convention, $\operatorname{Re}(\omega_m)$ measures propagation and $\operatorname{Im}(\omega_m)$ measures growth or damping. The linear stability condition is
\begin{equation}
\max_m|\lambda_m(\boldsymbol{k})|\le 1,
\label{eq:lsa_stability_condition}
\end{equation}
or, equivalently, $\operatorname{Im}(\omega_m)\le0$ for every branch.

The physical branch labels are identified by eigenvector projection, following the modal linear-stability analyses of \citet{lallemand2000} and \citet{wissocq2020}. For each nonzero $\boldsymbol{k}$, the eigenvectors are projected onto the theoretical shear and acoustic target directions in primitive-variable space ${(\delta\rho,\delta u_x,\delta u_y)}^{\mathrm{T}}$. Once the shear branch has been identified, its effective viscosity is computed from the logarithmic amplification factor for nonzero wavenumbers:
\begin{equation}
\nu_{\mathrm{eff}}
=
-
\frac{\operatorname{Re}[\log\lambda_s]}{|\boldsymbol{k}|^2}
=
-
\frac{\operatorname{Im}(\omega_s)}{|\boldsymbol{k}|^2}.
\label{eq:lsa_effective_viscosity}
\end{equation}

The algebraic construction of the amplification matrices, including the stencil symbols, equilibrium-moment Jacobians and eigenvector-projection scores used above, is summarized in \appref{app:amplification_matrix_construction}.

\subsubsection{Small-wavenumber shear-mode expansion}\label{sec:lsa_small_wavenumber}

Before examining the full spectra in wavenumber space, we first extract the low-wavenumber behaviour of a transverse shear wave. This branch is particularly useful because its logarithmic amplification factor gives a direct measure of viscous damping and phase error. We consider the one-dimensional configuration
\begin{equation}
\boldsymbol{U}_0=(U,0),
\qquad
\boldsymbol{k}=(k,0),
\qquad
\delta\boldsymbol{Q}=(0,0,\delta j_y)^{\mathrm{T}}.
\label{eq:lsa_transverse_shear_configuration}
\end{equation}
Here $k$ denotes the lattice-unit wavenumber in the streamwise direction. For this configuration, the transverse momentum perturbation is decoupled from the acoustic subsystem. The $3\times3$ amplification problem therefore reduces, for the shear branch, to a scalar amplification factor. This reduction is used here only to interpret the small-$k$ coefficients; the full amplification matrices are evaluated in \secref{sec:lsa_wavenumber_maps}.

We first use the original SLBM as a reference. Define
\begin{equation}
a(k)=\frac{2+\cos k}{3},
\qquad
b(k)=U\sin k.
\label{eq:lsa_slbm_ab_definitions}
\end{equation}
Then the full predictor-corrector shear eigenvalue of SLBM can be written as
\begin{equation}
\lambda_s^{\mathrm{SLBM}}(U)
=
a(k)-ib(k)
+
\left(3\nu-\frac12\right)
\left[
a^2(k)+b^2(k)-1
\right].
\label{eq:lsa_slbm_shear_eigenvalue}
\end{equation}
If only the predictor stage is examined, the stationary shear wave gives
\begin{equation}
\log\lambda_s^{(P)}(U=0)
=
-\frac{k^2}{6}
+
O(k^6),
\label{eq:lsa_slbm_predictor_log}
\end{equation}
so the predictor-only apparent viscosity is
\begin{equation}
\nu_P=\frac16,
\qquad
\nu_P-\nu=\frac16-\nu.
\label{eq:lsa_slbm_predictor_viscosity}
\end{equation}
This is the virtual-viscosity mechanism identified in the Taylor-expansion analysis in \secref{sec:methodology}. However, it is only a stage-level statement and should not be mistaken for the leading shear viscosity of the full SLBM.

For the complete SLBM, the stationary shear branch instead gives
\begin{equation}
\log\lambda_s^{\mathrm{SLBM}}(U=0)
=
-\nu k^2
-
\frac{(6\nu-1)^2}{72}k^4
+
O(k^6).
\label{eq:lsa_slbm_static_log}
\end{equation}
The corrector therefore restores the leading low-wavenumber shear viscosity to the physical value $\nu$, while leaving a fourth-order finite-wavenumber damping residual. Equivalently,
\begin{equation}
\nu_{\mathrm{eff}}^{\mathrm{SLBM}}(U=0)
=
\nu
+
\frac{(6\nu-1)^2}{72}k^2
+
O(k^4).
\label{eq:lsa_slbm_static_effective_viscosity}
\end{equation}
The stationary SLBM shear branch is thus correct in the long-wave limit, but it contains an additional $k^4$ damping residual. This leading-order recovery provides the spectral evidence for the compensation mechanism discussed in \secref{sec:methodology}: the corrector offsets the leading long-wave effect of the predictor-level virtual viscosity, but the compensation is not exact at finite wavenumbers. This residual becomes important at finite wavenumbers and explains why SLBM may remain strongly dissipative even when its leading shear viscosity is correct. In the low-viscosity limit,

\begin{equation}
\log\lambda_s^{\mathrm{SLBM}}(U=0,\nu=0)
=
-\frac{k^4}{72}
+
O(k^6),
\label{eq:lsa_slbm_zero_viscosity_limit}
\end{equation}
showing that finite-wavenumber stabilization is not supplied by physical viscosity alone.

With background flow, the complete SLBM shear branch may be expanded as
\begin{equation}
\begin{aligned}
\log\lambda_s^{\mathrm{SLBM}}(U)
=&
-\mathrm{i}Uk
-
\nu(1-3U^2)k^2\\
&+
i
\left[
U^3\left(3\nu-\frac16\right)
+
U\left(\frac16-\nu\right)
\right]k^3
+
C_4^{\mathrm{SLBM}}(U,\nu)k^4
+
O(k^5).
\end{aligned}
\label{eq:lsa_slbm_convected_log}
\end{equation}
The detailed expression for $C_4^{\mathrm{SLBM}}$ is not needed in the main text. The leading result is
\begin{equation}
\nu_{\mathrm{eff}}^{\mathrm{SLBM}}(U)
=
\nu(1-3U^2)
+
O(k^2).
\label{eq:lsa_slbm_convected_effective_viscosity}
\end{equation}
Therefore, the Mach-number-dependent viscosity reduction already appears in SLBM at leading order which corresponds to the Galilean invariance error in the conventional LBM. The strong damping of SLBM should not be interpreted as a simple leading-order viscosity increase of the full scheme. The shear expansion instead separates three effects: predictor-level virtual viscosity, leading-order compensation by the corrector, and residual finite-wavenumber damping. In the notation $\tau=3\nu+1/2$, the loss of robustness observed for $\tau>1$ is likewise associated with finite-wavenumber over-correction rather than with a negative leading viscosity of the complete scheme.

We now turn to RSLBM. For the second-order equilibrium and a stationary base flow, the full predictor-corrector shear branch gives
\begin{equation}
\log\lambda_s(U=0)
=
-\nu k^2
-
\left(
\frac{\nu^2}{2}
-
\frac{t_4}{3}
\right)k^4
+
O(k^6).
\label{eq:lsa_rslbm_static_log}
\end{equation}
This expression makes the role of $t_4$ explicit: it enters the fourth-order real part of the logarithmic amplification factor and therefore controls finite-wavenumber dissipation. The parameter $t_3$ does not appear in the leading dissipation of the stationary shear wave.

With a nonzero horizontal base flow, the same second-order equilibrium gives
\begin{equation}
\log\lambda_s(U)
=
-\mathrm{i}Uk
-
\nu(1-3U^2)k^2
+
iC_3(U,\nu,t_3)k^3
+
C_4(U,\nu,t_3,t_4)k^4
+
O(k^5),
\label{eq:lsa_rslbm_convected_log}
\end{equation}
where
\begin{equation}
C_3
=
\frac{U}{6}
\left(
18U^2\nu-U^2-18\nu^2-3\nu-6t_3
\right),
\label{eq:lsa_rslbm_c3}
\end{equation}
and the fourth-order real coefficient contains the parameter dependence
\begin{equation}
C_4
=
C_4^{(0)}(U,\nu)
+
\left(
3U^2\nu+\frac{U^2}{2}
\right)t_3
+
\frac{t_4}{3}.
\label{eq:lsa_rslbm_c4_parameter_dependence}
\end{equation}
The term $-\nu(1-3U^2)k^2$ is the Mach-number-dependent leading shear dissipation associated with the second-order equilibrium. The coefficient $C_3$ shows that $t_3$ first enters as a third-order phase correction and thus controls the dispersion of convected shear waves. The fourth-order real coefficient shows that, when $U\ne0$, $t_3$ also contributes to finite-wavenumber dissipation through terms proportional to $U^2t_3$, while $t_4$ continues to set the baseline fourth-order dissipative correction.

The convected-shear expansion also clarifies the role of equilibrium truncation. With the second-order D2Q9 equilibrium, the leading damping contains the factor $1-3U^2$. Including the relevant partially extended third- or fourth-order equilibrium contribution \citep{coreixas2017} in the transverse-shear branch removes this $O(Ma^2)$ correction, changing
\begin{equation}
\log\lambda_s(U)
=
-\mathrm{i}Uk-\nu(1-3U^2)k^2+O(k^3)
\label{eq:lsa_second_order_equilibrium_shear_log}
\end{equation}
to
\begin{equation}
\log\lambda_s(U)
=
-\mathrm{i}Uk-\nu k^2+O(k^3).
\label{eq:lsa_extended_equilibrium_shear_log}
\end{equation}
This observation is used only to interpret the leading convected-shear error; the remaining analysis keeps the second-order equilibrium as the common baseline, so that the effects of the predictor-corrector reconstruction and of $t_3,t_4$ can be isolated. A systematic extension to other equilibria or velocity sets would require a corresponding reconstruction of the equivalent macroscopic equations and is therefore outside the scope of the present article.

In summary, the above analysis refines the interpretation obtained from \secref{sec:methodology}. For SLBM, the predictor-only apparent viscosity explains the origin of the virtual-viscosity mechanism, but the full predictor-corrector shear branch restores the leading static viscosity to $\nu$. Its strong damping is therefore a finite-wavenumber effect associated with the residual $k^4$ term and with the sign and magnitude of the corrector residual at larger wavenumbers. For RSLBM, $t_4$ directly controls the fourth-order real part of $\log\lambda_s$, while $t_3$ first enters the convected shear branch as a third-order phase correction and, for $U\ne0$, also contributes to finite-wavenumber damping. The wavenumber-space scans in the next subsection evaluate these mechanisms beyond the asymptotic small-$k$ limit.

\subsection{Wavenumber-space spectra and stability maps}\label{sec:lsa_wavenumber_maps}

We now evaluate the amplification matrices over wavenumber space using the second-order equilibrium as the common baseline. Unless otherwise stated, the background flow is fixed in the horizontal direction, $\boldsymbol{U}_0=(U,0)$, with $Ma=U/c_s=0.1$, and the high-order predictor parameters of RSLBM are set to $t_3=0$ and $t_4=0$. Two low-viscosity relaxation times are considered: $\tau=0.503$, which serves as the main comparison point for the modal behaviour of RSLBM and SLBM, and $\tau=0.501$, which is closer to the stability boundary and therefore highlights finite-wavenumber stability risks.

For the spectral scans, we distinguish the lattice-unit wavenumber $\boldsymbol{k}$ from the dimensionless wavenumber normalized by $\delta x$ as
\[
\boldsymbol{\kappa}=\boldsymbol{k}\,\delta x,
\]
which is restricted to the first Brillouin zone $\kappa_x,\kappa_y\in[-\pi,\pi]$. In lattice units $\delta x=1$, so $\boldsymbol{\kappa}$ and $\boldsymbol{k}$ have the same numerical values, although they emphasize different interpretations.

\begin{figure}[!htbp]
\centering
\includegraphics[width=\linewidth]{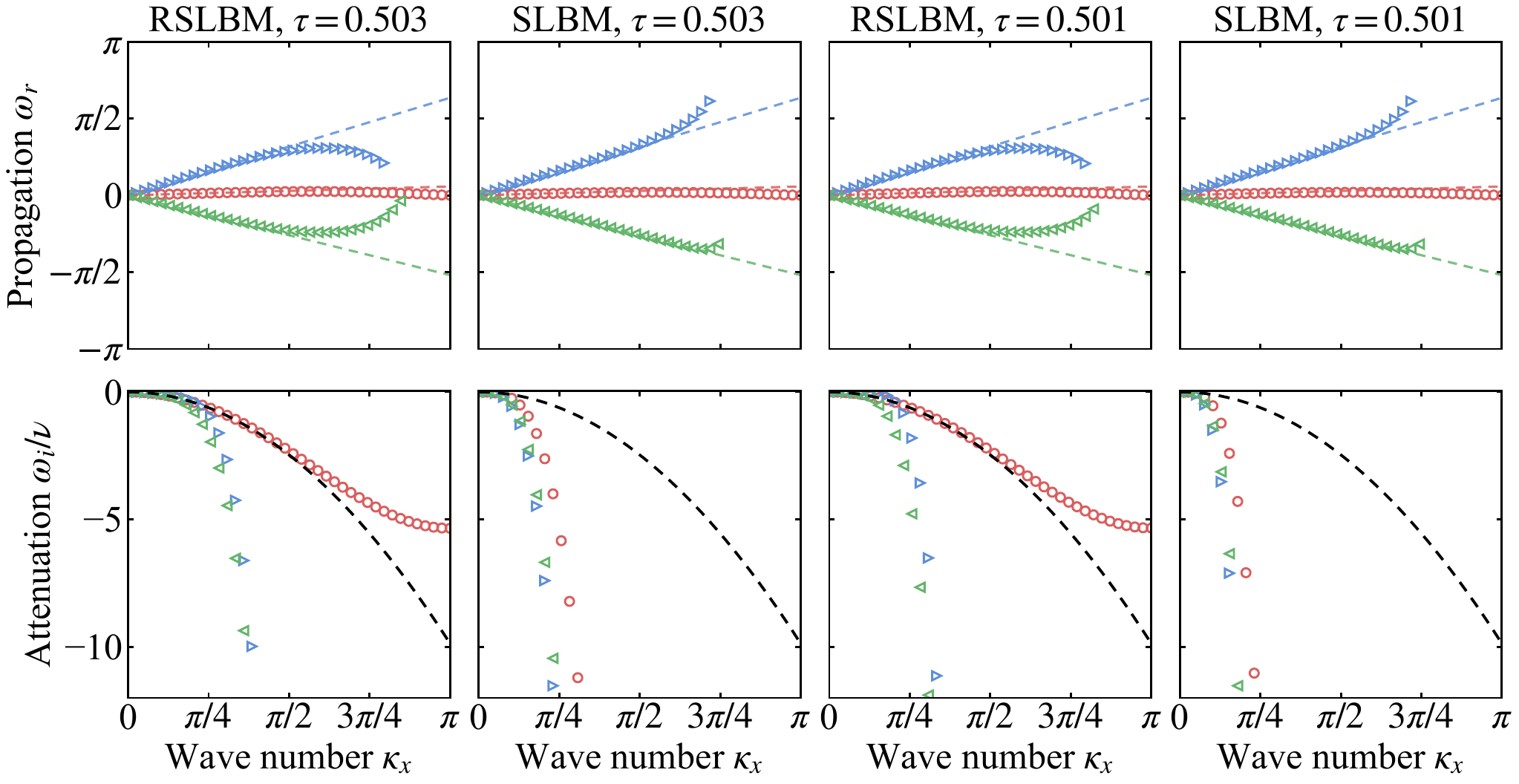}
\caption{One-dimensional LSA spectra for RSLBM and SLBM using the second-order equilibrium at $Ma=0.1$, with $t_3=t_4=0$ for RSLBM. The symbols \textcolor[HTML]{D95F5F}{$\circ$}, \textcolor[HTML]{5F8FD9}{$\triangleright$} and \textcolor[HTML]{63B56B}{$\triangleleft$} denote the shear, acoustic $+$ and acoustic $-$ branches, respectively; only modes with eigenvector-projection scores $\geq0.9$ are shown. The coloured and black dashed curves denote the Navier--Stokes references.}
\label{fig:lsa_1d_spectra_baseline}
\end{figure}

A one-dimensional cut along the streamwise discrete wavenumber, $\boldsymbol{\kappa}=(\kappa,0)$, first gives the branch-level view of dispersion and dissipation, as shown in \figref{fig:lsa_1d_spectra_baseline}. With $\omega=\omega_r+\mathrm{i}\omega_i$, the real part $\omega_r$ is used to assess modal propagation and dispersion, whereas the normalized imaginary part $\omega_i/\nu$ is used to compare modal damping. Since $\lambda_m=\exp(-\mathrm{i}\omega_m)$, negative values of $\omega_i$ correspond to damping and positive values correspond to linear amplification. In this representation, the shear and acoustic branches can be followed directly as functions of $\kappa$. The upper and lower rows in \figref{fig:lsa_1d_spectra_baseline} show propagation and attenuation, respectively; the first two columns compare RSLBM and SLBM at $\tau=0.503$, and the last two columns give the corresponding comparison at $\tau=0.501$.

The one-dimensional spectra show that the two schemes do not optimize the same modal properties. RSLBM reduces the excess shear-mode damping relative to SLBM over a broad wavenumber range, whereas SLBM retains a more accurate acoustic phase relation along this streamwise cut. This comparison is consistent with the small-wavenumber interpretation above: the full SLBM recovers the correct long-wave shear viscosity, but its finite-wavenumber residual still produces substantially stronger damping away from the asymptotic $k\to0$ limit. A streamwise cut, however, cannot characterize oblique discrete wavenumbers.

The LSA is therefore extended to the full two-dimensional dimensionless wavenumber plane. Following the setting used by \citet{wissocq2020}, the mean flow is kept fixed along the $x$-direction, $\boldsymbol{U}_0=(U,0)$, while $\boldsymbol{\kappa}=(\kappa_x,\kappa_y)$ is varied. At each corresponding $\boldsymbol{k}=\boldsymbol{\kappa}/\delta x$, we compute the eigenvalues of the amplification matrix and the spectral radius
\begin{equation}
\varrho(\mathsfbi{G}(\boldsymbol{k}))
=
\max_m|\lambda_m(\boldsymbol{k})|.
\label{eq:lsa_spectral_radius}
\end{equation}
The condition $\varrho(\mathsfbi{G})\le1$ for all resolved wavenumber vectors is the linear stability condition for the uniform base state.

\begin{figure}[!htbp]
\centering
\includegraphics[width=\linewidth]{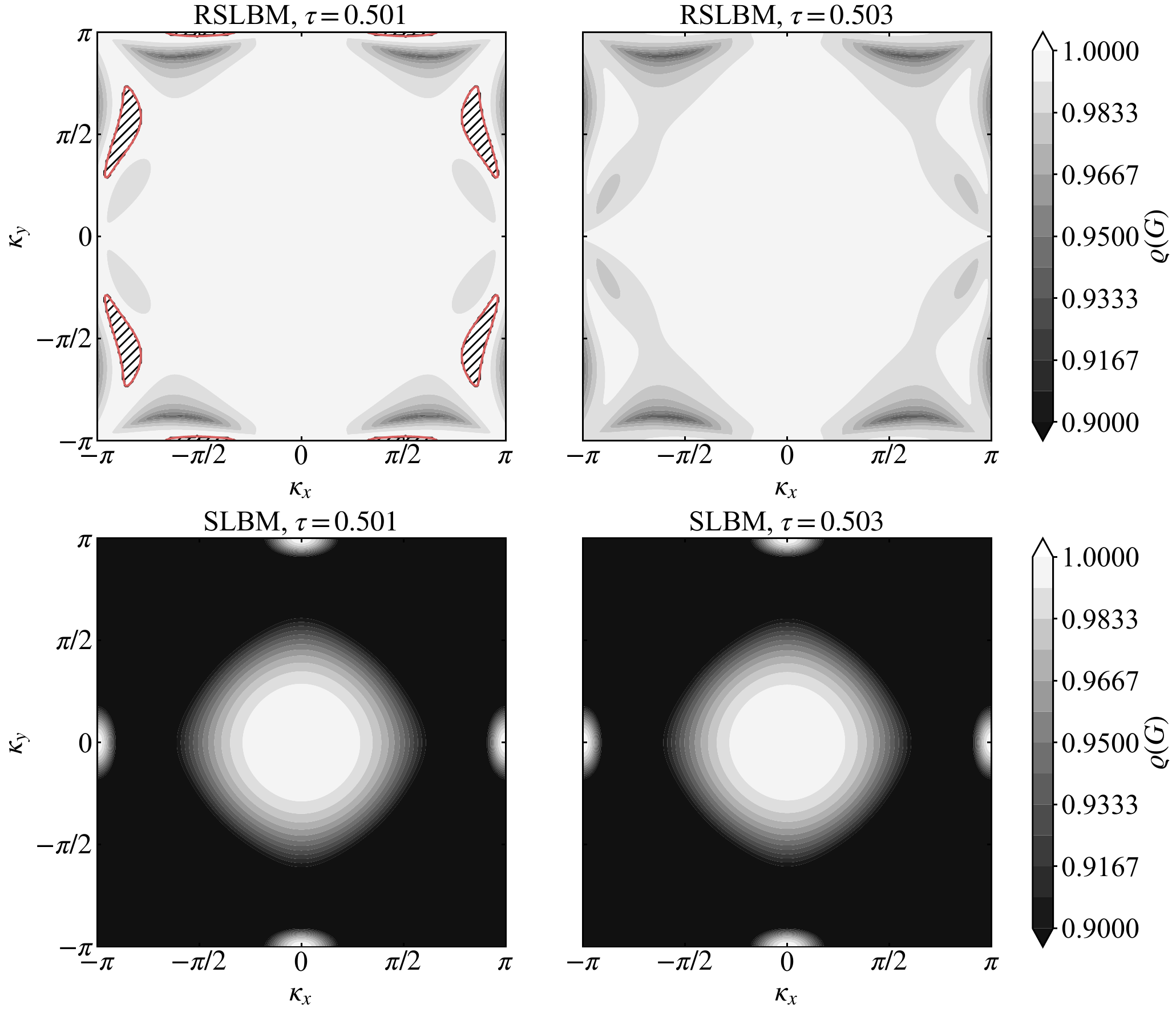}
\caption{Two-dimensional spectral-radius maps of the amplification matrix for RSLBM and SLBM using the second-order equilibrium at $Ma=0.1$, with $t_3=t_4=0$ for RSLBM. The red contour marks $\varrho(\mathsfbi{G})=1$; hatching indicates $\varrho(\mathsfbi{G})>1$, where at least one Fourier mode is linearly amplified by the numerical update.}
\label{fig:lsa_2d_spectral_radius_baseline}
\end{figure}

The two-dimensional spectral-radius maps in \figref{fig:lsa_2d_spectral_radius_baseline} show that, at $\tau=0.501$ with $t_3=t_4=0$, the RSLBM amplification matrix contains localized unstable areas that are not sufficiently characterized by the one-dimensional spectra alone. These regions should be interpreted as linear stability-risk indicators: they identify where the linearized update operator amplifies Fourier perturbations. The contrast with SLBM is also instructive. In the SLBM maps, the spectral radius is strongly suppressed over the high-wavenumber regions, especially near the edges and corners of the resolved dimensionless wavenumber plane. Thus high-frequency Fourier modes are heavily damped before they can grow, consistent with previous observations of high-wavenumber suppression in SLBM \citep{he2025}; this provides a linear explanation for the observed robustness of the original SLBM at low viscosity. The same mechanism, however, also anticipates its excessive finite-wavenumber dissipation.

\begin{figure}[!htbp]
\centering
\includegraphics[width=\linewidth]{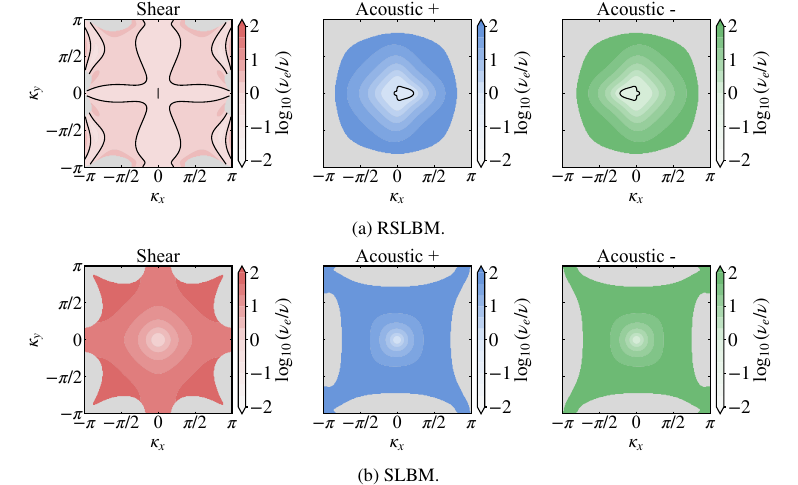}
\caption{Modal effective-viscosity ratios for RSLBM and SLBM at $\tau=0.503$, using the second-order equilibrium at $Ma=0.1$ and $t_3=t_4=0$ for RSLBM. Panels (a) and (b) show RSLBM and SLBM; in each panel, the columns show the shear, acoustic $+$ and acoustic $-$ branches. Colours represent $\log_{10}(\nu_{\mathrm{eff}}/\nu)$, where $\nu_{\mathrm{eff}}=-\operatorname{Im}(\omega_m)/|\boldsymbol{k}|^2$. Black contours mark unity; grey regions have eigenvector-projection scores below $0.9$; hatching denotes $\nu_{\mathrm{eff}}\leq0$.}
\label{fig:lsa_2d_modal_viscosity_tau0p503}
\end{figure}

At the less restrictive relaxation time $\tau=0.503$, the modal effective-viscosity maps in \figref{fig:lsa_2d_modal_viscosity_tau0p503} provide a cleaner comparison between RSLBM and SLBM. The shear-mode maps show that RSLBM reduces the excess finite-wavenumber damping relative to SLBM over most of the resolved dimensionless wavenumber plane. The acoustic-mode maps show more comparable damping between the two schemes. Together with the one-dimensional spectra, these results indicate a trade-off: RSLBM improves the shear-mode dissipation behaviour targeted by the reconstruction, whereas SLBM retains a more favorable acoustic phase relation along the streamwise cut.

\begin{figure}[!htbp]
\centering
\includegraphics[width=\linewidth]{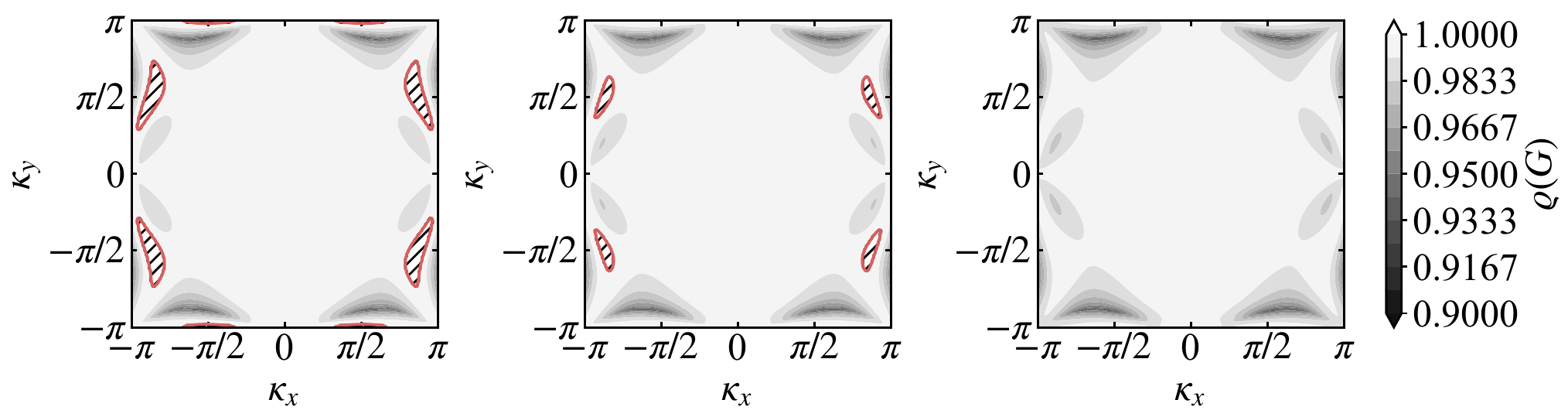}
\caption{Effect of the fourth-order predictor parameter $t_4$ on the two-dimensional spectral radius of RSLBM at $\tau=0.501$, using the second-order equilibrium at $Ma=0.1$ and $t_3=0$. From left to right, $t_4=0$, $-0.0002$ and $-0.0005$. The red contour marks $\varrho(\mathsfbi{G})=1$, and hatching indicates $\varrho(\mathsfbi{G})>1$, corresponding to linear amplification of at least one Fourier mode.}
\label{fig:lsa_t4_spectral_radius_control}
\end{figure}

\begin{figure}[!htbp]
\centering
\includegraphics[width=\linewidth]{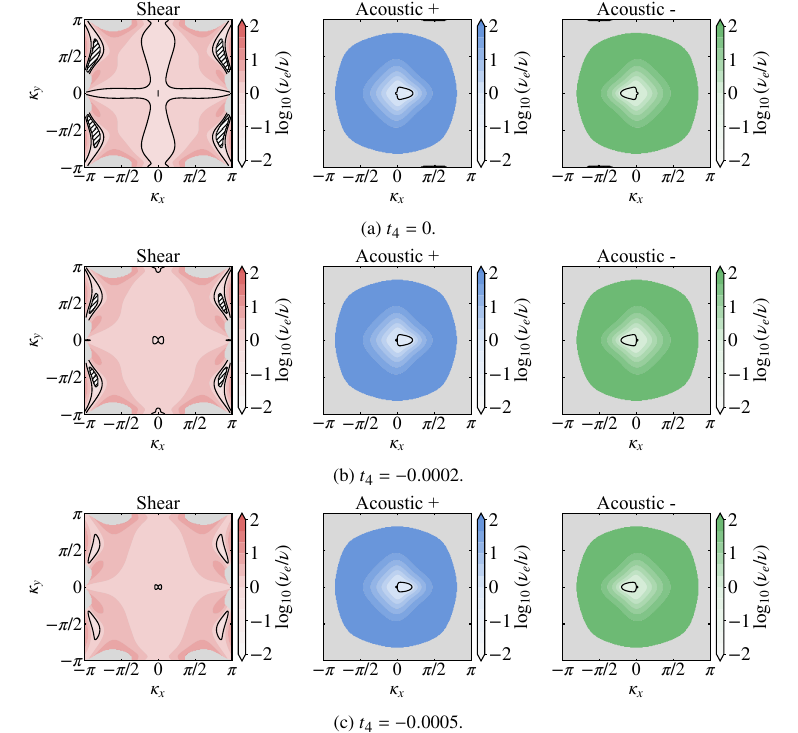}
\caption{Modal effective-viscosity ratios for RSLBM at $\tau=0.501$, using the second-order equilibrium at $Ma=0.1$ and $t_3=0$. Panels (a)--(c) correspond to $t_4=0$, $-0.0002$ and $-0.0005$; in each panel, the columns show the shear, acoustic $+$ and acoustic $-$ branches. Colours represent $\log_{10}(\nu_{\mathrm{eff}}/\nu)$, where $\nu_{\mathrm{eff}}=-\operatorname{Im}(\omega_m)/|\boldsymbol{k}|^2$. Black contours mark unity; grey regions have eigenvector-projection scores below $0.9$; hatching denotes $\nu_{\mathrm{eff}}\leq0$.}
\label{fig:lsa_t4_modal_viscosity_control}
\end{figure}

The unstable areas observed near $\tau=0.501$ can be controlled by the fourth-order predictor parameter $t_4$. With $t_3=0$, decreasing $t_4$ from zero to negative values reduces the spectral-radius excess and can suppress the unstable regions, as shown in \figref{fig:lsa_t4_spectral_radius_control}. The corresponding modal effective-viscosity maps in \figref{fig:lsa_t4_modal_viscosity_control} show the mechanism: a negative $t_4$ increases the effective viscosity of the shear and acoustic branches simultaneously. Thus, $t_4$ acts as a practical finite-wavenumber damping control, but it also introduces an accuracy trade-off. Mildly negative values reduce the most dangerous linear stability risks, whereas more negative values provide stronger robustness at the cost of over-damping physical modes.

This full-spectrum view also clarifies the role of the linear-wave validation below. The maps in this subsection are obtained from the amplification matrices over the continuous two-dimensional dimensionless wavenumber plane and include all branches. The linear-wave experiments in the next subsection instead excite one-dimensional transverse shear waves at the discrete wavenumbers allowed by finite periodic domains. They therefore validate the shear branch of the same LSA framework, but they do not replace the two-dimensional full-spectrum stability diagnosis.

\FloatBarrier

\subsection{Linear-wave validation of the shear mode}\label{sec:linear_wave_validation}

The one-dimensional linear-wave tests are used here to validate the shear branch predicted by the amplification-matrix analysis. The five RSLBM parameter sets listed in \tabref{tab:high_order_parameters} are used in this subsection to separate the roles of $t_3$ and $t_4$ in the fitted shear-wave dissipation and dispersion. The initial condition is a small-amplitude transverse shear wave imposed on a uniform streamwise background flow,
\begin{equation}
u_y=A\cos(\mathrm{k}x),
\qquad
u_x=U_{\mathrm{bg}},
\label{eq:wave_initial_condition}
\end{equation}
where $A$ is sufficiently small to ensure linearity. In the present tests, $A=10^{-4}$. Periodic boundary conditions are applied, and the domain length $L=N\delta x$ is chosen to be an integer multiple of the wavelength.

To control the wave resolution flexibly, the wavelength is written as
\begin{equation}
\lambda_w=\frac{p}{q}\delta x,
\label{eq:wave_wavelength}
\end{equation}
where $p$ and $q$ are integers. This means that $q$ wavelengths are resolved by $p$ mesh cells. The corresponding dimensionless wavenumber is
\begin{equation}
\kappa
=
k\delta x
=
\frac{2\pi\delta x}{\lambda_w}
=
\frac{2q\pi}{p}.
\label{eq:wave_wavenumber}
\end{equation}
Exact periodicity over the whole computational domain requires
\begin{equation}
\frac{L}{\lambda_w}
=
\frac{Nq}{p}
\in \mathbb N^+ .
\label{eq:wave_periodicity_condition}
\end{equation}
Thus, the numerical data do not sample the continuous two-dimensional dimensionless wavenumber plane considered in \secref{sec:lsa_wavenumber_maps}, but provide validation points at the finite set of streamwise discrete wavenumbers allowed by the periodic domain.

The quantities compared below are obtained from fits to the raw time series rather than from a single final-time amplitude. For pure decay with \(U_{\mathrm{bg}}=0\), the fitted decay rate gives the effective viscosity. For convected waves with \(U_{\mathrm{bg}}\ne0\), a damped phase-shifted waveform is fitted so that both the effective viscosity and the phase velocity can be extracted:
\begin{equation}
u_y(x,t)
=
A\exp(-\nu_{\mathrm{eff}}\mathrm{k}^2t)
\cos\left[\mathrm{k}(x-c_{\mathrm{num}}t)+\phi\right].
\label{eq:wave_fitted_form}
\end{equation}
Here \(c_{\mathrm{num}}\) is the fitted phase velocity of the shear wave, and \(\nu_{\mathrm{eff}}\) is the total effective viscosity inferred from the decay rate.

In the dissipation plots, the ordinate is expressed as the normalized numerical-viscosity contribution
\begin{equation}
\frac{\nu_{\mathrm{num}}}{\nu}
=
\frac{\nu_{\mathrm{eff}}-\nu}{\nu},
\qquad
\nu=\frac{\tau-1/2}{3},
\label{eq:wave_numerical_viscosity_ratio}
\end{equation}
where \(\nu\) is the physical lattice viscosity. Thus, \(\nu_{\mathrm{num}}\) measures the excess or deficit of damping relative to the target viscosity. In the logarithmic dissipation panels, \(|\nu_{\mathrm{num}}/\nu|\) is plotted, with positive and negative branches distinguished by the plotting convention. For convected waves, the dispersion panels report \(c_{\mathrm{num}}/U_{\mathrm{bg}}\), so that unity corresponds to the ideal advection speed imposed by the background flow.

\begin{figure}[!htbp]
\centering
\includegraphics[width=0.52\linewidth]{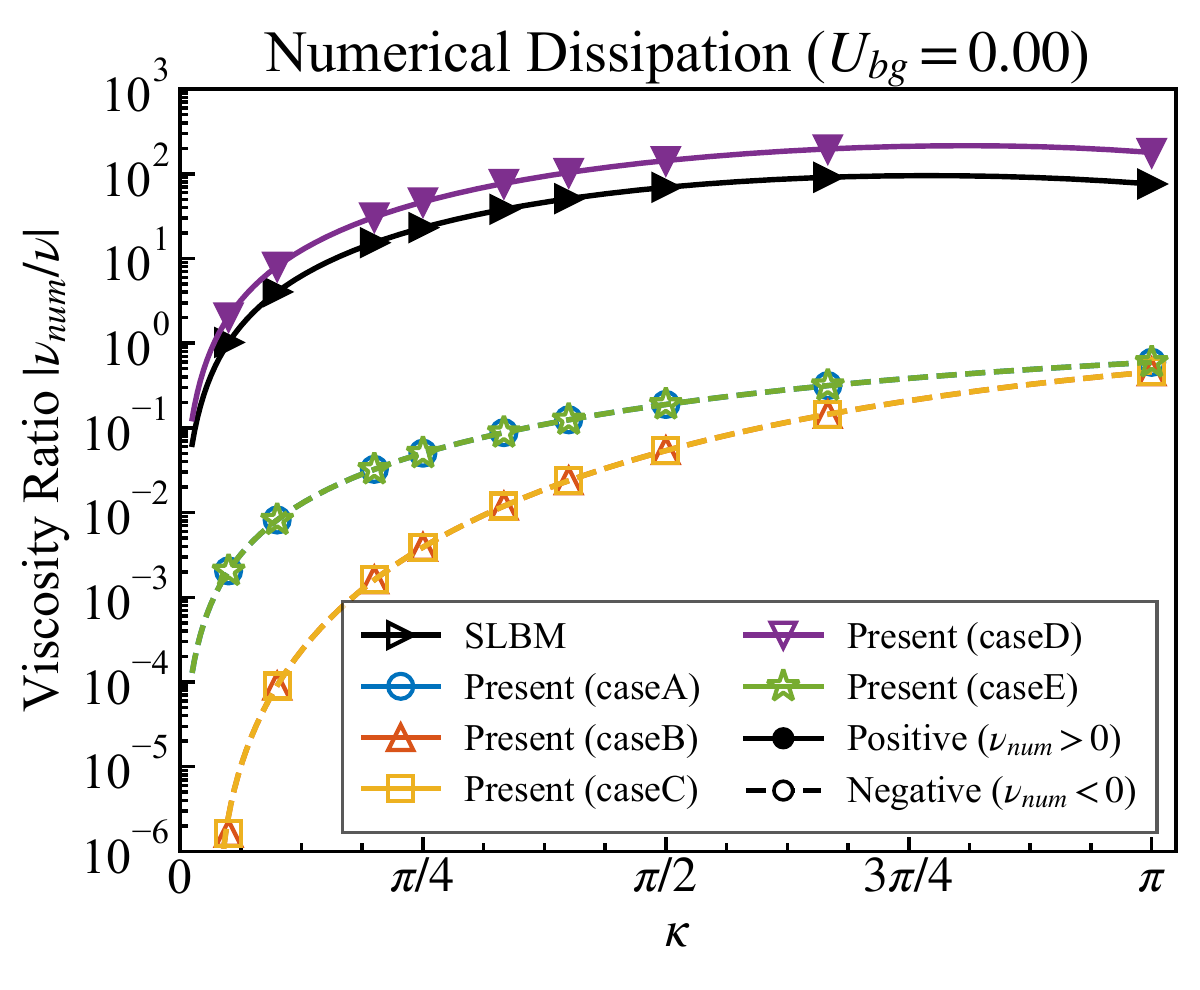}
\caption{Linear-wave validation of the LSA prediction for shear-mode dissipation at $\tau=0.501$ and $U_{\mathrm{bg}}=0$. The fitted numerical results are shown as markers, while the LSA predictions are shown as continuous curves.}
\label{fig:lwa_lsa_overlay_u0}
\end{figure}

\begin{figure}[!htbp]
\centering
\includegraphics[width=\linewidth]{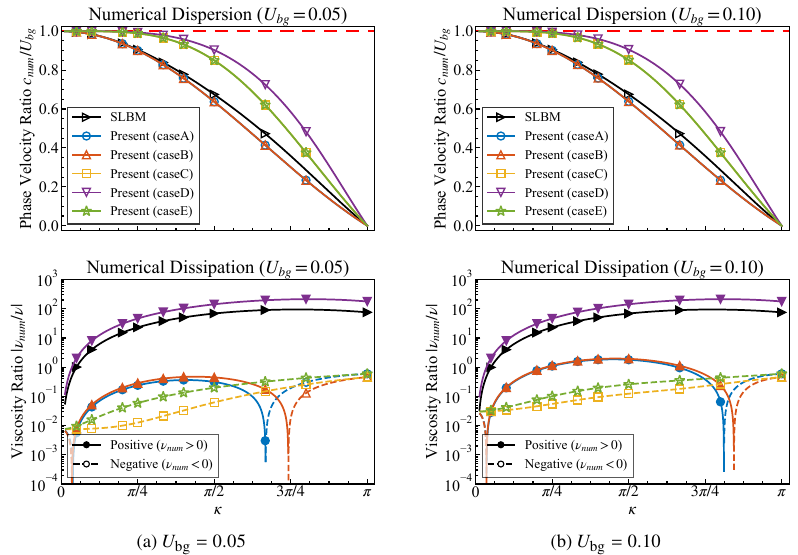}
\caption{Linear-wave validation of the LSA prediction for the convected shear branch at $\tau=0.501$ for (a) $U_{\mathrm{bg}}=0.05$ and (b) $U_{\mathrm{bg}}=0.10$. The fitted numerical results are shown as markers, while the LSA predictions are shown as continuous curves.}
\label{fig:lwa_lsa_overlay_convected}
\end{figure}

In the following comparison figures, e.g. \figref{fig:lwa_lsa_overlay_u0} and \figref{fig:lwa_lsa_overlay_convected}, the fitted numerical results are shown as markers, while the LSA predictions are shown as continuous curves. Only numerical markers associated with fits satisfying \(R^2>0.999\) are retained, so that the comparison is restricted to well-resolved linear-wave signals. The LSA curves are obtained by evaluating the identified shear eigenvalue on a dense wavenumber grid using the same amplification-matrix construction and eigenvector-projection procedure described above. Agreement between markers and curves therefore directly validates the shear eigenvalue predicted by the amplification-matrix analysis.

For the pure-decay case $U_{\mathrm{bg}}=0$, the normalized-dissipation markers lie on the LSA shear-mode curves, as shown in \figref{fig:lwa_lsa_overlay_u0}. This confirms that the finite-wavenumber dissipation measured from the time series is the same dissipation predicted by the logarithmic shear eigenvalue. The grouping of the curves is also consistent with the small-wavenumber expansion \eqnref{eq:lsa_rslbm_static_log}. At zero background velocity, $t_4$ enters the fourth-order real part of $\log\lambda_s$, while $t_3$ does not appear in the leading dissipation. The pure-decay data therefore isolate the role of $t_4$ as a finite-wavenumber damping control.

For convected shear waves, the same LSA branch predicts both the phase velocity and the effective viscosity. At $U_{\mathrm{bg}}=0.05$, the fitted phase-velocity markers follow the LSA curves, confirming that the observed dispersion is governed by the shear eigenvalue, as shown in \figpanelref{fig:lwa_lsa_overlay_convected}{a}. The grouping by $t_3$ follows from the third-order phase term in \eqnref{eq:lsa_rslbm_convected_log}. Since $C_3$ contains $t_3$, this parameter first appears as a dispersion control when the shear wave is convected by a background flow. At the same time, the fourth-order real coefficient contains contributions proportional to $U_{\mathrm{bg}}^2t_3$ as well as $t_4/3$, so $t_3$ can also affect the fitted dissipation of convected waves, while $t_4$ continues to control the baseline finite-wavenumber damping.

The comparison at $U_{\mathrm{bg}}=0.10$ reinforces this interpretation at a larger background velocity, as shown in \figpanelref{fig:lwa_lsa_overlay_convected}{b}. The dispersion grouping by $t_3$ becomes more visible, and the dissipation curves reflect the stronger velocity-dependent contribution of $t_3$. The numerical markers remain aligned with the LSA curves over the tested wavenumber range, indicating that the same shear-mode eigenvalue describes both the propagation and damping of the linear-wave experiments.

These results should be interpreted together with the two-dimensional maps in \secref{sec:lsa_wavenumber_maps}. The linear-wave tests validate the one-dimensional transverse-shear branch at discrete finite-domain wavenumbers. They do not test the acoustic branches, oblique wavenumber vectors or the full spectral-radius stability boundary. The role of the present subsection is therefore to show that the shear-mode quantities extracted from the numerical experiments are precisely those predicted by the amplification-matrix analysis, rather than to replace the full-spectrum LSA diagnosis.


\section{Numerical tests}\label{sec:numerical_tests}

The linear stability analysis and linear-wave validation in \secref{sec:linear_stability_analysis} provide a mechanistic understanding of the numerical dissipation, dispersion and stability of the original SLBM and the reformulated SLBM. In this section, we further assess how these properties manifest in nonlinear, multidimensional flow problems through three benchmarks. Specifically, the Taylor-Green vortex flow is used to examine accuracy and convergence. The double shear layer flow is employed to investigate numerical dissipation, dispersion, and stability in both resolved and under-resolved conditions. Finally, the lid-driven cavity flow is simulated to evaluate boundary treatment and the ability to resolve delicate flow structures on coarse meshes.

For clarity, the following abbreviations are used throughout this section. ``LBM'' denotes the standard BGK lattice Boltzmann method, while ``SLBM'' refers to the original simplified lattice Boltzmann method. ``MAME-2nd'' and ``MAME-4th'' denote solvers in which the predictor step of the MAME is discretized using second- and fourth-order finite-difference schemes, respectively, with the corrector step always solved by a fourth-order finite difference. ``Case A-E'' refers to the parametric combinations of the reformulated SLBM introduced in \tabref{tab:high_order_parameters}.

\subsection{Taylor-Green vortex flow}\label{sec:taylor_green_vortex}
To assess the accuracy of the reformulated SLBM, the two-dimensional Taylor-Green vortex flow is simulated using different numerical schemes. The analytical solution reads \citep{kruger2017}:
\begin{equation}
\begin{aligned}
\begin{cases}
u(x,y,t)
  &=-U_0\cos\left(\frac{2\pi x}{L}\right)
\sin\left(\frac{2\pi y}{L}\right)
\exp\left(-\frac{8\pi^2U_0t}{\mathrm{Re}\,L}\right),\\
v(x,y,t)
  &= U_0\sin\left(\frac{2\pi x}{L}\right)
\cos\left(\frac{2\pi y}{L}\right)
\exp\left(-\frac{8\pi^2U_0t}{\mathrm{Re}\,L}\right),\\
p(x,y,t)
  &= p_0-\frac{1}{4}\rho_0U_0^2
\left[
\cos\left(\frac{4\pi x}{L}\right)
        +\cos\left(\frac{4\pi y}{L}\right)
\right]
\exp\left(-\frac{16\pi^2U_0t}{\mathrm{Re}\,L}\right).
\end{cases}
\end{aligned}
\label{eq:tgv_analytical_solution}
\end{equation}
where $U_0$ denotes the amplitude of the initial velocity field; $\rho_0$ is the reference density; $p_0$ is the background pressure, taken as $p_0=\rho_0c_s^2$.

The size of the computational domain is set to be $[0,L]\times[0,L]$ and three uniform mesh sizes ($51\times51$, $76\times76$, $101\times101$) are employed to study grid convergence. The Reynolds number is defined as $\mathrm{Re}=U_0L/\nu$ and is fixed at 10, while $\tau=0.75$. Periodic boundary conditions are applied in both directions. Numerical solutions at the non-dimensional time $t^*=tU_0/L=0.5$ and $t^*=1.0$ are compared with the analytical solution. The relative error is quantified by the $L_2$ norm:
\begin{equation}
L_2
=\frac{
\sqrt{\sum_{i,j}\left(u_{i,j}^{\mathrm{numerical}}-u_{i,j}^{\mathrm{analytical}}\right)^2}
}{
\sqrt{\sum_{i,j}\left(u_{i,j}^{\mathrm{analytical}}\right)^2}
},
\label{eq:tgv_l2_error}
\end{equation}
where the superscripts ``numerical'' and ``analytical'' denote the numerical and the analytical solutions, respectively.

\begin{figure}[!htbp]
\centering
\includegraphics[width=\linewidth]{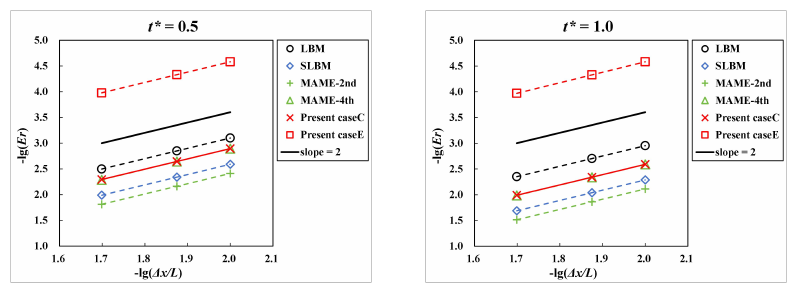}
\caption{Relative $L_2$ velocity errors versus mesh spacing for the Taylor--Green vortex at $\mathrm{Re}=10$, evaluated at $t^*=0.5$ (left) and $t^*=1.0$ (right). Slopes of the fitting lines for [LBM, SLBM, MAME-2nd, MAME-4th, present case C, present case E] are [1.9994, 1.9942, 1.9912, 2.0086, 1.9969, 2.0024] (left) and [1.9980, 1.9885, 1.9828, 2.0056, 1.9940, 2.0312] (right).}
\label{fig:tgv_relative_error_convergence}
\end{figure}

Figure~\ref{fig:tgv_relative_error_convergence} shows the convergence behaviour of all methods. As expected, all schemes exhibit second-order convergence, which is consistent with the second-order accuracy of the underlying BGK lattice Boltzmann framework. Among the tested methods, the original SLBM and MAME-2nd yield the largest relative errors. In contrast, both case C and case E of the reformulated SLBM show lower relative errors, while case E is consistently more accurate than case C on all mesh sizes. This improvement can be attributed to the reduced numerical dissipation associated with the fourth-order derivative term, as analysed in \secref{sec:linear_stability_analysis}. It is also observed that the error curves of case C almost coincide with those of MAME-4th. This agreement can be expected, since both approaches eliminate the third- and the fourth-order derivative error terms at the level of the reconstructed macroscopic equations. Nevertheless, the reformulated SLBM achieves this consistency with simpler formulations, avoiding the complex discretization required by fourth-order finite-difference schemes, especially for mixed derivatives such as $\partial_{xy}$. Although higher-order derivative terms ($O(\partial^n)$, $n>4$) differ between the two approaches, these differences do not affect the convergence behaviour in the present test. Their influence on stability, however, becomes evident in more challenging flow configurations and will be discussed in \secref{sec:double_shear_layer}.

\subsection{Double shear layer flow}\label{sec:double_shear_layer}

The double shear layer flow is employed to further examine the numerical dissipation, dispersion, and stability of the reformulated SLBM in a strongly convective and vortex-dominated flow. This problem involves sharp velocity gradients and rapid roll-up of shear layers, making it particularly sensitive to both numerical dissipation and dispersion, and is therefore well suited for validating the parametric effects identified in \secref{sec:linear_stability_analysis}.

The computational domain is $[0,L]\times[0,L]$ with periodic boundary conditions applied in both directions. The initial velocity field is prescribed as
\begin{equation}
\begin{aligned}
u &=
\begin{cases}
U_0\tanh\left[80\left(\dfrac{y}{L}-0.25\right)\right],
  & y/L\le 0.5,\\
U_0\tanh\left[80\left(0.75-\dfrac{y}{L}\right)\right],
  & y/L>0.5,
\end{cases}\\
v &= 0.05U_0\sin\left[2\pi\left(\frac{x}{L}+0.5\right)\right],
\end{aligned}
\label{eq:dsl_initial_condition}
\end{equation}
where $U_0$ is the characteristic velocity and is chosen as 0.04 here; the Reynolds number is set to be $\mathrm{Re}=U_0L/\nu=10\,000$. The simulations were run until $t^*=U_0t/L=2.0$. For clarity, the values of the parameters used in this section are summarized in \tabref{tab:dsl_high_order_values}. These parametric sets are designed to isolate the effects of the third- and the fourth-order derivative terms, enabling a direct comparison with the dissipation and dispersion characteristics predicted by the LSA and linear-wave validation.

\begin{table}
\centering
\renewcommand{\arraystretch}{1.22}
\begin{tabular*}{0.96\textwidth}{@{\extracolsep{\fill}}ccc@{}}
\toprule
    Case & $t_3$ & $t_4$\\
\midrule
    A & $-\frac{1}{6}$ & $\frac{\tau}{12}-\frac{1}{24}\approx2.56\times10^{-4}$\\
    B & $-\frac{1}{6}$ & $0$\\
    C & $0$ & $0$\\
    D & $0$ & $\frac{\tau}{12}-\frac{1}{8}\approx-0.0831$\\
    E & $0$ & $\frac{\tau}{12}-\frac{1}{24}\approx2.56\times10^{-4}$\\
\bottomrule
\end{tabular*}
\smallskip
\caption{Parameter values for RSLBM cases A--E in the double-shear-layer simulations on a $257\times257$ mesh. The numerical approximations are evaluated at $\tau=0.503072$.}
\label{tab:dsl_high_order_values}
\end{table}

\subsubsection{Well-resolved case on the mesh size of \texorpdfstring{$257\times257$}{257 by 257}: dissipation and dispersion}\label{sec:dsl_well_resolved}

\begin{figure}[!htbp]
\centering
\includegraphics[width=0.5\linewidth]{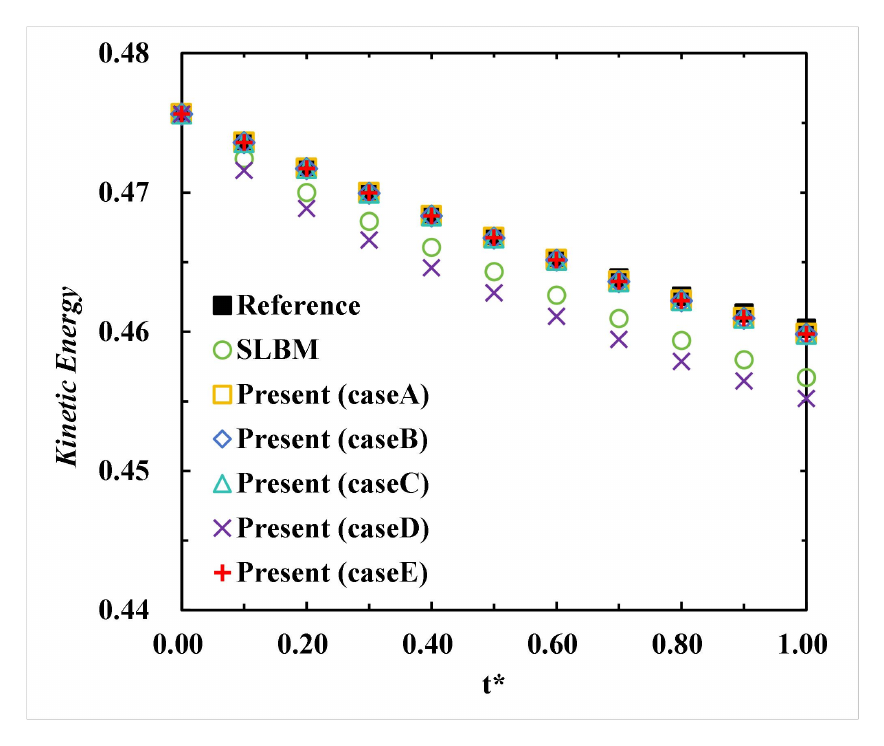}
\caption{Kinetic-energy decay obtained with different methods on a $257\times257$ mesh at $\mathrm{Re}=10\,000$. Reference data from the pseudo-spectral calculation of \citet{minion1997} are also shown.}
\label{fig:dsl_kinetic_energy_decay}
\end{figure}

The decay of kinetic energy is first examined to provide a global and quantitative measure of numerical dissipation. The kinetic energy is normalized and defined as
\begin{equation}
E_{k}=\frac{1}{U_0^2\Omega}
\iint_{\Omega}\frac{1}{2}\left(u^2+v^2\right)\,d\Omega,
\label{eq:dsl_kinetic_energy}
\end{equation}
and is compared with the reference results obtained by the pseudo-spectral method of \citet{minion1997}. As shown in \figref{fig:dsl_kinetic_energy_decay}, both the original SLBM and case D exhibit the fastest decay of the kinetic energy among all tested schemes. This behaviour is consistent with the strong finite-wavenumber dissipation identified in \secref{sec:linear_stability_analysis}. In contrast, the remaining cases of the present reformulated SLBM show significantly slower decay, indicating reduced numerical dissipation.

\begin{figure}[!htbp]
\centering
\includegraphics[width=\linewidth]{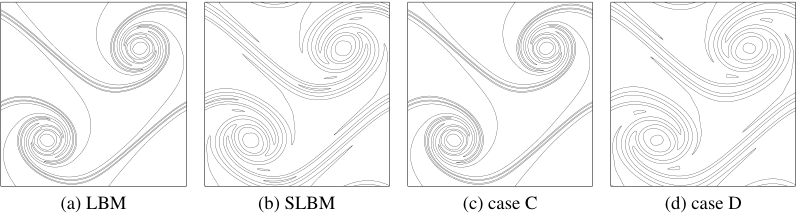}
\caption{Vorticity contours at $t^*=1.0$, with levels from $-70$ to $70$ in increments of 10. The reference LBM result in (a) uses a $1025\times1025$ mesh; the original SLBM in (b) and the present RSLBM cases C and D in (c,d) use a $257\times257$ mesh.}
\label{fig:dsl_vorticity_case_c_d}
\end{figure}

To further examine how numerical dissipation manifests in local flow structures, the vorticity contours at $t^*=1.0$ are first compared, as shown in \figref{fig:dsl_vorticity_case_c_d}. A clear qualitative difference can be observed among the tested methods. The original SLBM and case D exhibit nearly identical excessive dissipation, while case C preserves the finer flow structures. This behaviour is consistent with the dissipation characteristics predicted in \secref{sec:linear_stability_analysis}. It is worth noting that the dissipation of the original SLBM is so strong that even with a mesh resolution as fine as $1025\times1025$, the vortical structures remain severely damped.

\begin{figure}[!htbp]
\centering
\includegraphics[width=\linewidth]{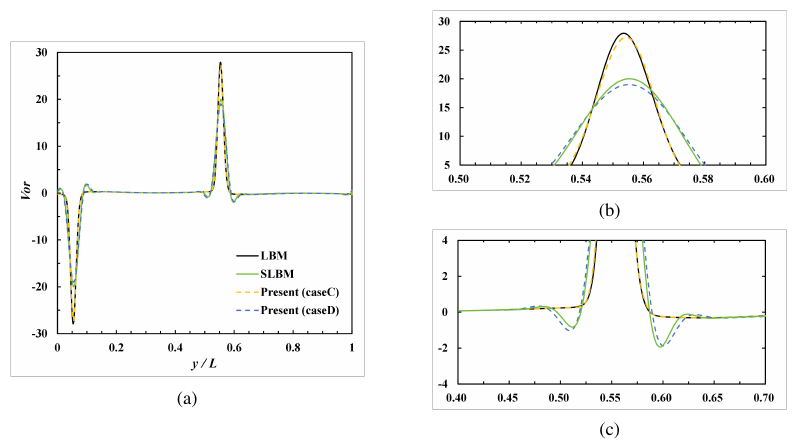}
\caption{Distribution of vorticity along the vertical centreline $x/L=0.5$ for the SLBM and the present RSLBM (cases C and D) at $t^*=1.0$. (a) Global view; (b,c) zoom-in views. The reference LBM result uses a $1025\times1025$ mesh.}
\label{fig:dsl_centerline_vorticity_case_c_d}
\end{figure}

To provide a more quantitative assessment, the vorticity distribution along the vertical centreline $x/L=0.5$ at $t^*=1.0$ is plotted in \figref{fig:dsl_centerline_vorticity_case_c_d} for the original SLBM and the reformulated SLBM (case C and case D). Compared with other methods, the vorticity peaks produced by the original SLBM and case D are significantly broader and lower, indicating both a reduction in peak intensity and an artificial diffusion of the shear layers. These features are typical signatures of excessive numerical dissipation and confirm, in a more quantitative manner, the observations of the vorticity contours.

\begin{figure}[!htbp]
\centering
\includegraphics[width=\linewidth]{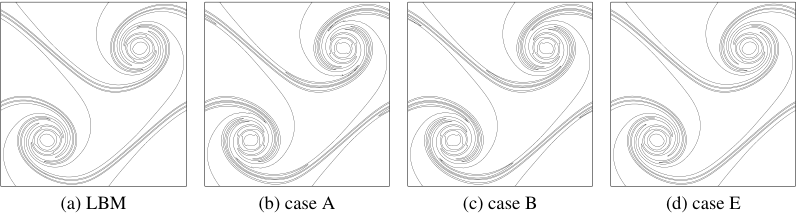}
\caption{Vorticity contours at $t^*=1.0$, with levels from $-70$ to $70$ in increments of 10. The reference LBM result in (a) uses a $1025\times1025$ mesh; the present RSLBM cases A, B and E in (b--d) use a $257\times257$ mesh.}
\label{fig:dsl_vorticity_case_a_b_e}
\end{figure}

The influence of numerical dispersion is examined next. Figure~\ref{fig:dsl_vorticity_case_a_b_e} presents the vorticity contours at $t^*=1.0$ for cases A, B, and E. In contrast to the strong dissipation observed in the original SLBM and case D, the overall vorticity structures produced by these three cases are close to the reference. This indicates that their numerical dissipation levels are comparable, consistent with their similar values of $t_4$. Only in some localized regions can slight deviations from the reference be identified by visual inspection, suggesting that the numerical dissipation alone is insufficient to distinguish their numerical behaviours at this level.

\begin{figure}[!htbp]
\centering
\includegraphics[width=\linewidth]{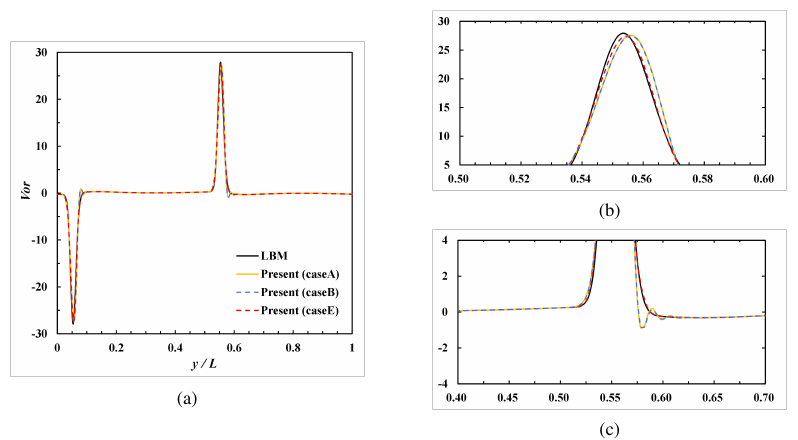}
\caption{Distribution of vorticity along the vertical centreline $x/L=0.5$ for the present RSLBM (cases A, B and E) at $t^*=1.0$. (a) Global view; (b,c) zoom-in views. The reference LBM result uses a $1025\times1025$ mesh.}
\label{fig:dsl_centerline_vorticity_case_a_b_e}
\end{figure}

To further quantify these differences, the vorticity distribution along the vertical centreline $x/L=0.5$ is plotted in \figref{fig:dsl_centerline_vorticity_case_a_b_e}. Physically, the vorticity structures should be convected from the right to the left in \figref{fig:dsl_centerline_vorticity_case_a_b_e}, corresponding to downward transport in the computational domain. All three cases produce peak amplitudes that are very close to the reference LBM solution, again confirming that their numerical dissipation is nearly the same. However, clear differences emerge in the phase behaviour. For case A and case B, noticeable phase delays are observed, together with small oscillations near the downstream side of the vorticity peak, particularly near its base. These features indicate that part of the high-wavenumber components propagate with an underestimated phase velocity, a characteristic manifestation of numerical dispersion. In contrast, when the third-order derivative term is eliminated by setting $t_3=0$, as in case E (and also case C shown previously in \figref{fig:dsl_centerline_vorticity_case_c_d}), both the phase lag and the spurious oscillations disappear. This observation is consistent with the shear-mode dispersion tendency predicted in \secref{sec:linear_wave_validation} and further confirms that the third-order derivative terms play a dominant role in controlling numerical dispersion.

\begin{figure}[!htbp]
\centering
\includegraphics[width=\linewidth]{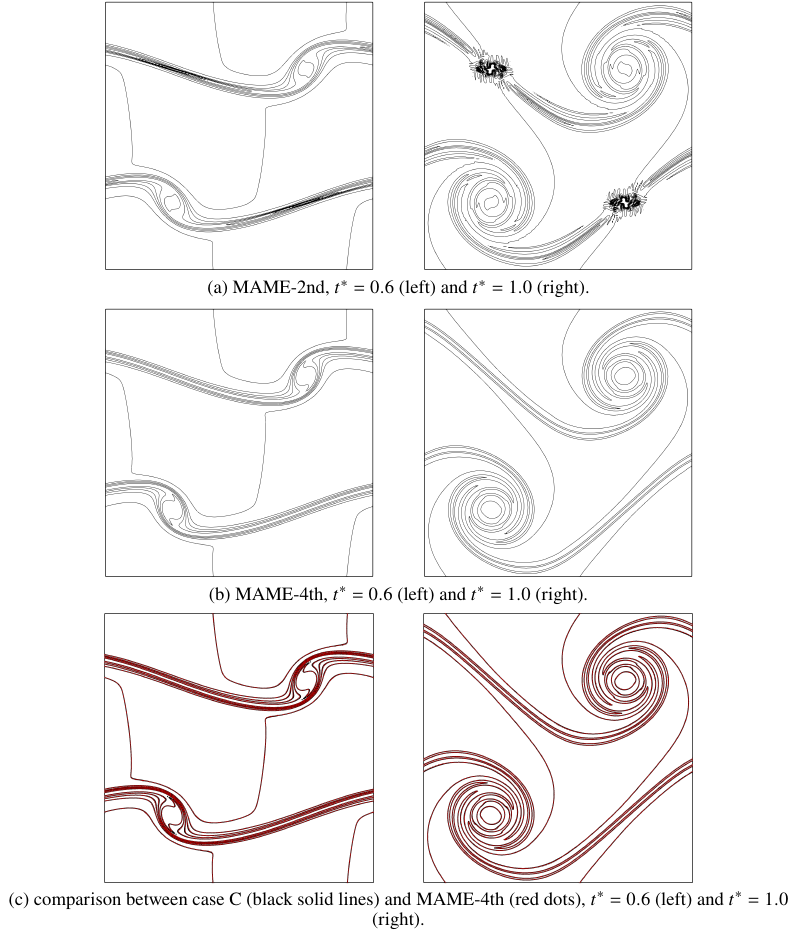}
\caption{Vorticity contours on the $257\times257$ mesh at $t^*=0.6$ (left) and $t^*=1.0$ (right), with levels from $-70$ to $70$ in increments of 10. (a) MAME with second-order finite differences; (b) MAME with fourth-order finite differences; (c) comparison between RSLBM case C (black solid lines) and MAME-4th (red dots).}
\label{fig:dsl_mame_vorticity_comparison}
\end{figure}

For comparison, the performance of finite-difference-based MAME solvers is also examined and the results are presented in \figref{fig:dsl_mame_vorticity_comparison}. Although both MAME-2nd and MAME-4th initially produce convergent solutions, the results obtained with MAME-2nd exhibit spurious oscillations in the upper-left and bottom-right regions of the domain. As the simulation proceeds beyond $t^*=1.0$, these oscillations develop into two non-physical secondary spurious vortices. Such behaviour is commonly associated with numerical dispersion \citep{coreixas2025} and could be attributed to the presence of the specific uncanceled third-order derivative terms of MAME-2nd, which inevitably arise when the second-order finite-difference schemes are employed. In contrast, MAME-4th does not suffer from this issue and produces physically consistent flow structures.

Finally, a direct comparison between case C of the reformulated SLBM and MAME-4th is presented in \figpanelref{fig:dsl_mame_vorticity_comparison}{c}. The vorticity contours obtained by the two approaches overlap almost perfectly, confirming that both methods effectively eliminate the third- and the fourth-order derivative errors at the macroscopic level. This observation is fully consistent with the results of the Taylor-Green vortex test in \secref{sec:taylor_green_vortex} and further demonstrates that the present reformulated SLBM can achieve the accuracy of the fourth-order finite-difference MAME solver.

\subsubsection{Under-resolved case on the mesh size of \texorpdfstring{$129\times129$}{129 by 129}: stability test}\label{sec:dsl_underresolved}
Having examined the dissipation and dispersion characteristics of different methods on a well-resolved mesh in the previous subsection, we now turn to a more stringent test that focuses on numerical stability. The under-resolved mesh of $129\times129$ is employed for the double shear layer flow, a configuration known to be highly sensitive to numerical errors due to the rapid development of small-scale vortical structures \citep{minion1997,coreixas2017,kramer2017}.

Since the reformulated SLBM and the MAME-nth formulations are designed to recover the same macroscopic equations, it is natural to directly compare their stability properties. In the present simulations, both MAME-2nd and MAME-4th fail to produce stable solutions and diverge at an early stage of $t^*\leq 0.2$. In contrast, the reformulated SLBM with case C remains stable throughout the simulation and yields physically meaningful flow structures.

As discussed in \secref{sec:taylor_green_vortex}, case C and MAME-4th both eliminate the leading third- and fourth-order derivative error terms and therefore exhibit nearly identical accuracy in well-resolved regimes. Their different behaviour in the under-resolved double-shear-layer test thus provides a relatively clean indication that the remaining stability gap is mainly associated with higher-order residuals beyond the fourth-order equivalent-equation level. In this sense, the distinctive high-order residual structure of the reformulated SLBM supplies an additional numerical stabilization mechanism, which becomes particularly relevant when the mesh resolution is insufficient.

\begin{figure}[!htbp]
\centering
\includegraphics[width=\linewidth]{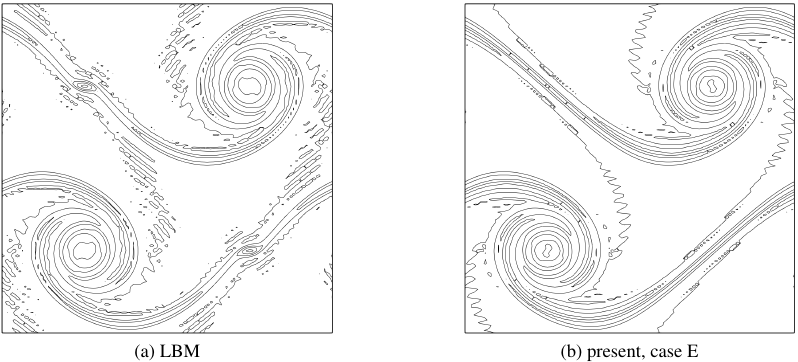}
\caption{Vorticity contours at $t^*=1.0$ on the under-resolved $129\times129$ mesh, with levels from $-70$ to $70$ in increments of 10: (a) LBM; (b) present RSLBM (case E).}
\label{fig:dsl_underresolved_vorticity}
\end{figure}

After clarifying this case-C/MAME-4th comparison, we further assess the practical under-resolved performance of the reformulated SLBM using case E, the low-dissipation setting identified in \secref{sec:linear_stability_analysis}. This test focuses on both numerical stability and the possible formation of secondary spurious vortices. The results are summarized in \tabref{tab:stability_methods_mesh}. Case E remains stable on all tested meshes, even down to $17\times17$, whereas the MAME-based solvers diverge under comparable coarse-resolution conditions. On the $129\times129$ mesh, case E also avoids the secondary spurious vortices observed in the LBM result, as shown in \figref{fig:dsl_underresolved_vorticity}. Taken together, these observations indicate that the reformulated SLBM can retain the low-dissipation behaviour of case E while benefiting from the additional stabilization provided by its higher-order residual structure.

\begin{table}
\centering
\renewcommand{\arraystretch}{1.08}
\begin{tabular*}{0.92\textwidth}{@{\extracolsep{\fill}}llcc@{}}
\toprule
    Mesh & Method & Spurious vortex & Status\\
\midrule
    $257\times257$ & Present, case E & N & Stable\\
    & LBM & N & Stable\\
    & MAME-2nd & Y & Stable\\
    & MAME-4th & N & Stable\\
\midrule
    $129\times129$ & Present, case E & N & Stable\\
    & LBM & Y & Stable\\
    & MAME-2nd & $\ldots$ & Diverged\\
    & MAME-4th & $\ldots$ & Diverged\\
\midrule
    $65\times65$ & Present, case E & Y & Stable\\
    & LBM & Y & Stable\\
    & MAME-2nd & $\ldots$ & Diverged\\
    & MAME-4th & $\ldots$ & Diverged\\
\midrule
    $33\times33$ & Present, case E & Y & Stable\\
    & LBM & $\ldots$ & Diverged\\
    & MAME-2nd & $\ldots$ & Diverged\\
    & MAME-4th & $\ldots$ & Diverged\\
\midrule
    $17\times17$ & Present, case E & Y & Stable\\
    & LBM & $\ldots$ & Diverged\\
    & MAME-2nd & $\ldots$ & Diverged\\
    & MAME-4th & $\ldots$ & Diverged\\
\bottomrule
\end{tabular*}
\smallskip
\caption{Stability and occurrence of spurious vortices for different methods at various mesh sizes. ``Stable'' denotes completion to $t^{*}=2.0$ without numerical divergence; ``Y'' and ``N'' indicate the presence and absence of spurious vortices, respectively; $\ldots$ denotes numerical divergence.}
\label{tab:stability_methods_mesh}
\end{table}

\subsection{Lid-driven cavity}\label{sec:lid_driven_cavity}

The two-dimensional lid-driven cavity flow is a widely used benchmark for incompressible flow solvers. A square cavity of size $L=1$ is considered, in which the top lid moves with a constant velocity $U=0.1$ while the remaining three walls are stationary. The Reynolds number is defined as $Re=UL/\nu $ and is set to $\mathrm{Re}=7500$, corresponding to a high-Reynolds-number case with steady-state solution. All simulations are advanced in time until convergence to the steady state. The convergence criterion here is defined as
\begin{equation}
\frac{
\sqrt{\sum_{i,j}\left(u_{i,j}^{t+\delta t}-u_{i,j}^{t}\right)^2}
}{
\sqrt{\sum_{i,j}\left(u_{i,j}^{t+\delta t}\right)^2}
}<10^{-8}.
\label{eq:cavity_convergence_criterion}
\end{equation}
The mesh spacing is denoted by $\delta x=1/N$, where $N$ is the number of mesh cells along each direction, and the mesh size is then $(N+1)\times(N+1)$. The performance of the present reformulated SLBM is compared with the original SLBM. Based on the stability maps in \secref{sec:lsa_wavenumber_maps} and the numerical tests in \secref{sec:double_shear_layer}, case C $(t_4=0)$ is employed for $N=100$ to ensure stability on the coarsest meshes, while case E $(t_4=\tau/12-1/24)$ is adopted in finer mesh resolutions with $N\geq 150$, where improved accuracy can be achieved without compromising robustness.

As a five-point-stencil scheme, the reformulated SLBM requires special treatment near the boundaries, where insufficient neighbouring nodes are available. Two possible strategies can be considered. The first is to switch locally to case A, which reduces the stencil to three points. Also, the continuity reconstruction can be reduced independently using the analogous coefficient constraints. The second is to adopt a hybrid formulation by coupling the present method with MAME-nth, which solves the same macroscopic equations and offers greater flexibility in handling boundary nodes.

Although case A is attractive from the implementation standpoint, the shear-mode dissipation analysis in \secref{sec:linear_wave_validation} indicates that, in the presence of relatively high velocities, case A introduces excessive numerical dissipation due to the contribution of the third-order derivative terms. This effect is particularly pronounced near the moving lid, where the local velocity is the highest. While such dissipation may not significantly affect short-time simulations, such as the double shear layer flow in \secref{sec:double_shear_layer}, it becomes detrimental in steady problems like the lid-driven cavity, which essentially corresponds to long-time integrations of a transient solver. Consequently, the use of case A near the boundary leads to overly diffused boundary layers and incorrect velocity distributions.

\begin{figure}[!htbp]
\centering
\includegraphics[width=0.5\linewidth]{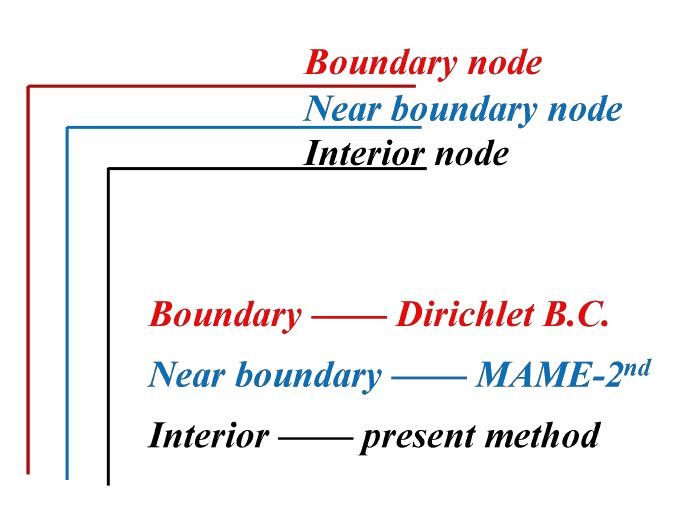}
\caption{Schematic of the hybrid boundary treatment. Dirichlet boundary conditions are imposed at boundary nodes, MAME-2nd is applied at near-boundary nodes, and the present RSLBM is used in the interior.}
\label{fig:cavity_boundary_treatment}
\end{figure}

For this reason, a hybrid scheme is adopted in which the interior flow is solved using the reformulated SLBM, while the near-boundary region is treated with MAME-2nd, which has been shown to be accurate and robust for the lid-driven cavity simulations in previous studies \citep[see][]{lu2020}. The coupling strategy is illustrated in \figref{fig:cavity_boundary_treatment}. This hybrid approach preserves the low-dissipation properties of the present method in the bulk flow while maintaining numerical robustness near the boundaries.

\begin{figure}[!htbp]
\centering
\includegraphics[width=\linewidth]{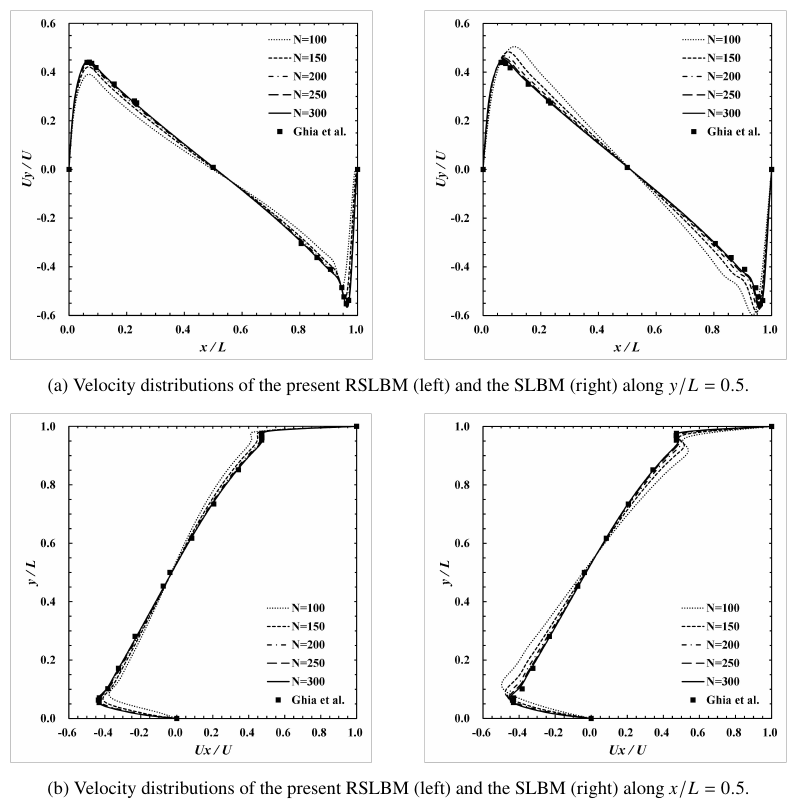}
\caption{Velocity distributions along centrelines for the lid-driven cavity flow at $\mathrm{Re}=7500$. (a) $U_y/U$ along $y/L=0.5$; (b) $U_x/U$ along $x/L=0.5$. In each row, the present RSLBM and original SLBM are shown on the left and right, respectively; symbols denote the reference data from \citet{ghia1982}.}
\label{fig:cavity_centerline_velocity}
\end{figure}

We first assess the accuracy of large-scale vortex structures by comparing velocity profiles along the horizontal and the vertical centrelines with the reference data of \citet{ghia1982}, as shown in \figref{fig:cavity_centerline_velocity}. The reformulated SLBM achieves satisfactory agreement with the reference solution at the resolution of $\delta x=1/200$, whereas the original SLBM requires a significantly finer mesh of $\delta x=1/300$ to obtain comparable accuracy. This difference is consistent with the dissipation analysis in \secref{sec:linear_stability_analysis}, which showed that the original SLBM introduces substantial numerical viscosity even at relatively low wavenumbers, leading to excessive damping of large-scale flow features.

\begin{figure}[!htbp]
\centering
\includegraphics[width=\linewidth]{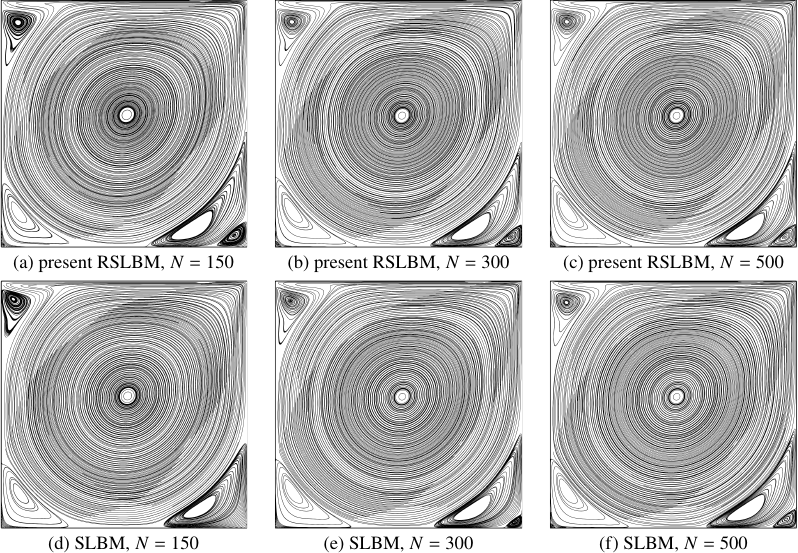}
\caption{Steady-state streamlines for the lid-driven cavity flow at $\mathrm{Re}=7500$ obtained by the present RSLBM and the original SLBM at three resolutions. The first row shows the present RSLBM, the second row shows the original SLBM, and the columns correspond to $N=150$, $N=300$ and $N=500$, respectively.}
\label{fig:cavity_streamlines_resolution}
\end{figure}

\begin{figure}[!htbp]
\centering
\includegraphics[width=\linewidth]{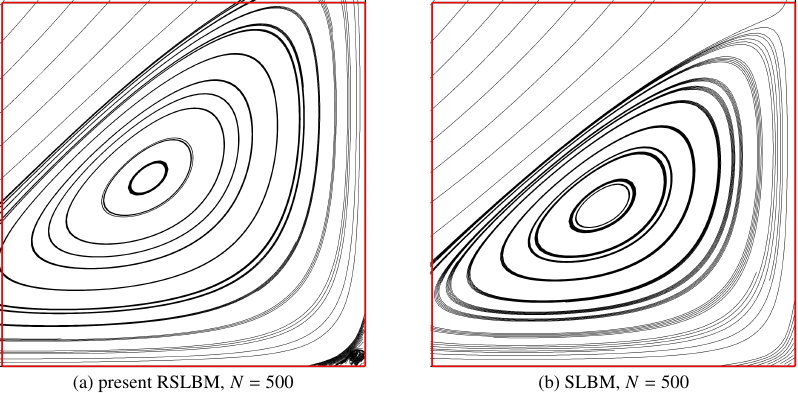}
\caption{Local streamline structures in the region of the bottom-right BR3 vortex for the lid-driven cavity flow at $\mathrm{Re}=7500$ and $N=500$: (a) present RSLBM; (b) original SLBM.}
\label{fig:cavity_br3_streamlines}
\end{figure}

A more stringent comparison is performed by examining the streamline patterns, which illustrate the method's ability to capture small-scale vortical structures. According to the reference solution of \citet{ghia1982}, the lid-driven cavity flow at $\mathrm{Re}=7500$ consists of one primary central vortex, one top-left vortex (TL1), two bottom-left vortices (BL1 and BL2), and three bottom-right vortices (BR1, BR2 and BR3). As shown in \figref{fig:cavity_streamlines_resolution}, at a relatively coarse resolution of $\delta x=1/150$, the present method successfully resolves the secondary vortices in both bottom corners, including the small-scale BL2 and BR2 structures. In contrast, the original SLBM completely omits these vortices at the same resolution due to its strong numerical dissipation. Although the BR2 vortex begins to appear in the original SLBM solution at $\delta x=1/300$, accurate resolution of both BL2 and BR2 is not achieved until the mesh is refined to $\delta x=1/500$. This enhanced accuracy is further quantified in \tabref{tab:cavity_vortex_locations}, which compares the locations of vortex centres with the reference data and shows good agreement for the reformulated SLBM at all tested resolutions. Although the BR3 vortex is not resolved accurately, the reformulated SLBM with $\delta x=1/500$ recovers it, as illustrated in \figpanelref{fig:cavity_br3_streamlines}{a}, whereas the original SLBM fails to capture it in \figpanelref{fig:cavity_br3_streamlines}{b}.

The performance of MAME-nth is comparable to the reformulated SLBM in terms of accuracy when sufficiently fine meshes are used. However, the finite-difference-based MAME-nth schemes remain stable only in resolutions finer than $\delta x\leq 1/200$, whereas the present hybrid scheme maintains stability on coarser meshes.

Overall, these results demonstrate that the proposed hybrid RSLBM--MAME-2nd implementation effectively combines the low-dissipation properties of the reformulated SLBM with the boundary robustness of MAME-2nd, enabling accurate and stable simulation of high-Reynolds-number flows with a substantially lower resolution requirement than the original SLBM.

\FloatBarrier
\begingroup
\makeatletter
\def\fps@table{!t}
\makeatother
\begin{table}
\centering
\scriptsize
\setlength{\tabcolsep}{3.5pt}
\renewcommand{\arraystretch}{1.05}
\begin{tabular*}{0.96\textwidth}{@{\extracolsep{\fill}}llrrrrrr@{}}
\toprule
    & &\multicolumn{2}{c}{Primary} &\multicolumn{2}{c}{TL1} &\multicolumn{2}{c}{BL1}\\
\cmidrule(lr){3-4}\cmidrule(lr){5-6}\cmidrule(l){7-8}
    Method & Mesh & $X$ & $Y$ & $X$ & $Y$ & $X$ & $Y$\\
\midrule
    Present & $151\times151$ & 0.5098 & 0.5354 & 0.0679 & 0.9091 & 0.0658 & 0.1428\\
    & $251\times251$ & 0.5128 & 0.5324 & 0.0674 & 0.9112 & 0.0650 & 0.1507\\
    & $301\times301$ & 0.5130 & 0.5321 & 0.0673 & 0.9115 & 0.0648 & 0.1514\\
    SLBM & $151\times151$ & 0.5137 & 0.5348 & 0.0493 & 0.9271 & 0.0743 & 0.1415\\
    & $251\times251$ & 0.5137 & 0.5321 & 0.0565 & 0.9236 & 0.0680 & 0.1506\\
    & $301\times301$ & 0.5137 & 0.5318 & 0.0620 & 0.9184 & 0.0668 & 0.1517\\
\multicolumn{2}{l}{Ghia et al.} & 0.5117 & 0.5322 & 0.0664 & 0.9141 & 0.0645 & 0.1504\\
\bottomrule
\end{tabular*}
\vspace{0.5em}
\begin{tabular*}{0.96\textwidth}{@{\extracolsep{\fill}}llrrrrrr@{}}
\toprule
    & &\multicolumn{2}{c}{BL2} &\multicolumn{2}{c}{BR1} &\multicolumn{2}{c}{BR2}\\
\cmidrule(lr){3-4}\cmidrule(lr){5-6}\cmidrule(l){7-8}
    Method & Mesh & $X$ & $Y$ & $X$ & $Y$ & $X$ & $Y$\\
\midrule
    Present & $151\times151$ & 0.0086 & 0.0091 & 0.7728 & 0.0693 & 0.9458 & 0.0477\\
    & $251\times251$ & 0.0105 & 0.0108 & 0.7888 & 0.0664 & 0.9521 & 0.0413\\
    & $301\times301$ & 0.0107 & 0.0111 & 0.7899 & 0.0660 & 0.9523 & 0.0411\\
    SLBM & $151\times151$ & $\ldots$ & $\ldots$ & 0.7982 & 0.0703 & $\ldots$ & $\ldots$\\
    & $251\times251$ & $\ldots$ & $\ldots$ & 0.7883 & 0.0651 & 0.9761 & 0.0210\\
    & $301\times301$ & $\ldots$ & $\ldots$ & 0.7881 & 0.0642 & 0.9691 & 0.0258\\
\multicolumn{2}{l}{Ghia et al.} & 0.0117 & 0.0117 & 0.7813 & 0.0625 & 0.9492 & 0.0430\\
\bottomrule
\end{tabular*}
\smallskip
\caption{Locations of vortex centres at $\mathrm{Re}=7500$ predicted by the present RSLBM and original SLBM, together with the reference data from \citet{ghia1982}. $X$ and $Y$ denote dimensionless coordinates in the $x$- and $y$-directions; $\ldots$ denotes an unresolved vortex.}
\label{tab:cavity_vortex_locations}
\end{table}
\endgroup


\section{Conclusion}\label{sec:conclusion}

This work establishes a comprehensive understanding of the numerical dissipation, dispersion, and stability properties of the simplified lattice Boltzmann method (SLBM). By introducing a generalized SLBM formulation, it is shown that the macroscopic equations recovered by the SLBM contain intrinsic physical deviations and numerical truncation errors compared to the physically consistent reference equations derived from the BGK lattice Boltzmann equation, which explains the numerical limitations of the SLBM exhibited in practical simulations.

To address these limitations, the reformulated SLBM (RSLBM) is proposed. The RSLBM utilizes the more actual macroscopic equation (MAME) as the target macroscopic equations to be recovered and thus successfully removes the physical deviations in the original SLBM. The predictor-corrector strategy is used to reconstruct macroscopic solutions to the target equations. The predictor step is constructed within the generalized SLBM framework with explicitly controllable high-order parameters $t_3$ and $t_4$, while the corrector step is based on the finite-difference discretization. Linear stability analysis and linear-wave validation reveal the distinct contributions of these parameters. The third-order derivative term governed by $t_3$ dominates the convected shear-mode dispersion and, in the presence of background flow, introduces additional velocity-dependent numerical dissipation, whereas the fourth-order derivative term associated with $t_4$ determines the baseline dissipation level and stability characteristics. These results provide a physically motivated and practically useful guideline for the parameter selection of the RSLBM.

Numerical experiments are then conducted to validate the analysis. The Taylor-Green vortex demonstrates accuracy and convergence comparable to the finite-difference schemes. The double shear layer problem confirms the predictions in LSA: the RSLBM shows reduced dispersion and dissipation errors and good stability, with high-order derivative terms providing additional stabilization. The lid-driven cavity flow further illustrates the capability of the RSLBM in capturing delicate vortical structures on relatively coarser grids than the original SLBM while maintaining numerical stability.

In summary, the RSLBM retains the simplicity and memory efficiency of the original SLBM while significantly reducing its dissipation and dispersion. The free parameters in the RSLBM allow flexible adjustment of the magnitude of the high-order derivative terms and thus contribute to fine tuning its numerical performance without sacrificing the second-order accuracy. Future work will be focused on adaptive parameter selection based on local flow conditions, extensions to various lattice stencils, complex geometries, and applications to mesh refinement strategies.


\begin{bmhead}[Funding]
This work was supported by the National Natural Science Foundation of China (grant number 52471333) and the Fundamental Research Funds for the Central Universities.
\end{bmhead}

\begin{bmhead}[Competing interests]
The authors report no conflict of interest.
\end{bmhead}

\begin{bmhead}[Data availability statement]
The data that support the findings of this study are available from the corresponding author upon reasonable request.
\end{bmhead}

\begin{bmhead}[Author contributions]
\textbf{Zhengwei He:} Conceptualization (equal); Data curation (lead); Investigation (lead); Methodology (equal); Validation (lead); Writing -- original draft (lead). \textbf{Zhen Chen:} Conceptualization (equal); Funding acquisition (lead); Methodology (equal); Project administration (lead); Supervision (lead); Writing -- review and editing (lead).
\end{bmhead}

\begin{appen}
\renewcommand{\theHsection}{appendix.\thesection}
\renewcommand{\theHsubsection}{appendix.\thesubsection}
\renewcommand{\theHsubsubsection}{appendix.\thesubsubsection}
\renewcommand{\theHequation}{appendix.\thesection.\arabic{equation}}

\section{Derivation of the generalized SLBM scheme}\label{app:gslbm_derivation}

Inspired by the simplified lattice Boltzmann method (SLBM), this appendix constructs the general mathematical formulation that preserves the equilibrium-based reconstruction principle underlying the SLBM and will be used to analyse the properties of SLBM-type approaches in \appref{app:slbm_analysis}. For the algebra below, the relaxation parameter $\tau=\frac{\nu}{c_s^2\delta t}+\frac{1}{2}$ is treated as a constant within each parameter set. Its variation across meshes is not prescribed here; any refinement analysis must additionally specify the joint scaling of $\delta x$, $\delta t$, $\nu$, and $\tau$, as discussed in \appref{app:slbm_error_scaling}.

We first focus on the predictor step and introduce two auxiliary functions used to reconstruct the intermediate macroscopic quantities from equilibrium information. The density and momentum auxiliary functions are given, respectively, by
\begin{equation}
\begin{aligned}
h_{\rho}=&A_{\rho}\cdot\sum_{i}f_{i}^{\mathrm{eq}}\left(\boldsymbol{r}-e_{i}\delta t,t-\delta t\right)+B_{\rho}\cdot\sum_{i}f_{i}^{\mathrm{eq}}\left(\boldsymbol{r}-2e_{i}\delta t,t-\delta t\right)\\
&\quad +C_{\rho}\cdot\sum_{i}f_{i}^{\mathrm{eq}}\left(\boldsymbol{r}+e_{i}\delta t,t-\delta t\right)+D_{\rho}\cdot\sum_{i}f_{i}^{\mathrm{eq}}\left(\boldsymbol{r}+2e_{i}\delta t,t-\delta t\right)
\end{aligned},
\label{eq:gslbm_density_auxiliary_function}
\end{equation}
\begin{equation}
\begin{aligned}
h_{m,\alpha}=&A_{m}\cdot\sum_{i}e_{i\alpha}f_{i}^{\mathrm{eq}}\left(\boldsymbol{r}-e_{i}\delta t,t-\delta t\right)+B_{m}\cdot\sum_{i}e_{i\alpha}f_{i}^{\mathrm{eq}}\left(\boldsymbol{r}-2e_{i}\delta t,t-\delta t\right)\\
&+C_{m}\cdot\sum_{i}e_{i\alpha}f_{i}^{\mathrm{eq}}\left(\boldsymbol{r}+e_{i}\delta t,t-\delta t\right)+D_{m}\cdot\sum_{i}e_{i\alpha}f_{i}^{\mathrm{eq}}\left(\boldsymbol{r}+2e_{i}\delta t,t-\delta t\right),
\end{aligned}
\label{eq:gslbm_momentum_auxiliary_function}
\end{equation}
where all quantities depend on the local macroscopic variables evaluated at the corresponding time step. To analyse the macroscopic behaviour implied by this construction, we perform the Taylor expansion in space around the current lattice node. Here, we denote $f_i^{\mathrm{eq}}(\boldsymbol{r},t-\delta t)=f_i^{\mathrm{eq}}$ for simplicity and use $s\in\{-2,-1,1,2\}$ for the stencil offset:
\begin{equation}
\begin{aligned}
f_{i}^{\mathrm{eq}}\left(\boldsymbol{r}-se_{i}\delta t,t-\delta t\right)=&f_{i}^{\mathrm{eq}}-s\delta t\left(e_{i\alpha}\partial_{\alpha}\right)f_{i}^{\mathrm{eq}}+\frac{1}{2}s^{2}\delta t^{2}\left(e_{i\alpha}\partial_{\alpha}\right)^{2}f_{i}^{\mathrm{eq}}\\
&-\frac{1}{6}s^{3}\delta t^{3}\left(\partial^{3}\right)+\frac{1}{24}s^{4}\delta t^{4}\left(\partial^{4}\right)+O\left(\delta t^{5}\right),
\end{aligned}
\end{equation}
with required derivative--moment relations given as
\begin{equation}
\begin{aligned}
\sum_{i}\left(e_{i\alpha}\partial_{\alpha}\right)f_{i}^{\mathrm{eq}}=&\partial_{\alpha}\rho u_{\alpha}\\
\sum_{i}e_{i\alpha}\left(e_{i\beta}\partial_{\beta}\right)f_{i}^{\mathrm{eq}}=&\partial_{\beta}\Pi_{\alpha\beta}^{\mathrm{eq}}\\
\sum_{i}\left(e_{i\alpha}\partial_{\alpha}\right)^{2}f_{i}^{\mathrm{eq}}=&\partial_{\alpha}\partial_{\beta}\Pi_{\alpha\beta}^{\mathrm{eq}}\\
\sum_{i}e_{i\alpha}\left(e_{i\beta}\partial_{\beta}\right)^{2}f_{i}^{\mathrm{eq}}=&\partial_{\beta}\partial_{\gamma}\Pi_{\alpha\beta\gamma}^{\mathrm{eq}}.
\end{aligned}
\end{equation}
These moment relations apply to any equilibrium-based reconstruction considered here. Based on the auxiliary functions, the predictor reconstruction operators and the resulting intermediate state are defined together as
\begin{equation}
\begin{aligned}
g_{\rho}&=h_{\rho}-N_{\rho}\rho,
&
g_{m,\alpha}&=h_{m,\alpha}-N_{m}\left(\rho u_{\alpha}\right),\\
\rho^{\ast}&=g_{\rho},
&
\left(\rho u_{\alpha}\right)^{\ast}&=g_{m,\alpha},
\qquad
u_{\alpha}^{\ast}=\frac{\left(\rho u_{\alpha}\right)^{\ast}}{\rho^{\ast}}.
\end{aligned}
\label{eq:gslbm_intermediate_state_definition}
\end{equation}
Here $\rho=\rho(\boldsymbol{r},t-\delta t)$ and $(\rho u_{\alpha})=(\rho u_{\alpha})(\boldsymbol{r},t-\delta t)$. Substituting the stencil definitions into these operators and applying the Taylor expansion above gives the intermediate density and momentum as
\begin{equation}
\begin{aligned}
\rho^{\ast}=&h_{\rho}-N_{\rho}\rho\\
=&A_{\rho}\cdot\sum_{i}f_{i}^{\mathrm{eq}}\left(\boldsymbol{r}-e_{i}\delta t,t-\delta t\right)+B_{\rho}\cdot\sum_{i}f_{i}^{\mathrm{eq}}\left(\boldsymbol{r}-2e_{i}\delta t,t-\delta t\right)\\
&+C_{\rho}\cdot\sum_{i}f_{i}^{\mathrm{eq}}\left(\boldsymbol{r}+e_{i}\delta t,t-\delta t\right)+D_{\rho}\cdot\sum_{i}f_{i}^{\mathrm{eq}}\left(\boldsymbol{r}+2e_{i}\delta t,t-\delta t\right)-N_{\rho}\rho\\
=&\tilde{t}_{0}\rho+\tilde{t}_{1}\delta t\partial_{\alpha}\rho u_{\alpha}+\tilde{t}_{2}\delta t^{2}\partial_{\alpha}\partial_{\beta}\Pi_{\alpha\beta}^{\mathrm{eq}}+\tilde{t}_{3}\delta t^{3}\left(\partial^{3}\right)+\tilde{t}_{4}\delta t^{4}\left(\partial^{4}\right)+O\left(\delta t^{5}\right),
\end{aligned}
\end{equation}
\begin{equation}
\begin{aligned}
\left(\rho u_{\alpha}\right)^{\ast}=&h_{m,\alpha}-N_{m}\cdot\left(\rho u_{\alpha}\right)\\
=&A_{m}\cdot\sum_{i}e_{i\alpha}f_{i}^{\mathrm{eq}}\left(\boldsymbol{r}-e_{i}\delta t,t-\delta t\right)+B_{m}\cdot\sum_{i}e_{i\alpha}f_{i}^{\mathrm{eq}}\left(\boldsymbol{r}-2e_{i}\delta t,t-\delta t\right)\\
&+C_{m}\cdot\sum_{i}e_{i\alpha}f_{i}^{\mathrm{eq}}\left(\boldsymbol{r}+e_{i}\delta t,t-\delta t\right)+D_{m}\cdot\sum_{i}e_{i\alpha}f_{i}^{\mathrm{eq}}\left(\boldsymbol{r}+2e_{i}\delta t,t-\delta t\right)-N_{m}\cdot\left(\rho u_{\alpha}\right)\\
=&t_{0}\rho u_{\alpha}+t_{1}\delta t\partial_{\beta}\Pi_{\alpha\beta}^{\mathrm{eq}}+t_{2}\delta t^{2}\partial_{\beta}\partial_{\gamma}\Pi_{\alpha\beta\gamma}^{\mathrm{eq}}+t_{3}\delta t^{3}\left(\partial^{3}\right)+t_{4}\delta t^{4}\left(\partial^{4}\right)+O\left(\delta t^{5}\right).
\end{aligned}
\end{equation}
Here, $\partial^{n}$ denotes the corresponding nth-order derivative tensor. For $q\in\{\rho,m\}$, we write $t_n^\rho=\tilde{t}_n$ and $t_n^m=t_n$. The density and momentum coefficients then satisfy the same linear system,
\begin{equation}
\begin{bmatrix}
1 & 1 & 1 & 1 & -1\\
-1 & -2 & 1 & 2 & 0\\
\frac{1}{2} & 2 & \frac{1}{2} & 2 & 0\\
-\frac{1}{6} & -\frac{4}{3} & \frac{1}{6} & \frac{4}{3} & 0\\
\frac{1}{24} & \frac{2}{3} & \frac{1}{24} & \frac{2}{3} & 0
\end{bmatrix}
\begin{bmatrix}
A_q\\ B_q\\ C_q\\ D_q\\ N_q
\end{bmatrix}
=
\begin{bmatrix}
t_0^q\\ t_1^q\\ t_2^q\\ t_3^q\\ t_4^q
\end{bmatrix}.
\label{eq:gslbm_predictor_coefficient_system}
\end{equation}
In the corrector step, the reconstruction uses an equilibrium distribution evaluated from the intermediate state:
\begin{equation}
f_{i}^{\mathrm{eq},\ast}\left(\boldsymbol{r},t\right)
=f_{i}^{\mathrm{eq}}\left(\rho^{\ast}\left(\boldsymbol{r},t\right),\left(u_{\alpha}\right)^{\ast}\left(\boldsymbol{r},t\right)\right).
\end{equation}
Due to the mathematical form of the equilibrium distribution, its moments satisfy
\begin{equation}
\begin{aligned}
\sum_{i}f_{i}^{\mathrm{eq},\ast}=&\rho^{\ast},\\
\sum_{i}e_{i\alpha}f_{i}^{\mathrm{eq},\ast}=&\left(\rho u_{\alpha}\right)^{\ast},\\
\sum_{i}e_{i\alpha}e_{i\beta}f_{i}^{\mathrm{eq},\ast}=&\Pi_{\alpha\beta}^{\mathrm{eq},\ast}=\left(\rho u_{\alpha}u_{\beta}+\rho c_{s}^{2}\delta_{\alpha\beta}\right)^{\ast},\\
\sum_{i}e_{i\alpha}e_{i\beta}e_{i\gamma}f_{i}^{\mathrm{eq},\ast}=&\Pi_{\alpha\beta\gamma}^{\mathrm{eq},\ast}=\left(\rho c_{s}^{2}\left(u_{\alpha}\delta_{\beta\gamma}+u_{\beta}\delta_{\alpha\gamma}+u_{\gamma}\delta_{\alpha\beta}\right)\right)^{\ast}.
\end{aligned}
\end{equation}

Since the predictor already recovers the continuity equation, only the momentum is corrected. Using the equilibrium distribution evaluated from the intermediate state, we define
\begin{equation}
\begin{aligned}
h_{m,\alpha}^{\ast}=&A_{m}^{\ast}\cdot\sum_{i}e_{i\alpha}f_{i}^{\mathrm{eq},\ast}\left(\boldsymbol{r}-e_{i}\delta t,t\right)+B_{m}^{\ast}\cdot\sum_{i}e_{i\alpha}f_{i}^{\mathrm{eq},\ast}\left(\boldsymbol{r}-2e_{i}\delta t,t\right)\\
&+C_{m}^{\ast}\cdot\sum_{i}e_{i\alpha}f_{i}^{\mathrm{eq},\ast}\left(\boldsymbol{r}+e_{i}\delta t,t\right)+D_{m}^{\ast}\cdot\sum_{i}e_{i\alpha}f_{i}^{\mathrm{eq},\ast}\left(\boldsymbol{r}+2e_{i}\delta t,t\right).
\end{aligned}
\label{eq:gslbm_corrector_momentum_auxiliary_function}
\end{equation}
Applying the same spatial Taylor expansion to this stencil gives the first four columns of the predictor matching system:
\begin{equation}
\begin{bmatrix}
1 & 1 & 1 & 1\\
-1 & -2 & 1 & 2\\
\frac{1}{2} & 2 & \frac{1}{2} & 2\\
-\frac{1}{6} & -\frac{4}{3} & \frac{1}{6} & \frac{4}{3}\\
\frac{1}{24} & \frac{2}{3} & \frac{1}{24} & \frac{2}{3}
\end{bmatrix}
\begin{bmatrix}
A_m^\ast\\ B_m^\ast\\ C_m^\ast\\ D_m^\ast
\end{bmatrix}
=
\begin{bmatrix}
t_0^\ast\\ t_1^\ast\\ t_2^\ast\\ t_3^\ast\\ t_4^\ast
\end{bmatrix}.
\label{eq:gslbm_corrector_coefficient_mapping}
\end{equation}
The corrector parameters must satisfy
\begin{equation}
\frac{t_0^\ast}{6}-\frac{5t_2^\ast}{12}+t_4^\ast=0.
\label{eq:gslbm_corrector_coefficient_compatibility}
\end{equation}
The coefficient $N_m^\ast$ does not enter this off-centre mapping; it multiplies the old-state momentum separately in the corrector reconstruction. The resulting $g^\ast$-operator is
\begin{equation}
g_{m,\alpha}^{\ast}=h_{m,\alpha}^{\ast}-N_{m}^{\ast}\cdot\left(\rho u_{\alpha}\right).
\end{equation}
By applying the Taylor expansion, we get
\begin{equation}
\begin{aligned}
g_{m,\alpha}^{\ast}=&t_{0}^{\ast}\left(\rho u_{\alpha}\right)^{\ast}-N_{m}^{\ast}\cdot\left(\rho u_{\alpha}\right)+t_{1}^{\ast}\delta t\partial_{\beta}\Pi_{\alpha\beta}^{\mathrm{eq},\ast}+t_{2}^{\ast}\delta t^{2}\partial_{\beta}\partial_{\gamma}\Pi_{\alpha\beta\gamma}^{\mathrm{eq},\ast}\\
&+t_{3}^{\ast}\delta t^{3}\left(\partial^{3}\right)+t_{4}^{\ast}\delta t^{4}\left(\partial^{4}\right)+O\left(\delta t^{5}\right).
\end{aligned}
\end{equation}

Combining the predictor and corrector reconstructions gives the generalized density reconstruction
\begin{equation}
\left(\rho\right)_{\mathrm{GSLBM}}\left(\boldsymbol{r},t\right)=h_{\rho}-N_{\rho}\cdot\rho,
\label{eq:gslbm_continuity_reconstruction}
\end{equation}
and the generalized momentum reconstruction
\begin{equation}
\left(\rho u_{\alpha}\right)_{\mathrm{GSLBM}}\left(\boldsymbol{r},t\right)=h_{m,\alpha}-N_{m}\cdot\left(\rho u_{\alpha}\right)+M\cdot\left(h_{m,\alpha}^{\ast}-N_{m}^{\ast}\cdot\left(\rho u_{\alpha}\right)\right)=g_{m,\alpha}+M\cdot g_{m,\alpha}^{\ast}.
\label{eq:gslbm_momentum_reconstruction}
\end{equation}
These equations represent the general equilibrium-based predictor--corrector reconstruction, with the predictor and corrector coefficients introduced above treated as parameters of the numerical scheme. Since the continuity equation in \eqnref{eq:gslbm_continuity_reconstruction} has already been addressed in \secref{sec:rslbm_equilibrium_formulation}, the subsequent analysis focuses on \eqnref{eq:gslbm_momentum_reconstruction}.

Combining the predictor expansion with the corrector expression above yields the general momentum equation recovered by the GSLBM:
\begin{equation}
\begin{aligned}
\left(\rho u_{\alpha}\right)_{\mathrm{GSLBM}}\left(\boldsymbol{r},t\right)=&t_{0}\rho u_{\alpha}+t_{1}\delta t\partial_{\beta}\Pi_{\alpha\beta}^{\mathrm{eq}}+t_{2}\delta t^{2}\partial_{\beta}\partial_{\gamma}\Pi_{\alpha\beta\gamma}^{\mathrm{eq}}\\
&+t_{3}\delta t^{3}\left(\partial^{3}\right)+t_{4}\delta t^{4}\left(\partial^{4}\right)+O\left(\delta t^{5}\right)\\
&+M\cdot\left[\begin{aligned}
&\left(t_{0}^{\ast}t_{0}-N_{m}^{\ast}\right)\left(\rho u_{\alpha}\right)\\
&\quad +\delta t\left(t_{1}^{\ast}\partial_{\beta}\Pi_{\alpha\beta}^{\mathrm{eq},\ast,\mathrm{GSLBM}}+t_{0}^{\ast}t_{1}\partial_{\beta}\Pi_{\alpha\beta}^{\mathrm{eq}}\right)\\
&\quad +\delta t^{2}\left(t_{2}^{\ast}\partial_{\beta}\partial_{\gamma}\Pi_{\alpha\beta\gamma}^{\mathrm{eq},\ast,\mathrm{GSLBM}}+t_{0}^{\ast}t_{2}\partial_{\beta}\partial_{\gamma}\Pi_{\alpha\beta\gamma}^{\mathrm{eq}}\right)\\
&\quad +\delta t^{3}\left(t_{3}^{\ast}\left(\partial^{3,\ast,\mathrm{GSLBM}}\right)+t_{0}^{\ast}t_{3}\left(\partial^{3}\right)\right)\\
&\quad +\delta t^{4}\left(t_{4}^{\ast}\left(\partial^{4,\ast,\mathrm{GSLBM}}\right)+t_{0}^{\ast}t_{4}\left(\partial^{4}\right)\right)+O\left(\delta t^{5}\right)
\end{aligned}\right]
\end{aligned}
\end{equation}
This equation contains the $O(M\delta t^2)$ contribution
\begin{equation}
M t_{0}^{\ast}t_{2}\delta t^{2}\partial_{\beta}\partial_{\gamma}\Pi_{\alpha\beta\gamma}^{\mathrm{eq}}.
\end{equation}
As shown below, $t_2$ is fixed by the target macroscopic equation and $t_0^\ast$ is nonzero. For a nonzero corrector prefactor $M$, this term cannot be eliminated by adjusting the remaining reconstruction coefficients. Division by $\delta t$ in the temporal update reduces its order by one power of $\delta t$. Its order under mesh refinement, however, cannot be inferred from this term alone, because the parameter scaling and other contributions with the same derivative structure must also be considered (see \appref{app:slbm_error_scaling}). This term therefore identifies a structural constraint of the generalized SLBM formulation rather than an unconditional global accuracy order. Its implications for the complete SLBM and SS-SLBM schemes are analysed in \appref{app:slbm_analysis}.

To connect this observation with the semi-discrete MAME, matching the low-order terms requires
\begin{equation}
\begin{aligned}
t_0&=1,
&
t_1&=-1,
&
t_2&=\frac{\nu}{c_s^2\delta t},\\
t_2^\ast&=0,
&
t_0^\ast t_0-N_m^\ast&=0,
&
t_1^\ast&=-t_0^\ast t_1.
\end{aligned}
\end{equation}
The first three relations follow from the MAME, while the remaining relations enforce the equilibrium-based corrector structure and a consistent approximation of the temporal derivative. Since $t_0=1$ and $t_1=-1$, these constraints give
\begin{equation}
t_0^\ast=t_1^\ast=N_m^\ast=t^\ast.
\end{equation}
The corrector construction considered here requires $t^\ast\ne0$.

The predictor coefficients are therefore obtained from \eqnref{eq:gslbm_predictor_coefficient_system}, with $q=m$ and
\begin{equation}
\left(t_0,t_1,t_2,t_3,t_4\right)
=\left(1,-1,\frac{\nu}{c_s^2\delta t},t_3,t_4\right).
\end{equation}
The corrector coefficients follow from \eqnref{eq:gslbm_corrector_coefficient_mapping}. Combining the low-order constraints above with \eqnref{eq:gslbm_corrector_coefficient_compatibility} gives
\begin{equation}
\left(t_0^\ast,t_1^\ast,t_2^\ast,t_3^\ast,t_4^\ast\right)
=\left(t^\ast,t^\ast,0,t_3^\ast,-\frac{t^\ast}{6}\right),
\qquad
N_m^\ast=t^\ast.
\end{equation}
Thus $t_4^\ast=-t^\ast/6\ne0$ is fixed by compatibility rather than being an independent higher-order parameter. The parameters $t_3$, $t_4$, and $t_3^\ast/t^\ast$ control the remaining higher-order freedom, whereas $t^\ast$ sets the common scale of the corrector coefficients. The semi-discrete MAME fixes only the product of this scale and the corrector prefactor:
\begin{equation}
M t^{\ast}=\frac{\nu}{c_s^2\delta t}-\frac{1}{2}.
\end{equation}
Moreover, because the predictor matches the MAME through second order, the divergence of the intermediate equilibrium-flux difference satisfies
\begin{equation}
\partial_\beta\left(
\Pi_{\alpha\beta}^{\mathrm{eq},\ast,\mathrm{GSLBM}}
-\Pi_{\alpha\beta}^{\mathrm{eq},\ast,\mathrm{MAME}}
\right)
\sim O\left(\delta t^{3}\right)^{\mathrm{pred}}.
\end{equation}
Substituting these relations into the general equation gives the recovered momentum directly relative to the semi-discrete MAME:
\begin{equation}
\begin{aligned}
\left(\rho u_{\alpha}\right)_{\mathrm{GSLBM}}^{\mathrm{MAME}}\left(\boldsymbol{r},t\right)
=&\left(\rho u_{\alpha}\right)_{\mathrm{MAME}}\left(\boldsymbol{r},t\right)
+\left(\frac{\nu}{c_s^2}-\frac{\delta t}{2}\right)O\left(\delta t^{3}\right)^{\mathrm{pred}}\\
&+\left[t_{3}\delta t^{3}\left(\partial^{3}\right)+t_{4}\delta t^{4}\left(\partial^{4}\right)+O\left(\delta t^{5}\right)\right]\\
&+\left(\frac{\nu}{c_{s}^{2}\delta t}-\frac{1}{2}\right)\cdot\left[\begin{aligned}
&\delta t^{2}\left(\frac{\nu}{c_{s}^{2}\delta t}\partial_{\beta}\partial_{\gamma}\Pi_{\alpha\beta\gamma}^{\mathrm{eq}}\right)
\\
&\quad +\delta t^{3}\left(\frac{t_{3}^{\ast}}{t^{\ast}}\left(\partial^{3,\ast,\mathrm{GSLBM}}\right)+t_{3}\left(\partial^{3}\right)\right)\\
&\quad +\delta t^{4}\left(\frac{t_{4}^{\ast}}{t^{\ast}}\left(\partial^{4,\ast,\mathrm{GSLBM}}\right)+t_{4}\left(\partial^{4}\right)\right)+O\left(\delta t^{5}\right)
\end{aligned}\right].
\end{aligned}
\label{eq:gslbm_mame_recovered_momentum}
\end{equation}
For a nonzero corrector prefactor, the recovered equation contains two structural contributions that cannot be removed by adjusting the remaining stencil parameters: the second-derivative term associated with $t_2$ and the nonzero fourth-order corrector term implied by $t_4^\ast=-t^\ast/6$. The special case $M t^\ast=0$ is discussed separately in \appref{app:slbm_analysis_tau_one}.

In summary, the generalized formulation reconstructs the macroscopic variables from equilibrium moments through a predictor--corrector procedure. Taylor matching fixes the low-order coefficients required by the target macroscopic equations, while the remaining stencil parameters control higher-order truncation effects. The structural terms identified above provide the basis for the SLBM and SS-SLBM analyses in \appref{app:slbm_analysis}.


\section{Theoretical analysis of the SLBM and SS-SLBM}\label{app:slbm_analysis}

Building upon the generalized SLBM formulation in \appref{app:gslbm_derivation}, this appendix applies the resulting analytical framework to the simplified lattice Boltzmann method (SLBM) and its variant (SS-SLBM). This analysis provides a concrete illustration of how the intrinsic limitations identified in \appref{app:gslbm_derivation} manifest in specific SLBM-type schemes.

\subsection{Analysis of the simplified lattice Boltzmann method (SLBM)}\label{app:slbm_analysis_slbm}

\subsubsection{Recovery of the macroscopic equation}\label{app:slbm_macroscopic_equation}

For the original SLBM, the reconstruction coefficients take the specific values
\begin{equation}
\begin{gathered}
\left \{A_m=1,B_m=0,C_m=0,D_m=0,N_m=0\right\}^{\mathrm{SLBM}},\\
\left \{A_m^{\ast}=0,B_m^{\ast}=0,C_m^{\ast}=1,D_m^{\ast}=0,N_m^{\ast}=1\right\}^{\mathrm{SLBM}}.
\end{gathered}
\end{equation}
The corresponding parameters in the generalized SLBM formulation are given by
\begin{equation}
\begin{gathered}
\left \{t_0=1,t_1=-1,t_2=\frac{1}{2},t_3=-\frac{1}{6},t_4=\frac{1}{24}\right\}^{\mathrm{SLBM}},\\
\left \{t_0^{\ast}=1,t_1^{\ast}=1,t_2^{\ast}=\frac{1}{2},t_3^{\ast}=\frac{1}{6},t_4^{\ast}=\frac{1}{24}\right\}^{\mathrm{SLBM}},\\
M^{\mathrm{SLBM}}=\frac{\nu}{c_s^2\delta t}-\frac{1}{2}
=\tau-1.
\end{gathered}
\end{equation}
It is evident that the coefficients $\{t_n,t_n^{\ast}\mid n\le2\}^{\mathrm{SLBM}}$ differ from those required by the MAME. This discrepancy reflects an essential physical difference between the SLBM and the MAME, rather than the choice of numerical parameters.

To write the repeated Taylor contributions compactly, define
\begin{equation}
\begin{aligned}
\mathcal{P}_{\alpha}^{\mathrm{SLBM}}
={}&-\frac{1}{6}\delta t^{2}\left(\partial^{3}\right)
+\frac{1}{24}\delta t^{3}\left(\partial^{4}\right)
+O\left(\delta t^{4}\right),\\
\mathcal{C}_{\alpha}^{\mathrm{SLBM}}
={}&\frac{1}{2}\partial_{\beta}\partial_{\gamma}
\left(
\Pi_{\alpha\beta\gamma}^{\mathrm{eq},\ast,\mathrm{SLBM}}
+\Pi_{\alpha\beta\gamma}^{\mathrm{eq}}
\right)\\
&+\frac{1}{6}\delta t
\left[
\left(\partial^{3,\ast,\mathrm{SLBM}}\right)
-\left(\partial^{3}\right)
\right]\\
&+\frac{1}{24}\delta t^{2}
\left[
\left(\partial^{4,\ast,\mathrm{SLBM}}\right)
+\left(\partial^{4}\right)
\right]
+O\left(\delta t^{3}\right).
\end{aligned}
\label{eq:appb_slbm_taylor_contributions}
\end{equation}
Here $\mathcal{P}_{\alpha}^{\mathrm{SLBM}}$ collects the third- and fourth-order predictor terms, whereas $\mathcal{C}_{\alpha}^{\mathrm{SLBM}}$ collects the second- through fourth-order corrector Taylor terms. Substituting the above SLBM parameters into the generalized formulation derived in \appref{app:gslbm_derivation} then gives
\begin{equation}
\begin{aligned}
\left(\rho u_{\alpha}\right)_{\mathrm{GSLBM}}^{\mathrm{SLBM}}
\left(\boldsymbol{r},t\right)
={}&\rho u_{\alpha}
-\delta t\partial_{\beta}\Pi_{\alpha\beta}^{\mathrm{eq}}
+\frac{1}{2}\delta t^{2}
\partial_{\beta}\partial_{\gamma}\Pi_{\alpha\beta\gamma}^{\mathrm{eq}}
+\delta t\mathcal{P}_{\alpha}^{\mathrm{SLBM}}\\
&+M^{\mathrm{SLBM}}
\left[
\delta t\partial_{\beta}
\left(
\Pi_{\alpha\beta}^{\mathrm{eq},\ast,\mathrm{SLBM}}
-\Pi_{\alpha\beta}^{\mathrm{eq}}
\right)
+\delta t^{2}\mathcal{C}_{\alpha}^{\mathrm{SLBM}}
\right].
\end{aligned}
\label{eq:appb_slbm_reconstructed_momentum}
\end{equation}

Let $\nu=\nu_{\mathrm{phy}}$ denote the physical viscosity. To quantify the discrepancy between the SLBM predictor step and that implied by the MAME, we define the difference between the equilibrium momentum-flux divergences evaluated from the corresponding intermediate states as
\begin{equation}
\mathcal{T}_{\alpha}^{\ast}
=\frac{\partial_{\beta}\left(
\Pi_{\alpha\beta}^{\mathrm{eq},\ast,\mathrm{SLBM}}
-\Pi_{\alpha\beta}^{\mathrm{eq},\ast,\mathrm{MAME}}
\right)}{\delta t}.
\label{eq:appb_slbm_intermediate_flux_difference}
\end{equation}
With the leading difference between the SLBM and MAME momentum predictors reads
\begin{equation*}
\begin{aligned}
\left(\rho u_\alpha\right)^{\ast}_{\mathrm{SLBM}}
-\left(\rho u_\alpha\right)^{\ast}_{\mathrm{MAME}}
={}&\left(\frac{1}{2}-\frac{\nu}{c_s^2\delta t}\right)
\delta t^2\partial_\beta\partial_\gamma
\Pi_{\alpha\beta\gamma}^{\mathrm{eq}}
+O\left(\delta t^3\right)\\
={}&-M^{\mathrm{SLBM}}\delta t^2
\partial_\beta\partial_\gamma
\Pi_{\alpha\beta\gamma}^{\mathrm{eq}}
+O\left(\delta t^3\right),
\end{aligned}
\end{equation*}
\eqnref{eq:appb_slbm_intermediate_flux_difference} satisfies $\mathcal{T}_{\alpha}^{\ast}\propto O(M^{\mathrm{SLBM}}\delta t)$. This quantity represents the effective deviation in the temporal derivative approximation induced by the SLBM predictor step when compared with the semi-discrete MAME. This difference originates from the parameter choices inherent to the SLBM formulation, rather than from the numerical discretization in space. In particular, the flux-divergence difference in \eqnref{eq:appb_slbm_reconstructed_momentum} can be separated as
\begin{equation}
\begin{aligned}
\partial_{\beta}\left(
\Pi_{\alpha\beta}^{\mathrm{eq},\ast,\mathrm{SLBM}}
-\Pi_{\alpha\beta}^{\mathrm{eq}}
\right)
={}&
\partial_{\beta}\left(
\Pi_{\alpha\beta}^{\mathrm{eq},\ast,\mathrm{MAME}}
-\Pi_{\alpha\beta}^{\mathrm{eq}}
\right)
+\delta t\mathcal{T}_{\alpha}^{\ast} .
\end{aligned}
\label{eq:appb_slbm_flux_decomposition}
\end{equation}

The effective viscosity associated with the SLBM predictor and its virtual contribution satisfy
\begin{equation}
\nu_{\mathrm{eff}}
=\nu_{\mathrm{phy}}+\nu_{\mathrm{virtual}},
\qquad
\nu_{\mathrm{virtual}}
=c_s^2\left(1-\tau\right)\delta t
=-c_s^2M^{\mathrm{SLBM}}\delta t.
\label{eq:appb_slbm_viscosities}
\end{equation}
The non-physical virtual viscosity is introduced by the equilibrium-based predictor reconstruction of the SLBM. It arises from the mismatch between the SLBM predictor parameters and those required by the MAME, and is therefore not associated with any physical dissipation mechanism.

Following the decomposition introduced in \eqnref{eq:slbm_error_decomposition}, the fully continuous form of \eqnref{eq:appb_slbm_reconstructed_momentum} can be organized as
\begin{equation}
\begin{aligned}
\partial_t\left(\rho u_{\alpha}\right)
\approx{}&
\frac{\left(\rho u_{\alpha}\right)\left(\boldsymbol{r},t\right)
-\left(\rho u_{\alpha}\right)\left(\boldsymbol{r},t-\delta t\right)}
{\delta t}\\
={}&\mathcal{M}_{\alpha}^{\mathrm{MAME}}
+\mathcal{D}_{\alpha}^{\mathrm{phys}}
+\mathcal{E}_{\alpha}^{\mathrm{num}},
\end{aligned}
\label{eq:appb_slbm_error_decomposition}
\end{equation}
where the physical-deviation and numerical-error terms are
\begin{equation}
\begin{aligned}
\mathcal{D}_{\alpha}^{\mathrm{phys}}
={}&
\frac{\nu_{\mathrm{virtual}}}{c_s^2}
\partial_{\beta}\partial_{\gamma}
\Pi_{\alpha\beta\gamma}^{\mathrm{eq}}
+M^{\mathrm{SLBM}}\delta t\,\mathcal{T}_{\alpha}^{\ast},\\
\mathcal{E}_{\alpha}^{\mathrm{num}}
={}&
\mathcal{P}_{\alpha}^{\mathrm{SLBM}}
+M^{\mathrm{SLBM}}\delta t\,
\mathcal{C}_{\alpha}^{\mathrm{SLBM}}.
\end{aligned}
\label{eq:appb_slbm_error_components}
\end{equation}

By comparing the recovered momentum equation with the MAME, the additional terms can be systematically classified into these two categories. The physical-deviation term $\mathcal{D}_{\alpha}^{\mathrm{phys}}$ consists of two contributions. The first is associated with the virtual viscosity introduced by the SLBM predictor step, whereas the second originates from the discrepancy between the SLBM predictor and the MAME-consistent predictor. Both terms arise from the mismatch in the coefficient $t_2$ between the SLBM and the MAME, and therefore represent physical deviations rather than numerical discretization errors.

In contrast, $\mathcal{E}_{\alpha}^{\mathrm{num}}$ contains the higher-order predictor contribution and the numerical errors introduced by the corrector. These numerical error terms originate from the structural limitation of the generalized SLBM formulation identified in \appref{app:gslbm_derivation}. Although \eqnref{eq:appb_slbm_error_components} distinguishes the physical deviations from the numerical errors, their orders must be determined after all terms in the recovered equation are combined.

\subsubsection{Error analysis under diffusive and acoustic scalings}\label{app:slbm_error_scaling}

We consider the same physical problem on all meshes by fixing $\mathrm{Re}$, the geometry, and the nondimensional initial and boundary conditions. The diffusive and acoustic scalings follow the standard lattice Boltzmann conventions \citep{kruger2017}.
From $c_s^2=\delta x^2/(3\delta t^2)$ and $\nu=c_s^2(\tau-1/2)\delta t$, one obtains
\begin{equation*}
\tau-\frac{1}{2}=\frac{3\nu\delta t}{\delta x^2}.
\end{equation*}
Equivalently, in lattice units, $\nu=(\tau-1/2)/3$ and $\mathrm{Re}=3U_0N/(\tau-1/2)$, where $N=L/\delta x$ and $U_0$ denotes the lattice-unit value of the characteristic velocity used in the numerical tests.

Under diffusive scaling, $\tau$ is fixed. At fixed $\mathrm{Re}$, this requires $U_0=O(N^{-1})=O(\delta x)$ and gives
\begin{equation*}
\delta t=O(\delta x^2),
\qquad
M^{\mathrm{SLBM}}=O(1),
\qquad
c_s^2=O(\delta x^{-2}).
\end{equation*}
Under acoustic scaling, the lattice-unit $U_0$ and $Ma$ are fixed. Keeping $\mathrm{Re}$ unchanged then requires $\tau-1/2=O(N)=O(\delta x^{-1})$, and hence
\begin{equation*}
\delta t=O(\delta x),
\qquad
M^{\mathrm{SLBM}}=O(\delta x^{-1}),
\qquad
c_s^2=O(1).
\end{equation*}

With these two scalings specified, the physical-deviation and numerical-error terms in \eqnref{eq:appb_slbm_error_components} must be combined before their orders are assessed. Using \eqnref{eq:appb_slbm_viscosities}, the recovered momentum equation can be regrouped as
\begin{equation}
\begin{aligned}
\partial_t\left(\rho u_{\alpha}\right)
\approx{}&
\frac{\left(\rho u_{\alpha}\right)\left(\boldsymbol{r},t\right)
-\left(\rho u_{\alpha}\right)\left(\boldsymbol{r},t-\delta t\right)}
{\delta t}\\
={}&\mathcal{M}_{\alpha}^{\mathrm{MAME}}
+M^{\mathrm{SLBM}}\delta t\,\mathcal{T}_{\alpha}^{\ast}
+\mathcal{P}_{\alpha}^{\mathrm{SLBM}}\\
&+M^{\mathrm{SLBM}}\delta t
\left[
\mathcal{C}_{\alpha}^{\mathrm{SLBM}}
-\partial_{\beta}\partial_{\gamma}
\Pi_{\alpha\beta\gamma}^{\mathrm{eq}}
\right].
\end{aligned}
\label{eq:appb_slbm_compact_recovered_equation}
\end{equation}
The virtual-viscosity contribution is therefore partially compensated by the leading numerical corrector term. After factoring out the common prefactor $M^{\mathrm{SLBM}}\delta t$, the second-derivative part of $M^{\mathrm{SLBM}}\delta t\left[\mathcal{C}_{\alpha}^{\mathrm{SLBM}}-\partial_{\beta}\partial_{\gamma}\Pi_{\alpha\beta\gamma}^{\mathrm{eq}}\right]$ in \eqnref{eq:appb_slbm_compact_recovered_equation} is
\begin{equation}
\begin{aligned}
&\frac{1}{2}\partial_{\beta}\partial_{\gamma}
\left(
\Pi_{\alpha\beta\gamma}^{\mathrm{eq},\ast,\mathrm{SLBM}}
+\Pi_{\alpha\beta\gamma}^{\mathrm{eq}}
\right)
-\partial_{\beta}\partial_{\gamma}
\Pi_{\alpha\beta\gamma}^{\mathrm{eq}}=
\frac{1}{2}\partial_{\beta}\partial_{\gamma}
\left(
\Pi_{\alpha\beta\gamma}^{\mathrm{eq},\ast,\mathrm{SLBM}}
-\Pi_{\alpha\beta\gamma}^{\mathrm{eq}}
\right).
\end{aligned}
\end{equation}
Since the difference scales as $(\rho u_{\alpha})^{\ast}-\rho u_{\alpha}\propto O(\delta t)$, the above residual is $O(\delta t)$. The orders of the three additional terms in \eqnref{eq:appb_slbm_compact_recovered_equation} are summarized in table~\ref{tab:appb_slbm_scaling_orders}.

Under diffusive scaling, all three additional terms are $O(\delta x^2)$, and the original SLBM is therefore formally second-order accurate under the smooth, low-Mach assumptions used here. This conclusion is consistent with the approximately second-order convergence observed in the Taylor--Green vortex test in \secref{sec:taylor_green_vortex}, where both $\tau$ and $\mathrm{Re}$ are held fixed during mesh refinement. Under acoustic scaling, the physical-deviation term $M^{\mathrm{SLBM}}\delta t\,\mathcal{T}_{\alpha}^{\ast}$ remains $O(1)$, while the term containing the compensation between the virtual viscosity and the numerical corrector is $O(\delta x)$; only $\mathcal{P}_{\alpha}^{\mathrm{SLBM}}$ remains $O(\delta x^2)$. Consequently, the original SLBM does not generally retain second-order accuracy under acoustic scaling. The second-order accuracy often observed in practice can be attributed to the fact that the correction terms are typically small \citep{chen202037} and therefore do not significantly degrade the overall convergence behaviour.

\FloatBarrier
\begingroup
\makeatletter
\def\fps@table{!htp}
\makeatother
\begin{table}
\begin{center}
\renewcommand{\arraystretch}{1.35}
\begin{tabular}{p{0.62\textwidth}cc}
Additional term in \eqnref{eq:appb_slbm_compact_recovered_equation}
& Diffusive scaling & Acoustic scaling\\
\midrule
$M^{\mathrm{SLBM}}\delta t\,\mathcal{T}_{\alpha}^{\ast}$
& $O(\delta x^2)$ & $O(1)$\\
$\mathcal{P}_{\alpha}^{\mathrm{SLBM}}$
& $O(\delta x^2)$ & $O(\delta x^2)$\\
$\displaystyle M^{\mathrm{SLBM}}\delta t
\left[
\mathcal{C}_{\alpha}^{\mathrm{SLBM}}
-\partial_{\beta}\partial_{\gamma}
\Pi_{\alpha\beta\gamma}^{\mathrm{eq}}
\right]$
& $O(\delta x^2)$ & $O(\delta x)$\\
\end{tabular}
\smallskip
\caption{Orders of the three additional terms in \eqnref{eq:appb_slbm_compact_recovered_equation} under diffusive and acoustic scalings at fixed $\mathrm{Re}$.}
\label{tab:appb_slbm_scaling_orders}
\end{center}
\end{table}
\endgroup
\FloatBarrier

\subsection{Analysis of the single-step simplified lattice Boltzmann method (SS-SLBM)}\label{app:slbm_analysis_ss_slbm}

For the SS-SLBM, the corresponding coefficients are given by
\begin{equation}
\left \{
A_m=\tau,B_m=0,C_m=\tau-1,D_m=0,N_m=2\left(\tau-1\right)
\right\}^{\mathrm{SS\text{-}SLBM}}.
\end{equation}
which lead to the parameter set
\begin{equation}
\left \{
t_0=1,t_1=-1,t_2=\tau-\frac{1}{2}
=\frac{\nu}{c_s^2\delta t},
t_3=-\frac{1}{6},
t_4=\frac{\tau}{12}-\frac{1}{24}
\right\}^{\mathrm{SS\text{-}SLBM}}.
\end{equation}
Since the SS-SLBM does not employ an explicit correction step, we formally assume that a set of parameters $\{t_0^{\ast},\allowbreak t_1^{\ast},\allowbreak t_2^{\ast},\allowbreak t_3^{\ast},\allowbreak t_4^{\ast}\}^{\mathrm{SS\text{-}SLBM}}$ can be introduced to recover the predictor--corrector approximation of the temporal derivative, with the additional constraint $t_0^{\ast}=t_1^{\ast}=t^{\ast}$. With this assumption, the relation
\begin{equation}
M^{\mathrm{SS\text{-}SLBM}}t^{\ast}
=\frac{\nu_{\mathrm{eff}}}{c_s^2\delta t}-\frac{1}{2}=0.
\end{equation}
is obtained. And, the reconstructed momentum at time $t$ can then be expressed as
\begin{equation}
\begin{aligned}
\left(\rho u_{\alpha}\right)_{\mathrm{SS\text{-}SLBM}}\left(\boldsymbol{r},t\right)
=&\rho u_{\alpha}-\delta t\partial_{\beta}\Pi_{\alpha\beta}^{\mathrm{eq}}
+\frac{\nu}{c_s^2\delta t}\delta t^2
\partial_{\beta}\partial_{\gamma}\Pi_{\alpha\beta\gamma}^{\mathrm{eq}}\\
&+t_3\delta t^3\left(\partial^3\right)
+t_4\delta t^4\left(\partial^4\right)
+O\left(\delta t^5\right)\\
&+\left(\frac{\nu_{\mathrm{eff}}}{c_s^2\delta t}-\frac{1}{2}\right)
\cdot\left[\begin{aligned}
&\delta t\left(\partial_{\beta}\Pi_{\alpha\beta}^{\mathrm{eq},\ast}
-\partial_{\beta}\Pi_{\alpha\beta}^{\mathrm{eq}}\right)\\
&\quad+\delta t^2\left(
t_2^{\ast}\partial_{\beta}\partial_{\gamma}
\Pi_{\alpha\beta\gamma}^{\mathrm{eq},\ast}
+t_0^{\ast}\frac{\nu}{c_s^2\delta t}
\partial_{\beta}\partial_{\gamma}
\Pi_{\alpha\beta\gamma}^{\mathrm{eq}}\right)\\
&\quad+\delta t^3\left(
\frac{t_3^{\ast}}{t^{\ast}}\left(\partial^{3,\ast}\right)
-\frac{1}{6}\left(\partial^3\right)\right)\\
&\quad+\delta t^4\left(
\frac{t_4^{\ast}}{t^{\ast}}\left(\partial^{4,\ast}\right)
+\left(\frac{\tau}{12}-\frac{1}{24}\right)\left(\partial^4\right)\right)
+O\left(\delta t^5\right)
\end{aligned}\right].
\end{aligned}
\end{equation}
Consequently, the recovered momentum equation reads
\begin{equation}
\begin{aligned}
\partial_t\left(\rho u_{\alpha}\right)
\approx&\frac{\left(\rho u_{\alpha}\right)\left(\boldsymbol{r},t\right)
-\left(\rho u_{\alpha}\right)\left(\boldsymbol{r},t-\delta t\right)}{\delta t}\\
=&-\partial_{\beta}\Pi_{\alpha\beta}^{\mathrm{eq}}
+\frac{\nu}{c_s^2}
\partial_{\beta}\partial_{\gamma}\Pi_{\alpha\beta\gamma}^{\mathrm{eq}}+\left(\frac{\nu}{c_s^2}-\frac{1}{2}\delta t\right)
\partial_t\partial_{\beta}\Pi_{\alpha\beta}^{\mathrm{eq}}\\
&+\frac{\nu_{\mathrm{virtual}}}{c_s^2}\partial_t\partial_{\beta}\Pi_{\alpha\beta}^{\mathrm{eq}}
+O\left(\delta t^{m},\partial^{l}\mid m\ge2,l\ge3\right)\\
&+\left(\frac{\nu_{\mathrm{eff}}}{c_s^2}-\frac{1}{2}\delta t\right)
O\left(\delta t^{m},\partial^{l},\partial^{l,\ast}\mid m\ge1,l\ge3\right)^{\ast}.
\end{aligned}
\end{equation}
Unlike the original SLBM, the SS-SLBM introduces a virtual viscosity to the temporal derivative term,
\begin{equation}
\nu_{\mathrm{eff}}=\nu+\nu_{\mathrm{virtual}}, \nu_{\mathrm{virtual}}=c_s^2\left(1-\tau\right)\delta t.
\end{equation}

\subsection{Discussion of the special case of \texorpdfstring{$\tau=1$}{tau=1}}\label{app:slbm_analysis_tau_one}
It is worth noting that when $\tau=1$, all physical deviations and leading-order numerical discrepancies vanish. In this case, the coefficient associated with the temporal derivative term becomes zero at second-order accuracy, which implies that the correction step is numerically eliminated. As a result, with second-order truncation, the SLBM, the SS-SLBM, and the semi-discrete MAME recover the same macroscopic equation.

Moreover, owing to the specific construction of the SLBM and SS-SLBM, their formulations reduce to a form that is formally identical to the BGK-LBM at $\tau=1$. Specifically,
\begin{equation}
\begin{aligned}
\left(\rho u_{\alpha}\right)_{\mathrm{LBM}}\left(\boldsymbol{r},t\right)
=&\sum_i e_{i\alpha}f_i\left(\boldsymbol{r},t\right)\\
=&\sum_i e_{i\alpha}\left[
f_i^{\mathrm{eq}}\left(\boldsymbol{r}-e_i\delta t,t-\delta t\right)
+\left(1-\frac{1}{\tau}\right)
f_i^{\mathrm{neq}}\left(\boldsymbol{r}-e_i\delta t,t-\delta t\right)
\right]\\
=&\sum_i e_{i\alpha}
f_i^{\mathrm{eq}}\left(\boldsymbol{r}-e_i\delta t,t-\delta t\right)\\
=&\left(\rho u_{\alpha}\right)_{\mathrm{SLBM}}\left(\boldsymbol{r},t\right)
=\left(\rho u_{\alpha}\right)_{\mathrm{SS\text{-}SLBM}}\left(\boldsymbol{r},t\right).
\end{aligned}
\end{equation}
This observation provides additional support for the viewpoint proposed by \citep{lu2020}, namely that the macroscopic equations recovered from the LBM inherently contain small temporal correction terms
\begin{equation}
-\frac{1}{2}\delta t\partial_t\partial_{\beta}
\left(\rho u_{\alpha}u_{\beta}+\rho c_s^2\delta_{\alpha\beta}\right).
\end{equation}
Since the SLBM (also the SS-SLBM) coincides with the BGK-LBM in form at $\tau=1$, the corresponding macroscopic equations must necessarily include the same higher-order temporal contribution. Consequently, in this special condition, the SS-SLBM, the SLBM, the MAME, and the BGK-LBM are theoretically equivalent and recover the same macroscopic equations.


\section{Derivation of the unified predictor-corrector scheme for the MAME}\label{app:mame_unified_scheme}

If only second-order accuracy is required, the $O(\partial^2)$ term in \eqnref{eq:gslbm_mame_recovered_momentum} can in principle be eliminated by introducing an additional support term. Specifically, one may consider the modified reconstruction
\begin{equation}
\left(\rho u_{\alpha}\right)^{\mathrm{new}}\left(\boldsymbol{r},t\right)
=h_{m,\alpha}-N_m\cdot\left(\rho u_{\alpha}\right)
+M\cdot\left(h_{m,\alpha}^{\ast}-t_0^{\ast}\rho u_{\alpha}\right)
-\left(h_{m,\alpha}^{s}-N_m^s\cdot\left(\rho u_{\alpha}\right)\right),
\end{equation}
where $h_{m,\alpha}^s$ is associated with a set of undetermined coefficients $\{A_m^s,\allowbreak B_m^s,\allowbreak C_m^s,\allowbreak D_m^s,\allowbreak N_m^s\}$. These coefficients cancel the leading second-order temporal error while retaining the higher-order parameters $t_3^s$ and $t_4^s$. Using the coefficient matrix in \eqnref{eq:gslbm_predictor_coefficient_system}, with coefficient order $(A_m^s,B_m^s,C_m^s,D_m^s,N_m^s)$, this requirement is specified by the right-hand side
\begin{equation}
\begin{bmatrix}
0\\
0\\
\dfrac{\nu}{c_s^2\delta t}
\left(\dfrac{\nu}{c_s^2\delta t}-\dfrac{1}{2}\right)\\
t_3^s\\
t_4^s
\end{bmatrix}.
\end{equation}
This preliminary support set is not carried unchanged into the unified construction below; after the support term is placed inside the factor $M$, its coefficients are re-determined. Nevertheless, the preliminary modification does not eliminate the contribution involving $t_4^{\ast}\partial^{4,\ast}$. To control the second-order and higher-order terms within one construction, we therefore consider the following predictor--corrector structure:
\begin{equation}
\left(\rho u_{\alpha}\right)^{\mathrm{unified}}\left(\boldsymbol{r},t\right)
=h_{m,\alpha}-N_m\cdot\left(\rho u_{\alpha}\right)
+M\cdot\left[
h_{m,\alpha}^{\ast}-N_m^{\ast}\cdot\left(\rho u_{\alpha}\right)^{\ast}
-\left(h_{m,\alpha}^{s}-N_m^s\cdot\left(\rho u_{\alpha}\right)\right)
\right],
\end{equation}
where $h_{m,\alpha}$, $h_{m,\alpha}^{\ast}$, and $h_{m,\alpha}^s$ correspond to three independent sets of coefficients $\{A_m,\allowbreak B_m,\allowbreak C_m,\allowbreak D_m,\allowbreak N_m\}$, $\{A_m^{\ast},\allowbreak B_m^{\ast},\allowbreak C_m^{\ast},\allowbreak D_m^{\ast},\allowbreak N_m^{\ast}\}$ and $\{A_m^s,\allowbreak B_m^s,\allowbreak C_m^s,\allowbreak D_m^s,\allowbreak N_m^s\}$, respectively. Here,
\begin{equation}
M=\frac{\nu}{c_s^2\delta t}-\frac{1}{2}.
\end{equation}
The parameters $\{A_m,B_m,C_m,D_m,N_m\}$ associated with the predictor step are constrained by the physical requirements of the MAME as in \appref{app:gslbm_derivation}. For the unified construction, the starred and support coefficient sets are re-determined independently using the common coefficient matrix in \eqnref{eq:gslbm_predictor_coefficient_system}. With coefficient orders $(A_m^\ast,B_m^\ast,C_m^\ast,D_m^\ast,N_m^\ast)$ and $(A_m^s,B_m^s,C_m^s,D_m^s,N_m^s)$, respectively, their right-hand sides are
\begin{equation}
\begin{aligned}
\text{starred set:}\quad
&
\begin{bmatrix}
0\\ 1\\ 0\\ t_3^\ast\\ t_4^\ast
\end{bmatrix},
&
\text{support set:}\quad
&
\begin{bmatrix}
0\\ 1\\ 0\\ t_3^s\\ t_4^s
\end{bmatrix}.
\end{aligned}
\end{equation}
In each set, the leading zero includes the centre subtraction; for example, it imposes $N_m^\ast=A_m^\ast+B_m^\ast+C_m^\ast+D_m^\ast$ and does not set the off-centre sum to zero. The second entry fixes the first-derivative coefficient to unity. Thus these are full $5\times5$ systems, not the off-centre $5\times4$ corrector mapping used in \appref{app:gslbm_derivation}.

Once the higher-order parameters $\{t_3,\allowbreak t_4,\allowbreak t_3^{\ast},\allowbreak t_4^{\ast},\allowbreak t_3^s,\allowbreak t_4^s\}$ are specified, all coefficients are uniquely fixed and the resulting numerical scheme is fully determined. The modified formulation introduces alternative combinations of higher-order terms, whose relative contributions are governed by the chosen parameters. 

It should be emphasized that an accurate implementation of the corrector step within this unified predictor--corrector formulation entails a substantial increase in computational cost, memory usage, and algorithmic complexity. For the sake of numerical efficiency and clarity, the present study restricts this formulation to theoretical analysis, while finite-difference discretization is employed in the numerical experiments.

Finally, this scheme may be interpreted as a general second-order accurate predictor--corrector approach applicable to a class of macroscopic equations characterized by different parametric sets $\{t_n,t_n^{\ast}\mid n\le2\}$. Among these, only the MAME is considered to be the physically consistent macroscopic equation in this article, as its derivation is firmly rooted in the underlying BGK lattice Boltzmann equation.


\section{Construction of the amplification matrices}\label{app:amplification_matrix_construction}

This appendix gives the algebraic construction of the amplification matrices used in \secref{sec:linear_stability_analysis}. The algebraic expressions were checked symbolically. The construction is carried out in the conserved macroscopic variables $\boldsymbol{Q}={(\rho,j_x,j_y)}^{\mathrm{T}}$, while the equilibrium distribution and equilibrium moment tensors enter through their Jacobians at the uniform base state.

\subsection{Linearization and equilibrium-based Jacobians}\label{app:lsa_equilibrium_jacobians}

We use the conserved vector in \eqnref{eq:lsa_macroscopic_variables} and its uniform base state, together with the Fourier perturbation in \eqnref{eq:lsa_fourier_perturbation}. The one-step amplification problem and the $\lambda_{m}-\omega_{m}$  convention are those of \eqnref{eq:lsa_amplification_problem} and \eqnref{eq:lsa_lambda_omega_convention}, respectively. With this convention, non-amplification corresponds to $\operatorname{Im}(\omega_m)\le0$, or equivalently $|\lambda_m|\le1$. For each lattice velocity $\boldsymbol{e}_i$, the Fourier phase required below is
\begin{equation}
\theta_i=\boldsymbol{k}\cdot\boldsymbol{e}_i .
\end{equation}

The equilibrium distribution is first written as a function of the conserved variables. For the D2Q9 lattice, let $\boldsymbol{e}_i=(e_{ix},e_{iy})$ and define the equilibrium-distribution Jacobian and its linearized perturbation by
\begin{equation}
\begin{aligned}
J_{i\ell}
&=
\left.
\frac{\partial f_i^{\mathrm{eq}}}{\partial Q_\ell}
\right|_{\boldsymbol{Q}=\boldsymbol{Q}_0},
&
\ell&\in\{\rho,x,y\},\\
\delta f_i^{\mathrm{eq}}
&=
\sum_\ell J_{i\ell}\delta Q_\ell .
\end{aligned}
\end{equation}
All amplification matrices below are therefore constructed in the macroscopic-variable space, not in the $f_i^{\mathrm{eq}}$ space.

The finite-difference corrector also requires the equilibrium second-moment tensor and its equilibrium-moment Jacobian at the base state:
\begin{equation}
\begin{aligned}
\Pi_{\alpha\beta}^{\mathrm{eq}}
&=
\sum_i e_{i\alpha}e_{i\beta}f_i^{\mathrm{eq}},\\
B_{\alpha\beta,\ell}
&=
\left.
\frac{\partial \Pi_{\alpha\beta}^{\mathrm{eq}}}
{\partial Q_\ell}
\right|_{\boldsymbol{Q}=\boldsymbol{Q}_0}.
\end{aligned}
\end{equation}

\subsection{Amplification matrices of the RSLBM and SLBM}

\subsubsection{RSLBM}\label{app:lsa_rslbm_matrices}

Under the lattice-unit convention of \secref{sec:linear_stability_analysis}, $c_s^2=1/3$ and $\delta t=1$. The RSLBM momentum-predictor coefficient and corrector factor are, respectively,
\begin{equation}
\begin{aligned}
t_2^m&=3\nu,\\
3\nu-\frac{1}{2}&=\tau-1.
\end{aligned}
\end{equation}

The generalized predictor uses the offsets $s\in\{-2,-1,1,2\}$ together with the centre subtraction introduced in \appref{app:gslbm_derivation}. For each reconstructed quantity $q\in\{\rho,m\}$, where $q=\rho$ denotes density and $q=m$ denotes momentum, define the off-centre Fourier stencil symbol by
\begin{equation}
\chi_i^q(\boldsymbol{k})
=
A_q\exp(-\mathrm{i}\theta_i)
+
B_q\exp(-2\mathrm{i}\theta_i)
+
C_q\exp(\mathrm{i}\theta_i)
+
D_q\exp(2\mathrm{i}\theta_i).
\label{eq:appd_chi_definition}
\end{equation}

The coefficients $(A_q,B_q,C_q,D_q,N_q)$ satisfy the common moment-matching system in \eqnref{eq:gslbm_predictor_coefficient_system}. Solving that system gives

\begin{equation}
\begin{aligned}
A_q
&=
-\frac{2t_1^q-4t_2^q-3t_3^q+12t_4^q}{3},\\
B_q
&=
\frac{t_1^q-t_2^q-6t_3^q+12t_4^q}{12},\\
C_q
&=
\frac{2t_1^q+4t_2^q-3t_3^q-12t_4^q}{3},\\
D_q
&=
-\frac{t_1^q+t_2^q-6t_3^q-12t_4^q}{12},\\
N_q
&=
\frac{-2t_0^q+5t_2^q-12t_4^q}{2}.
\end{aligned}
\label{eq:appd_abcdn_solution}
\end{equation}
The density and momentum predictors use different moment-matched coefficient sets. In the notation of \secref{sec:linear_stability_analysis}, they are
\begin{equation}
\begin{aligned}
(t_0^\rho,t_1^\rho,t_2^\rho,t_3^\rho,t_4^\rho)
&=
\left(1,-1,\frac{1}{2},0,0\right),\\
(t_0^m,t_1^m,t_2^m,t_3^m,t_4^m)
&=
(1,-1,3\nu,t_3,t_4).
\end{aligned}
\end{equation}
With these definitions, $\chi_i^\rho$ and $\chi_i^m$ contain only the off-centre shifts, while $N_\rho$ and $N_m$ are the centre-subtraction coefficients. The RSLBM predictor matrix is
\begin{equation}
\begin{aligned}
P_{\rho\ell}(\boldsymbol{k})
&=
\sum_i \chi_i^\rho(\boldsymbol{k})J_{i\ell}
-N_\rho\delta_{\rho\ell},\\
P_{\alpha\ell}(\boldsymbol{k})
&=
\sum_i e_{i\alpha}\chi_i^m(\boldsymbol{k})J_{i\ell}
-N_m\delta_{\alpha\ell},
\qquad
\alpha\in\{x,y\}.
\end{aligned}
\end{equation}
Equivalently, the linearized predictor relation is
\begin{equation}
\widehat{\boldsymbol{Q}}^{\,*}
=
\mathsfbi{P}_{\mathrm{RSLBM}}(\boldsymbol{k})\widehat{\boldsymbol{Q}}^{\,n}.
\end{equation}

For the fourth-order centred first derivative used in the corrector, the Fourier symbol is
\begin{equation}
\mathcal D_\beta(\boldsymbol{k})
=
\frac{\mathrm{i}}{3}\sin(k_\beta)\left[4-\cos(k_\beta)\right].
\end{equation}
The corrector matrix has no density row, while its momentum rows use the same finite-difference symbol:
\begin{equation}
\begin{aligned}
R_{\rho\ell}
&=0,\\
R_{\alpha\ell}(\boldsymbol{k})
&=
\left(3\nu-\frac{1}{2}\right)
\sum_{\beta=x,y}
\mathcal D_\beta(\boldsymbol{k})B_{\alpha\beta,\ell},
\qquad
\alpha\in\{x,y\}.
\end{aligned}
\end{equation}
The linearized corrector acts on the difference between the intermediate and current states:
\begin{equation}
\widehat{\boldsymbol{Q}}^{\,n+1}
=
\widehat{\boldsymbol{Q}}^{\,*}
+
\mathsfbi{R}_{\mathrm{RSLBM}}(\boldsymbol{k})
\left(\widehat{\boldsymbol{Q}}^{\,*}-\widehat{\boldsymbol{Q}}^{\,n}\right).
\end{equation}
Substituting $\widehat{\boldsymbol{Q}}^{\,*}=\mathsfbi{P}_{\mathrm{RSLBM}}\widehat{\boldsymbol{Q}}^{\,n}$ gives
\begin{equation}
\mathsfbi{G}_{\mathrm{RSLBM}}(\boldsymbol{k})
=
\mathsfbi{P}_{\mathrm{RSLBM}}(\boldsymbol{k})
+
\mathsfbi{R}_{\mathrm{RSLBM}}(\boldsymbol{k})
\left[
\mathsfbi{P}_{\mathrm{RSLBM}}(\boldsymbol{k})-\mathsfbi{I}
\right].
\label{eq:appd_rslbm_amplification}
\end{equation}
This is the compact form used in \secref{sec:lsa_formulation}.

\subsubsection{SLBM}\label{app:lsa_slbm_matrix}

The original SLBM can be written in the same macroscopic-variable setting. Its predictor is the backward equilibrium reconstruction. With
\begin{equation}
E_i^-(\boldsymbol{k})=\exp(-\mathrm{i}\theta_i),
\end{equation}
the predictor matrix is
\begin{equation}
\begin{aligned}
P_{\rho\ell}^{\mathrm{SLBM}}(\boldsymbol{k})
&=
\sum_i J_{i\ell}E_i^-(\boldsymbol{k}),
\\
P_{\alpha\ell}^{\mathrm{SLBM}}(\boldsymbol{k})
&=
\sum_i e_{i\alpha}J_{i\ell}E_i^-(\boldsymbol{k}),
\qquad
\alpha\in\{x,y\}.
\end{aligned}
\end{equation}
The corrector uses a forward equilibrium reconstruction from the intermediate state. With
\begin{equation}
E_i^+(\boldsymbol{k})=\exp(+\mathrm{i}\theta_i),
\end{equation}
define
\begin{equation}
\begin{aligned}
C_{\rho\ell}^{\mathrm{SLBM}}(\boldsymbol{k})
&=
\sum_i J_{i\ell}E_i^+(\boldsymbol{k}),
\\
C_{\alpha\ell}^{\mathrm{SLBM}}(\boldsymbol{k})
&=
\sum_i e_{i\alpha}J_{i\ell}E_i^+(\boldsymbol{k}),
\qquad
\alpha\in\{x,y\}.
\end{aligned}
\end{equation}
The opposite signs in $E_i^-$ and $E_i^+$ are the Fourier expression of the backward predictor and forward corrector. Since the SLBM corrector updates only the momentum components, its selector and corrector coefficient are
\begin{equation}
\mathsfbi{S}=\operatorname{diag}(0,1,1),
\qquad
\sigma_\nu=3\nu-\frac{1}{2}=\tau-1.
\end{equation}
The complete SLBM amplification matrix is
\begin{equation}
\mathsfbi{G}_{\mathrm{SLBM}}(\boldsymbol{k})
=
\mathsfbi{P}_{\mathrm{SLBM}}(\boldsymbol{k})
+
\sigma_\nu\mathsfbi{S}
\left[
\mathsfbi{C}_{\mathrm{SLBM}}(\boldsymbol{k})
\mathsfbi{P}_{\mathrm{SLBM}}(\boldsymbol{k})
-
\mathsfbi{I}
\right].
\label{eq:appd_slbm_amplification}
\end{equation}
The order $\mathsfbi{C}_{\mathrm{SLBM}}\mathsfbi{P}_{\mathrm{SLBM}}$ is essential: the forward reconstruction in the corrector acts on the intermediate state, not directly on the current state. The scalar $\sigma_\nu$ is only the SLBM corrector coefficient; it should not be read as an additional physical viscosity.

\subsection{Eigenvector-based modal identification}\label{app:lsa_modal_identification}

The eigenvalues of $\mathsfbi{G}$ have no intrinsic physical ordering. The shear and acoustic branches are therefore identified by projecting the eigenvectors onto target directions. For an eigenvector $\boldsymbol{v}_m={(\delta\rho,\delta j_x,\delta j_y)}^{\mathrm{T}}$, the corresponding primitive perturbation is
\begin{equation}
\delta u_x
=
\frac{\delta j_x-U_0\delta\rho}{\rho_0},
\qquad
\delta u_y
=
\frac{\delta j_y-V_0\delta\rho}{\rho_0}.
\end{equation}
Thus $\widetilde{\boldsymbol{v}}_m$ is obtained from $\boldsymbol{v}_m$ by the base-state-dependent linear transformation
\begin{equation}
\widetilde{\boldsymbol{v}}_m
=
\begin{pmatrix}
1 & 0 & 0\\
-U_0/\rho_0 & 1/\rho_0 & 0\\
-V_0/\rho_0 & 0 & 1/\rho_0
\end{pmatrix}
\boldsymbol{v}_m .
\label{eq:appd_conserved_to_primitive_eigenvector}
\end{equation}
The two vectors represent the same eigenmode in different variable sets: $\boldsymbol{v}_m$ is expressed in conserved variables ${(\delta\rho,\delta j_x,\delta j_y)}^{\mathrm{T}}$, whereas $\widetilde{\boldsymbol{v}}_m$ is expressed in primitive perturbations ${(\delta\rho,\delta u_x,\delta u_y)}^{\mathrm{T}}$.

For nonzero $\boldsymbol{k}$, define the longitudinal and transverse directions and the corresponding shear and acoustic target vectors by
\begin{equation}
\begin{aligned}
\boldsymbol{n}&=\frac{\boldsymbol{k}}{|\boldsymbol{k}|},
&
\boldsymbol{t}&=(-n_y,n_x),\\
\widehat{\boldsymbol{v}}_{s}
&={(0,t_x,t_y)}^{\mathrm{T}},\\
\widehat{\boldsymbol{v}}_{a+}
&={\left(1,\frac{c_s n_x}{\rho_0},\frac{c_s n_y}{\rho_0}\right)}^{\mathrm{T}},
&
\widehat{\boldsymbol{v}}_{a-}
&={\left(1,-\frac{c_s n_x}{\rho_0},-\frac{c_s n_y}{\rho_0}\right)}^{\mathrm{T}}.
\end{aligned}
\end{equation}
The projection score for target $a\in (s,a+,a-)$ and numerical eigenvector $m$ is
\begin{equation}
s_{am}
=
\frac{|\langle \widehat{\boldsymbol{v}}_a,\widetilde{\boldsymbol{v}}_m\rangle|}
{\|\widehat{\boldsymbol{v}}_a\|\,\|\widetilde{\boldsymbol{v}}_m\|},
\end{equation}
where $\widetilde{\boldsymbol{v}}_m$ denotes the primitive-variable representation of $\boldsymbol{v}_m$. The physical labels are assigned by the one-to-one matching that maximizes the total score over the shear, acoustic-plus and acoustic-minus targets.

The projection score is also used as a branch-assignment quality measure. In the branch-resolved scans and plots, an assigned physical label is treated as reliable only when the corresponding assigned score satisfies
\begin{equation}
s_{a,m(a)}\ge 0.9 .
\end{equation}
Here $m(a)$ is the eigenmode assigned to target $a$. Thus the identification procedure is: first find the one-to-one assignment that maximizes the total projection score, and then accept the assigned branch label for physical interpretation only when its score meets the confidence threshold.

\end{appen}

\bibliographystyle{preprint}
\bibliography{references}

\end{document}